\documentclass[floatfix,prd,aps,twocolumn,a4paper,showkeys,nofootinbib,showpacs,10pt]{revtex4-1}

\usepackage[colorlinks=true,linkcolor=blue,linktocpage=true,anchorcolor=black,citecolor=blue,filecolor=black,menucolor=black, urlcolor=blue]{hyperref}
\usepackage{graphicx,psfrag}
\usepackage{grffile} 
\usepackage{mathrsfs}
\usepackage{amsmath,amsfonts,amssymb}
\usepackage{tasks} 
\usepackage{multirow}
\usepackage{comment}
\usepackage{bm}
\usepackage{paralist}
\usepackage{ulem}
\usepackage{booktabs}
\usepackage{xcolor}
\usepackage{caption}
\usepackage{subcaption}
\usepackage{multirow} 
\usepackage{enumitem}
\usepackage{bm} 
\usepackage[T1]{fontenc}
\usepackage{soul}

\newcommand{\be}{\begin{equation}}
\newcommand{\ee}{\end{equation}}
\newcommand{\ba}{\begin{align}}
\newcommand{\ea}{\end{align}}

\newcommand{\ord}{\mathcal{O}}
\newcommand{\f}{\frac}

\newcommand{\nn}{\nonumber}
\newcommand{\mbf}[1]{\mathbf{#1}}

\newcommand{\Lhat}{\!\hat{\,\mathbf{L}}_\text{N}}

\newcommand{\Sa}{\mbf{S}_1}
\newcommand{\Sb}{\mbf{S}_2}
\newcommand{\Lhatdot}{\dot{\hat{\,\mathbf{L}}}_\text{N}}

\newcommand{\LN}{\mbf{L}_\text{N}}

\newcommand{\J}{\mbf{J}}

\renewcommand{\L}{\mbf{L}}

\newcommand{\Qone}{1.\bar{1}}

\def\i{{\rm i}}

\def\Msun{M_{\odot}}
\def\Ml{37.5M_{\odot}}
\def\Mh{150M_{\odot}}
\def\Mdet{M^\text{det}}
\def\Mtot{M}
\def\GMc2{G M_{\odot} c^{-2}}

\def\M{{\cal M}}
\def\MM{{\bar{\cal M}}}

\def\R{{\mathcal{R}}}

\def\git{\texttt{git}}

\def\Msun{M_\odot}
\def\bajes{\texttt{bajes}}

\def\TEOBResumS{\texttt{TEOBResumS}}

\def\TEOB{\texttt{TEOB}}

\def\PhenomPv3HM{\texttt{PhenomPv3HM}}
\def\PhenomXPHM{\texttt{IMRPhenomXPHM}}
\def\XPHM{\texttt{XPHM}}
\def\PhenomTPHM{\texttt{IMRPhenomTPHM}}
\def\TPHM{\texttt{TPHM}}

\def\SEOB{{\texttt{SEOB}}}

\def\NRsurP{{\texttt{NRSur7dq4}}}

\def\NR{{\texttt{NRSur}}}

\def\SXS{\texttt{SXS}}

\newcommand{\MMo}{\MM_\text{opt}}
\newcommand{\MMno}{\MMo^\text{noMR}}
\newcommand{\io}{\vartheta_\text{LN}}
\newcommand{\ioo}{\vartheta_{\text{LN},0}}
\newcommand{\noMR}{noMR}
\newcommand{\fifty}{\texttt{0050}}
\newcommand{\ste}{\texttt{0628}}

\DeclareSymbolFontAlphabet{\mathrsfs}{rsfs}
\DeclareMathAlphabet{\mathcal}{OMS}{cmsy}{m}{n}
\DeclareSymbolFontAlphabet{\mathrsfs}{rsfs}
\DeclareMathAlphabet\mathbfcal{OMS}{cmsy}{b}{n}

\usepackage{color}
\definecolor{cyan}{rgb}{0,0.9,0.9}
\definecolor{orange}{rgb}{0.9,0.5,0}
\definecolor{magenta}{rgb}{1,0,1}
\definecolor{purple}{rgb}{0.8,0.4,0.8}
\definecolor{gray}{rgb}{0.8242,0.8242,0.8242}
\definecolor{dodgerblue}{rgb}{0.12, 0.56, 1.0}

\newcommand{\SA}[1]{{\textcolor{dodgerblue}{{SA: #1}} }}

\newcommand{\todo}[1]{\textcolor{orange}{\texttt{TODO: #1}}}

\newcommand\B{\rule[-1.2ex]{0pt}{0pt}} 
\newcommand\T{\rule{0pt}{2.6ex}}       

\newcommand{\chp}{\chi_\perp}
\newcommand{\chip}{\chi_\text{p}}
\newcommand{\chipGen}{\chi_\text{p}^\text{Gen}}
\newcommand{\chiperp}{\chi_{\perp,\text{J}}}

\newcommand{\chieff}{\chi_{\mathrm{eff}}}

\newcommand{\Caltech}{Theoretical Astrophysics Group, California Institute of Technology, Pasadena, CA 91125, U.S.A.}

\begin{document}

\title{A Survey of Four Precessing Waveform Models for Binary Black Hole Systems}

\author{Jake \surname{Mac Uilliam}${}^{1}$}
\author{Sarp \surname{Ak\c{c}ay}$^{1}$}
\author{Jonathan E. \surname{Thompson}$^{2}$}

\affiliation{${}^1$University College Dublin, Belfield, D4, Dublin, Ireland}
\affiliation{${}^2$\Caltech}

\normalem 

\begin{abstract}
Angular momentum and spin precession are expected to be generic features of a significant fraction of binary black hole systems.
As such, it is essential to have waveform models that faithfully incorporate the effects of precession.
Here, we assess how well the current state of the art models achieve this
for waveform strains constructed only from the $\ell=2$ multipoles.
Specifically, we conduct a survey on the faithfulness of the waveform models 
\texttt{SEOBNRv5PHM}, \texttt{TEOBResumS}, \texttt{IMRPhenomTPHM}, \texttt{IMRPhenomXPHM}
to the numerical relativity (NR) surrogate \texttt{NRSur7dq4}
and to NR waveforms from the \texttt{SXS} catalog.
The former assessment involves systems with mass ratios up to six and dimensionless spins up to 0.8.
The latter employs $317$ short and $23$ long \texttt{SXS} waveforms.
For all cases, we use reference inclinations of zero and $90^\circ$.
We find that all four models become more faithful as the mass ratio approaches unity
and when the merger-ringdown portion of the waveforms are excluded.
We also uncover a correlation between the co-precessing $(2,\pm2)$ multipole mismatches and 
the overall strain mismatch.
We additionally find that for high inclinations, precessing $(2,\pm 1)$ multipoles that are more faithful than their $(2,\pm2)$ counterparts, and comparable in magnitude, improve waveform faithfulness.
As a side note, we show that use of uniformly-filled parameter spaces may lead to an overestimation of
precessing model faithfulness.
We conclude our survey with a parameter estimation study in which
we inject two precessing \texttt{SXS} waveforms (at low and high masses)
and recover the signal with \texttt{SEOBNRv5PHM}, \texttt{IMRPhenomTPHM} and \texttt{IMRPhenomXPHM}.
As a bonus, we present preliminary multidimensional fits to model unfaithfulness for Bayesian model
selection in parameter estimation studies.
\end{abstract}


\maketitle

\tableofcontents

\section{Introduction}
\label{Sec:introduction}
Over the course of three observing periods,
the terrestrial network of gravitational wave interferometers has detected roughly 100 compact binary inspiral-merger events
\cite{TheLIGOScientific:2016wfe, LIGOScientific:2016vbw, LIGOScientific:2016fbo,
TheLIGOScientific:2016pea, LIGOScientific:2016anc, Abbott:2016izl, Abbott:2016apu,
Abbott:2016nmj, Abbott:2016wiq, LIGOScientific:2016wyt, Abbott:2017gyy, Abbott:2017oio,
Abbott:2017vtc, LIGOScientific:2018mvr, Venumadhav:2019tad,Venumadhav:2019lyq, Nitz:2018imz,
Nitz:2019hdf, Abbott:2020khf, LIGOScientific:2020stg, Abbott:2020tfl, LIGOScientific:2021djp, Nitz:2021zwj, TheLIGOScientific:2017qsa,Abbott:2020uma, Abbott:2020mjq, LIGOScientific:2021qlt,
Olsen:2022pin, Williams:2024tna}.
An overwhelming majority of these involved binary black hole systems where
each compact body is expected to have non-negligible spin angular momentum,
which has been confirmed by the analysis of the binary black hole population thus far detected
\cite{LIGOScientific:2018mvr, Abbott:2020niy, LIGOScientific:2021djp}.
Depending on the binary formation scenario, a significant subpopulation of these binaries can have spins misaligned with the orbital angular momentum \cite{Mandel:2018hfr, Mapelli:2021taw, Gerosa:2013laa,
Vitale:2015tea, Rodriguez:2016vmx, Stevenson:2017dlk}.

The leading-order {general relativistic} effect {of such misaligned spins}
is the precession of the orbital angular momentum vector around the
total angular momentum of the system, which leaves a faint,
but detectable imprint on the gravitational waveform as amplitude and
phase modulations on a timescale 1.5 post-Newtonian (PN) orders longer than
the orbital timescale \cite{Apostolatos:1994mx, Kidder:1995zr},
and at an order 1.5PN higher in the phase than the leading order terms \cite{Cutler:1994ys, Poisson:1995ef}.
As such, spin effects are harder to infer, or even detect, than the chirp mass and the [symmetric] mass ratio of the binary system.
Nonetheless, information about black hole spins has been obtained successfully
\cite{TheLIGOScientific:2016htt, Abbott:2016izl, LIGOScientific:2016sjg, LIGOScientific:2017vox,
Abbott:2017vtc, LIGOScientific:2018mvr, Nitz:2018imz, Venumadhav:2019lyq, Abbott:2020khf,
LIGOScientific:2020stg, Abbott:2020mjq, Abbott:2020niy, Abbott:2020uma, LIGOScientific:2021djp,
LIGOScientific:2021psn, Nitz:2021zwj, Olsen:2022pin}.
One way to convey this information is to provide distributions
for the inferred spin magnitudes and tilts, the latter with respect to some reference frame.
However, thus far, such posteriors have been mostly uninformative with a few exceptions
as illustrated by Figs.~6, 10, and 11 of the three Gravitational Wave Transient Catalogs, respectively \cite{LIGOScientific:2018mvr, Abbott:2020niy, LIGOScientific:2021djp}.

A more fruitful way of gleaning spin information has
been via the construction of specific projections of the spins parallel and perpendicular to the orbital angular momentum at a reference frequency, usually 20\,Hz.
These projections reduce the seven-dimensional
intrinsic parameter space (mass ratio and six spin components) to three dimensions, which was shown to essentially capture the phenomenology of precessing waveforms \cite{Hannam:2013oca}.
The parallel scalar has come to be known as the effective spin parameter $\chi_\text{eff}$ \cite{Damour:2001tu, Racine:2008qv}.
Though the term effective (or reduced or PN) spin has been used for similar scalars \cite{Ajith:2011ec, Schmidt:2010it},
we employ the phenomenological definition given in our Eq.~\eqref{eq:chi_eff} \cite{Hannam:2013oca, Purrer:2013ojf} which has become standard.
It was shown in Ref.~\cite{Racine:2008qv} that $\chi_\text{eff}$ is a conserved quantity up to 1.5PN, i.e., neglecting spin-spin and higher-order interactions.
The magnitude of $\chi_\text{eff}$ changes very little even with the inclusion of these interactions.
As such, it has become a very useful quantity in parameter estimation, especially since it emerges
in the PN series for the waveform phase at a more dominant order than all other parameters except for the chirp mass
and the symmetric mass ratio \cite{Cutler:1994ys, Poisson:1995ef, Baird:2012cu}.
In fact, in the strong field regime, the contribution of the $\chi_\text{eff}$ term is large enough to
cause a well known partial degeneracy between the symmetric mass ratio and the parallel component of the total spin
\cite{Cutler:1994ys, Poisson:1995ef, Baird:2012cu, Hannam:2013uu, Purrer:2013ojf}.

$\chi_\text{eff}$ also provides a way to gather information about the
properties of the binary black hole (BBH) population since isotropically distributed spins would
result in a normal distribution for $\chi_\text{eff}$ centered on zero.
Several studies have already looked at this over the ensemble of the detected
BBHs and more or less agree that the distribution is somewhat asymmetric and peaks at values
slightly above zero \cite{Galaudage:2021rkt, Golomb:2022bon, Adamcewicz:2022hce, Tong:2022iws, Biscoveanu:2022qac, Callister:2022qwb, Baibhav:2022qxm}.

The perpendicular scalar is known as the effective precession parameter (or spin) $\chip$ \cite{Hannam:2013oca, Schmidt:2014iyl}
and is given below in Eq.~\eqref{eq:chi_p}.
It was shown by Refs.~\cite{Hannam:2013oca, Schmidt:2014iyl} that the dominant effects of precession on the
GW waveform can be characterized by mapping the four perpendicular components of the binary's spin vectors to just one parameter, $\chip$,
assigned to be the sole perpendicular component of the spin of the larger (primary) component of the binary.
Since $\chip$ provides information about the perpendicular component of the spins, specifically, the perpendicular component of the primary's spin, it is taken as an indication of spin precession.

The inference of $\chip$ is more challenging than that of $\chi_\text{eff}$
because, for instance, in the limit that the orbital angular momentum is much larger than the total spin,
the perpendicular components can be considered to be 0.5PN order higher than the parallel components \cite{Arun:2008kb}.
This can also be understood in terms of the respective contributions of the parallel and the perpendicular components to the waveform phase.
Comparing the contribution of the $\chi_\text{eff}$-term to the overall phase
with that of the accumulated precession phase (cf. Eq.~(45) of Ref.~\cite{Apostolatos:1994mx}) for a canonical
neutron star binary inspiralling from 10\,Hz, we see that the former is roughly two orders of magnitude larger.
Accordingly, at moderate signal-to-noise ratios (SNRs), it is not possible to obtain a clear
measurement of the individual spins \cite{Schmidt:2014iyl}.
Moreover, the inclination of the orbit also matters; it is harder to detect precession in binaries
where the line of sight is aligned with the total angular momentum as the precession-induced modulations are minimized \cite{Schmidt:2012rh}.
Therefore, it is not surprising that there have thus far been only three GW events for which there is evidence for precession: GW190412 \cite{LIGOScientific:2020stg},
GW190521 \cite{Abbott:2020tfl,Abbott:2020mjq,Miller:2023ncs},
and GW200129\_065458 \cite{LIGOScientific:2021djp, Hannam:2022pit, Payne:2022spz,Macas:2023wiw}.

Discerning precession in compact binary inspirals (and mergers) is important for breaking parameter degeneracies and understanding binary formation scenarios.
To this end, accurate modeling of spin effects has become crucial in GW astronomy.
Though nonprecessing, i.e., aligned-spin, quasi-circular waveform models such as those given by
Refs.~\cite{Cotesta:2018fcv, Varma:2018mmi, Estelles:2020twz, Pratten:2020fqn, Garcia-Quiros:2020qpx, Riemenschneider:2021ppj, Pompili:2023tna, Nagar:2023zxh} 
have matured to the faithfulness level of $\lesssim 10^{-3}$
with respect to numerical relativity waveforms\footnote{We quote the approximate median value for a specific mismatch with respect to numerical relativity simulations over a large sample.},
the precessing models are approximately one half to an order of magnitude worse in faithfulness and much more sensitive to modeling systematics.
As the detectors' sensitivities improve steadily through O4, O5 and beyond,
systematic errors in parameter estimation due to mismodelling the effects of precession will dominate over the statistical error, especially for signals with large SNR
\cite{Cutler:1994ys, Flanagan:1997kp}.
In fact, recent research has shown that even the current accuracy of the NR waveforms
may not be enough for bias-free parameter estimation at the projected sensitivities of the third
generation ground-based interferometers \cite{Purrer:2019jcp}.
For this reason, the waveform modeling community has been endeavoring toward
building ever more faithful precessing models.
There are now several precessing quasi-spherical waveform  models borne out of various waveform ``families'' and all have achieved $\lesssim 10^{-2}$ faithfulness.
The state of the art among these is \NRsurP{} \cite{Varma:2019csw},
the latest precessing member of the numerical relativity surrogate family \cite{Blackman:2014maa, Blackman:2017dfb, Varma:2018mmi, Williams:2019vub, Walker:2022zob, Yoo:2023spi} with an NR faithfulness of $\sim 10^{-4}$. It was recently employed in a re-analysis
of most O3 events \cite{Islam:2023zzj}.

Another set of models is the NR-informed effective-one-body (\textsc{EOB}) family
\cite{Buonanno:1998gg, Buonanno:2000ef,Damour:2000we,Damour:2001tu,Buonanno:2005xu,Damour:2015isa}.
This contains two similar, but distinct subfamilies: \textsc{SEOBNR}~\cite{Bohe:2016gbl,Babak:2016tgq,Cotesta:2018fcv,Hinderer:2016eia,Steinhoff:2016rfi,Lackey:2018zvw,Matas:2020wab,Ossokine:2020kjp, Gadre:2022sed, Thomas:2022rmc,Ramos-Buades:2023ehm, vandeMeent:2023ols, Pompili:2023tna, Khalil:2023kep, Mihaylov:2023bkc}
and \textsc{TEOBResumS}~\cite{Damour:2014yha, Bernuzzi:2014owa, Nagar:2017jdw,Nagar:2018zoe, Akcay:2018yyh, Akcay:2020qrj, Nagar:2020pcj, Chiaramello:2020ehz, Gamba:2021ydi, Nagar:2021xnh, Albertini:2021tbt, Nagar:2022fep, Albertini:2022rfe, Gonzalez:2022prs, Nagar:2023zxh, Andrade:2023trh}.
A third major family is the phenomenological inspiral-merger-ringdown, \textsc{IMRPhenom}
waveforms
\cite{Ajith:2007qp, Ajith:2007kx, Ajith:2009bn, Santamaria:2010yb, Hannam:2013oca, Schmidt:2014iyl, Husa:2015iqa, Khan:2015jqa,Dietrich:2017aum, London:2017bcn, Khan:2018fmp, Dietrich:2019kaq, Khan:2019kot, Pratten:2020ceb, Pratten:2020fqn, Thompson:2020nei, Estelles:2020osj, Garcia-Quiros:2020qpx, Estelles:2020twz, Estelles:2021gvs,Hamilton:2021pkf, Yu:2023lml},
which have mostly been built in the frequency domain. This has enabled faster parameter estimation runs compared with the time domain \textsc{EOB} models.
However, there are now also a few time domain \textsc{IMRPhenom} models \cite{Estelles:2020twz, Estelles:2021gvs}.

Specific waveform models from each family are usually referred to as approximants.
The state of the art approximant from each of the families listed in the previous paragraph
are, respectively, \texttt{SEOBNRv5PHM} \cite{Ramos-Buades:2023ehm, Khalil:2023kep, Mihaylov:2023bkc, SEOBNRv5_repo}
(recent upgrade from \texttt{SEOBNRv4PHM} \cite{Ossokine:2020kjp, Gadre:2022sed, Thomas:2022rmc}),
\TEOBResumS\footnote{Nonprecessing and precessing approximants in this family are all called by the same name.}\cite{Akcay:2020qrj, Gamba:2021ydi},
\PhenomTPHM~\cite{Estelles:2021gvs} (time domain) and
\PhenomXPHM{} \cite{Pratten:2020ceb} (frequency domain).
Note, we do not employ here the more recent \texttt{IMRPhenomXODE} \cite{Yu:2023lml} or
\texttt{IMRPhenomXO4a} \cite{Thompson:2023ase}, nor the upgraded \PhenomXPHM{} \cite{Colleoni:2023}
though we do include a brief comparison involving these two models in App.~\ref{app:MSA_vs_4_vs_SpinTaylor}.
Essentially, these models are ``too new'' as the bulk of our work was already completed by the time
they appeared.

When a new waveform model is complete, it may undergo a review where it is extensively compared with  numerical relativity simulations and with various other approximants as well as undergoing parameter estimation tests.
Such studies are also detailed in the articles introducing the specific models (see, e.g., the works cited above).
Recently, the LIGO-Virgo-KAGRA Collaboration (LVK) also conducted a detailed study on the
faithfulness of the waveforms generated by the models \NRsurP, \texttt{SEOBNRv5PHM}, \texttt{TEOBResumS},
\texttt{IMRPhenomTPHM}, \texttt{IMRPhenomXPHM} and \texttt{IMRPhenomXO4a}
using a set of $\gtrsim 1500$ NR simulations, but this is not publicly available.
However, some of this work has been documented in Ref.~\cite{Ramos-Buades:2023ehm}
(also see Ref.~\cite{Thompson:2023ase} for a similar study using BAM waveforms \cite{Hamilton:2023qkv, Husa:2007hp, Brugmann:2008zz}.
A separate study focusing on the parameter estimation performance of the precessing models was
conducted in Ref.~\cite{Puecher:2023rxw}.
This work compared the system parameters inferred by \NRsurP, \PhenomTPHM, \PhenomXPHM{} and \texttt{SEOBNRv4PHM} for an ensemble of nearly $60$ ``pure'' BBH O3 events.

Here, we undertake a more systematic survey based on simulated data.
First, in Sec.~\ref{Sec:faith_survey}, using the NR surrogate \NRsurP{} (henceforth \NR)
as a proxy for NR, we compute the unfaithfulnesses of
$\{$\texttt{SEOBNRv5PHM}, \texttt{TEOBResumS}, \PhenomTPHM, \PhenomXPHM$\}$ to it for
a discretely spaced and a random-uniformly filled set of intrinsic parameters
with the binary mass fixed to both a light and a heavy value, and inclination fixed to $0$ and $\pi/2$.
We focus solely on the $\ell=2$ waveform strain faithfulness throughout our work.
For the discrete parameter set,
we document in detail the deterioration of faithfulness with increasing mass asymmetry, which
is well known, (see e.g., in Refs.~\cite{Estelles:2021gvs, Hamilton:2021pkf, Ramos-Buades:2023ehm}),
but we also uncover multimodalities in the unfaithfulness distributions coming from strongly precessing cases.
We further reveal a correlation involving the unfaithfulness of the co-precessing $(2,\pm2)$ multipoles.
We additionally contrast the unfaithfulness results from the discrete set with those of
a uniformly-filled parameter set. 
Using this, we show that the latter type of parameter set may result in the overestimation of
model faithfulness.

We then move on to direct comparisons with NR waveforms
from the \SXS{} catalog \cite{Boyle:2019kee},
choosing simulations that were not used in the calibration of
the models of interest here.
We employ a set of 317 \SXS{} simulations with a mass ratio grouping similar to the discrete set above so that we can re-apply the same analyses which result in our reaching very similar conclusions.
We further employ a smaller set consisting of 23 \SXS{} waveforms containing more than 100 GW cycles each
in order to briefly assess the models' faithfulness for longer inspirals.

Finally, to complete our survey,
we perform zero-noise injections of a moderately and a strongly precessing numerical relativity waveform
and recover their parameters with \texttt{SEOBNRv5PHM}, \PhenomTPHM{} and \PhenomXPHM{}.
As we explain in detail in Sec.~\ref{Sec:PE}, we are unable to present PE results for
\texttt{TEOBResumS} for these injections.

Our faithfulness survey provides us
with enough data to construct fits for unfaithfulness over the intrinsic parameter space.
These fits can be used to generate a weighted categorical
prior to inform model choice in a joint Bayesian analysis \cite{Hoy:2022tst}.
We reduce the dimensionality of the parameter space by employing the parallel and perpendicular effective spin
projections for our fits.
We then construct three dimensional fits
to the logarithm of the unfaithfulness as functions of the mass ratio and the spin projections.
These fits can subsequently be employed to assign weights to
each waveform model in various regions of the parameter space.
Our results shown in App.~\ref{Sec:fits} are preliminary as we intend to pursue this line
of research elsewhere.

The remaining sections are organized as follows. In Sec.~\ref{Sec:precession},
we briefly review precession dynamics and precessing waveform construction,
and introduce the parallel and perpendicular spin projections.
Sec.~\ref{Sec:faithfulness} introduces the various metrics which we employ to assess
waveform faithfulness.
We summarize our work in Sec.~\ref{Sec:Summary} and
conclude with a retrospective discussion in Sec.~\ref{Sec:Discussion}.
Appendices \ref{app:SEOBv4_vs_v5} and \ref{app:MSA_vs_4_vs_SpinTaylor} contain
additional model comparisons. 
Finally in App.~\ref{app:benchmarks}, we present timing benchmarks for the main models 
that we consider here. We provide our data and the notebooks/codes/scripts
used to obtain it in a \git{} repository\footnote{\url{https://github.com/akcays2/Survey_Precessing_Models}}.

We work with geometrized units where $G=c=1$.
$m_1$ and $m_2$ denote the masses of the primary and secondary components of the compact binary system, with $m_1\ge m_2$.
Accordingly, we define small and large mass ratios as $q= m_2/m_1 \le 1$ and $Q=1/q\ge1$,
the symmetric mass ratio $\eta = q/(1+q)^2$, and the total mass $M=m_1+m_2$.
$\Sa$ and $\Sb$ denote the spin vectors of each binary component with the respective dimensionless
spin vectors given by $\bm{\chi}_i = \mbf{S}_i/m_i^2$ and $\chi_i:= |\bm{\chi}_i|$, $i=1,2$.
Unless otherwise noted, we set $M=1$. Overdots denote time derivatives whereas hats denote
unit vectors.
Throughout this article, we use the terms aligned and parallel interchangeably, as well as perpendicular/in-plane/planar where the former direction is along the orbital angular momentum
vector and the latter is in the orbital plane.

\section{Review of Precessing Waveform Construction}
\label{Sec:precession}

For this article, we focus only on quasi-spherical binary systems
as most binaries are expected to have circularized by the time they enter the LIGO-Virgo-KAGRA band \cite{Peters:1963ux} which is supported by the gravitational wave (GW) data so far \cite{Romero-Shaw:2019itr, Nitz:2019spj, Yun:2020aow, Ramos-Buades:2020eju, LIGOScientific:2020kqk, Lenon:2020oza, Wu:2020zwr, Iglesias:2022xfc} (see Refs.~\cite{Romero-Shaw:2020thy, Romero-Shaw:2022xko} for a few exceptions).
However, there are binary formation scenarios in which a moderate amount of eccentricity
survives beyond the decihertz regime
\cite{Breivik:2016ddj, Samsing:2017rat, Samsing:2017oij, Rodriguez:2017pec, Gondan:2018khr, Banerjee:2018pmh,
Liu:2019gdc, Zevin:2020gbd, Michaely:2020ogo, Hamers:2021olp, Spera:2022byb, Samsing:2022fxi, Hellstrom:2022qaj, DallAmico:2023neb}.
Therefore, it is important to have template banks of sufficiently faithful
eccentric and precessing waveforms in the near future.

In general relativity, GWs have only two propagating degrees of freedom, $+$ and $\times$ polarizations,
which can be obtained from the following multipole sum in terms of spin weight $= -2$ harmonics
\be
h_+ - i h_\times = \sum_{\ell=2}^{\infty}\sum_{m=-\ell}^{\ell} {}_{-2}Y^{\ell m}(\ioo,\varphi_\text{ref})h_{\ell m}(t) \label{eq:GW_polarizations},
\ee
where $\ioo:=\cos^{-1}(\hat{\,\mathbf{L}}_{\text{N},0}\cdot \hat{\mathbf{N}})$ is the orbital inclination
with $\hat{\mathbf{N}}$ being the line of sight vector from the binary's center of mass
to the observer.
$\varphi_\text{ref}=\pi/2 - \varphi_c$ is a constant reference phase \cite{Schmidt:2017btt},
where $\varphi_\text{c}$ denotes the phase at coalescence.

The polarizations $h_{+,\times}$ couple with the detectors' antenna patterns on Earth.
For an L-shaped interferometer, the GW strain in the time domain is given by
\begin{align}
h(t) =  &F_{+} (\theta_s,\phi_s,\psi_s) h_{+}(t,\ioo,\varphi_\text{ref})\nn\\ & + F_{\times} (\theta_s,\phi_s,\psi_s) h_{\times}(t,\ioo,\varphi_\text{ref})\label{eq:strain} ,
\end{align}
where $ \theta_s,\phi_s$ are the source sky location angles,
$\psi_s$ is the source polarization with respect to the detector and
$F_{+,\times}$ are the detector antenna pattern functions that can be found in, e.g., Ref.~\cite{Maggiore:1900zz}.
The GW strain at the detector (henceforth just the strain) given by Eq.~\eqref{eq:strain} is
the quantity which we use in determining the faithfulness of the EOB and the phenomenological approximants via a certain sky-maximized mismatch defined
in Sec.~\ref{Sec:faith_survey}.

It was shown by Refs.~\cite{Schmidt:2010it, Schmidt:2012rh, Boyle:2011gg} that
precessing waveform multipoles, $h_{\ell m}$, can be built to a good approximation from the
Euler rotation of aligned-spin (AS) multipoles via the following expression
\be
h_{\ell m} = \sum_{m'=-\ell}^\ell   D^{(\ell)\ast}_{m',m}(-\gamma,-\beta,-\alpha)\,
h^\text{AS}_{\ell m'},
\label{eq:hlm_Twist1}
\ee
where $ D^{(\ell)}_{m',m}$ are Wigner's D matrices \cite{Sakurai:1167961, Brown:2007jx}.
$\alpha, \beta$ and $\gamma$ are the Euler angles of the frame rotation with $\alpha,\beta$
being the spherical angles of the Newtonian orbital angular momentum vector $\LN$ with respect to a chosen
frame, e.g., $\L_{\text{N},0}:= \LN(f_0)$ where $f_0$ is a reference frequency.
The third rotation by $\gamma$ uniquely fixes
the frame to the so-called minimal-rotation frame \cite{Boyle:2011gg}.

The Euler angles $\alpha,\beta$ are obtained from the time evolution of $\LN =\text{L}_\text{N}\Lhat$ governed
by the PN precession equations. At next-to-leading order (NLO),
the orbit-averaged evolution equations
can be written in the following form
\begin{subequations}
\begin{align}
 \dot{\mbf{S}}_i &= \bm{\Omega}_i \times \mbf{S}{}_i,  \label{eq:Sidot_NLO}\\
 \Lhatdot &= \bm{\Omega}_\text{NLO} \times \Lhat \label{eq:LNdot_NLO},
 \end{align}
\end{subequations}
with $i=1,2 $ denoting the primary and the secondary, respectively.
The precession frequencies are given by
\begin{subequations}
\begin{align}
&\f{\bm{\Omega}_i}{v^5}= \eta\left(2+\f{3}{2}q_i\right)\Lhat
 +\f{v}{2}\left\{\mbf{S}_j-3 [(q_i \mbf{S}_i+\mbf{S}_j)\cdot\Lhat]\Lhat\right\} \label{eq:Omega_i},\\
& \bm{\Omega}_\text{NLO}=-\f{v}{\eta} \left(\bm{\Omega}_1+\bm{\Omega}_2  \right) \label{eq:Omega_NLO},
 \end{align}
\end{subequations}
where $v$ is the relative speed of the binary components in the center of mass frame,
$q_1=1/q,\, q_2=q$, and $j=3-i$.
Radiation reaction is incorporated through the decay of
the magnitude of the angular momentum, $\dot{\,\text{L}}_\text{N}$, while the direction of the total angular momentum vector
$\J = \LN+\Sa+\Sb$ is kept fixed at a desirable PN order.
It is standard to rewrite $\dot{\,\text{L}}_\text{N}$ as $\dot{v}(v)$ or $\dot{\omega}(\omega)$ using $\text{L}_\text{N}=\eta/v=\eta \omega^{-1/3}$ where $\omega= v^3$ is the orbital frequency.
Further details can be
found in, e.g., Refs.~\cite{Apostolatos:1994mx, Kidder:1995zr, Gerosa:2016sys, Akcay:2020qrj, Gerosa:2023xsx, Khalil:2023kep}.

The coupled system of ODEs consisting of the decay of $\text{L}_\text{N}$ and Eqs.~\eqref{eq:Sidot_NLO}, \eqref{eq:LNdot_NLO} can be numerically solved straightforwardly.
Analytic solutions have also been found at this PN order via the multi scale approach \cite{Kesden:2014sla, Gerosa:2015tea, Chatziioannou:2017tdw, Gerosa:2023xsx}.
PN information exists up to 3.5PN in the radiation reaction sector
\cite{Buonanno:2002fy, Buonanno:2009zt, Chatziioannou:2013dza}
and next$^5$LO in the precession dynamics \cite{Khalil:2023kep}.
Once the solution $\LN(t)$ is known, $\alpha(t)$ and $\beta(t)$ can be computed immediately,
and, subsequently, $\gamma$ from \cite{Boyle:2011gg}
\be
\dot{\gamma} = \pm \dot{\alpha}\cos\beta,\label{eq:gamma_dot}
\ee
where there is a sign freedom in the right hand side.

Note that Eq.~\eqref{eq:hlm_Twist1} is an approximation for the true precessing multipoles,
$h_{\ell m}^\text{prec}$ which can be extracted from NR simulations for example.
One can then obtain the co-precessing multipoles via inverse of the transformation in Eq.~\eqref{eq:hlm_Twist1}
\be
h^\text{coprec}_{\ell m} = \sum_{m'=-\ell}^\ell D^{(\ell)}_{m',m}(\alpha,\beta,\gamma)\,
h_{\ell m'}.
\label{eq:coprec_modes}
\ee
Though $h_{\ell m}^\text{coprec}\ne h_{\ell m}^\text{AS}$, the AS multipoles are, in general,
a good approximation for the co-precessing multipoles \cite{Ramos-Buades:2020noq} as we will show
in Sec.~\ref{sec:ASmode_MMs}.

The AS multipoles satisfy the relation $h^\text{AS}_{\ell,-m}=(-1)^\ell h^{\text{AS}\ast}_{\ell m}$,
which does not hold for $h_{\ell m}^\text{coprec}$
due to the asymmetric emission of GWs above and below the orbital plane leading to the ``bobbing'' of the
binary and eventually resulting in the well-known kick of the final black
hole \cite{Keppel:2009tc,Bruegmann:2007bri, Ramos-Buades:2020noq, Kalaghatgi:2020gsq}.
Although the difference in the $\pm m$ multipoles may be small, it seems to be non-negligible
for unbiased parameter estimation of precessing binaries even at moderate SNRs \cite{Kalaghatgi:2020gsq},
and most definitely so at high SNRs \cite{Kolitsidou:2024vub}.
The only model used in this work that does not neglect this multipole asymmetry is
\NRsurP{} \cite{Varma:2019csw}, though we note that the newest phenomenological approximant \textsc{PhenomXO4}a
also contains this feature for the dominant, quadrupolar multipole \cite{Ghosh:2023mhc, Thompson:2023ase}.

Another approximation made above is the use of orbit-averaged ODEs to describe the precession dynamics
which removes the nutation of the spins from the evolution of the binary\footnote{For non-orbit-averaged versions, see Eqs.~(2.8) and (2.10) of Ref.~\cite{Racine:2008qv}.}.
Though wide nutation angles may leave a unique signature on future GW events \cite{Gerosa:2018mwg},
the work of Ref.~\cite{Gangardt:2022ltd} has found no strong evidence of nutation in any detected event thus far.

Returning to the binary with the orbital and spin angular momenta vectors
$\{\LN,\Sa,\Sb\}$,
let $\theta_i = \cos^{-1}(\Lhat\cdot \hat{\mbf{S}}_i )$ denote the polar (tilt)
and $\phi_i$ the azimuthal angle of each spin vector with $i=1,2$.
The components of the spins parallel and perpendicular to $\Lhat$ are given by
$\mbf{S}_{i,||}=(\mbf{S}_i\cdot\Lhat)\Lhat= m_i^2 \chi_i \cos\theta_i \Lhat$ and
$\mbf{S}_{i,\perp}= \mbf{S}_i -\mbf{S}_{i,||}$ with magnitude $ m_i^2 \chi_i \sin\theta_i $.

We have already discussed the most commonly employed parallel and perpendicular scalars,
$\chi_\text{eff} $ and $\chip$, in Sec.~\ref{Sec:introduction}.
$\chi_\text{eff} $ is given by \cite{Damour:2001tu}
\be
\chi_\text{eff} = \f{1}{1+q}(\chi_1\cos\theta_1+q\chi_2\cos\theta_2)
\label{eq:chi_eff}
\ee
and its perpendicular counterpart by \cite{Hannam:2013oca, Schmidt:2014iyl}
\begin{equation}
	\chip=\max\left(\chi_1\sin\theta_1,q\frac{4q+3}{4+3q}\chi_2\sin\theta_2\right) \label{eq:chi_p}.
\end{equation}
The Kerr spin limit $\chi_i \le 1$ imposes the condition $ 0\le \chip \le 1$ where
$\chip = 0$ corresponds to a non-precessing (aligned spin or spinless) and $\chip = 1$ to a maximally precessing
binary.
There is an elegant discussion in Ref.~\cite{Gerosa:2020aiw} on the physical meaning of this definition
which leads to a generalized version of this parameter
\begin{equation}
\begin{split}
  \chipGen=&\left[(\chi_1\sin\theta_1)^2+\left(q\frac{4q+3}{4+3q}\chi_2\sin\theta_2\right)^2 \right.\\
  &+\left.2q\frac{4q+3}{4+3q}\chi_1\chi_2\sin\theta_1\sin\theta_2\cos(\Delta\phi)\right]^{1/2},\label{eq:chipgen}
\end{split}
\end{equation}
where
\begin{equation}
  \cos(\Delta\phi)=({\hat{\mbf{S}}_1\times\Lhat})\cdot({\hat{\mbf{S}}_2\times\Lhat})
  \label{eq:Cos_Phi}
\end{equation}
which is the span of the planar angle between $\Sa$ and $\Sb$.
$\chipGen$ is in fact equal to $|\Lhatdot|/\Omega_1$ \cite{Gerosa:2020aiw}
and a comparison of Eq.~\eqref{eq:chipgen} with Eq.~\eqref{eq:chi_p} reveals that $\chipGen$ can
exceed 1 ($\chipGen \le 2$) which can only be achieved by systems in which both spins are large and mostly planar \cite{Gerosa:2020aiw}.
This fact has already been used for a simulated study of O4 events showing that for moderate to high SNRs,
double-spin precession can be inferred \cite{DeRenzis:2022vsj}.

There are several other recently-proposed perpendicular scalars.
For example, Ref.~\cite{Akcay:2020qrj} introduced the following perpendicular scalar
\begin{equation}
  \chiperp=\frac{|\mbf{S}_{1,\perp}+\mbf{S}_{2,\perp}|}{M^2}. \label{eq:chiperp}
\end{equation}
Ref. \cite{Hamilton:2021pkf} introduced a modification to $\chip$ for improved single-spin mapping in the merger-ringdown regime for binaries where $Q\lesssim 1.5$.
Ref.~\cite{Thomas:2020uqj} replaced $\chip$ with a two-dimensional {vector}
in the phenomenological waveform mapping
to improve the faithfulness of the precessing $(2,\pm 1), (3,\pm3), (4,\pm 4)$ multipoles.
Here, we concern ourselves only with $\chip,\chipGen$ and $\chiperp$ all of which
we collectively refer to as the $\chp$'s.

As can be seen from the equations defining the $\chp$'s, each one depends
on quantities that evolve in time (in other words, in frequency).
Traditionally, the values for these quantities are usually quoted at a reference frequency of $20$\,Hz,
but this choice is somewhat arbitrary.
Alternative reference times (or frequencies) have been suggested such as $t_\text{p-100}:=t_\text{peak}-100M$ and $t_\text{ISCO} := t(f=6^{-3/2}/(M\pi))$, where $t_\text{peak}$
denotes the time at which the co-precessing $(2,2)$ multipole amplitude peaks,
and $t_\text{ISCO}$ is
the time at which the GW frequency equals twice the Schwarzschild innermost stable circular
orbit (ISCO) frequency.
Refs.~\cite{Johnson-McDaniel:2021rvv, Mould:2021xst} have recently adopted yet another alternative:  the
$t \to -\infty\, (f \to 0\,\text{Hz})$ limit where the spin tilt angles
$\theta_1, \theta_2$ can be unambigiously obtained. However, the angle
$\Delta\phi$ can not be uniquely determined which means that we can not compute
$\chipGen$ or $\chiperp$ at past infinity.
One can also work with averaged quantities instead. For example,
Ref.~\cite{Gerosa:2020aiw} introduces a precession cycle averaged version of $\chip$
which they further extend to precession- and RMS-averaged versions of
$\chipGen$ in Ref.~\cite{Gerosa:2023xsx}.
Unless otherwise noted, we quote the values of the various spin scalars
at the initial time (frequency) $t_0 (f_0)$.


\section{Quantifying Waveform Faithfulness}\label{Sec:faithfulness}

\begin{figure}[t]
    \centering
    \includegraphics[width=0.48\textwidth]{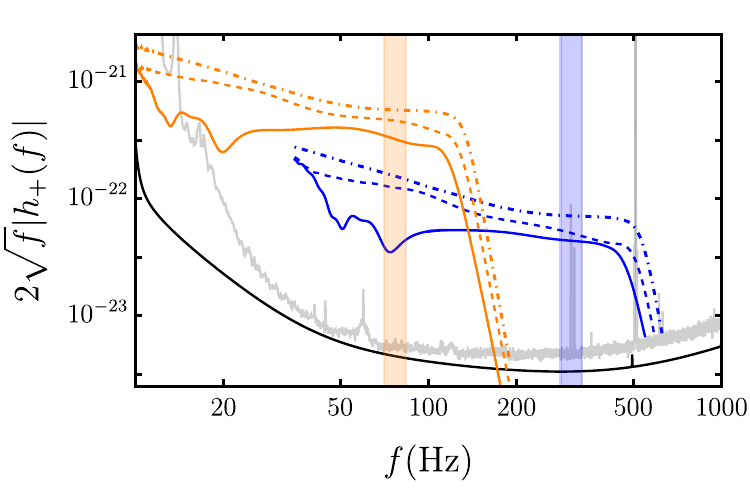}
\caption{Typical waveform amplitudes in the frequency domain from the sample of BBHs
generated for the survey in Sec.~\ref{sec:NRsur_survey_discrete}.
In particular, we display in blue (orange) a trio of light (heavy) precessing binaries
starting from $f_0=37.5\,$Hz (12\,Hz) with
$Q=\Qone,2,4$ (dot dashed, dotted, solid)
at a fiducial luminosity distance of 500\,Mpc and initial inclination of $\ioo=0$
shown against both Advanced LIGO design sensitivity
\cite{aLIGODesign_PSD} (black solid curve) and LIGO Livingston detector strain
sensitivity during O3 \cite{aLIGO03b_60Hz_Clean_PSD} (faint gray curve).
We used the approximant \texttt{IMRPhenomXPHM} to generate these waveforms.
For each case, we have $\max(\chi_\text{p}) =0.8$.
The orange and blue vertical bands mark the regions of $f_\text{peak}$
for each triplet of waveforms. Since $f_\text{peak}$ varies for each set of
parameters, we opted to show bands here instead of six additional vertical lines
marking individual $f_\text{peak}$.
}
\label{fig:sample_waveforms}
\end{figure}

The faithfulness of a given template waveform [strain] $h'(t)$ to a target waveform $h_0(t)$ is
measured in terms of the waveform match given by
\be
\M = \max\limits_{t_\text{c},\varphi_\text{c}} \f{\langle h_0 | h'\rangle}{\sqrt{\langle h_0 | h_0 \rangle \langle h' | h'\rangle}} \label{eq:standard_match}.
\ee
The match is maximized over constant
time and phase shifts which we take to be the time and phase shifts at coalescence $t_\text{c},\varphi_\text{c}$.
The angular brackets denote the noise-weighted inner product
\be
\hspace{20mm} \langle h_0 | h'\rangle:= 4 \Re
\int_{f_i}^{f_{f}}\f{\tilde{h}_0(f)\,\tilde{h}'^\ast(f)}{S_\text{n}(f)} d f,
\label{eq:inner_prod}
\ee
where $\tilde{h}(f)$ are the Fourier transforms of the time domain GW strains.
The inner product is weighted by the one-sided power spectral density (PSD) $S_\text{n}(f) $
of the detector noise for which we use the Advanced LIGO \cite{LIGOScientific:2014pky}
zero-detuned high-power design sensitivity \cite{aLIGODesign_PSD}.
The integration is performed from the initial frequency
$f_i$ to the final frequency $f_{f}$ which we set equal to 1024\,Hz.
As can be seen from Fig.~\ref{fig:sample_waveforms}, this {choice of final} frequency is large enough
to {encompass} the merger-ringdown portions of the waveforms for BBHs with $M=37.5\Msun, 150\Msun$ which will be our chosen values for much of this article.
To mediate the effects caused by finite signal duration,
we choose $f_i$ such that there are at least three
waveform cycles between it and $f_0$, the initial co-precessing \((2,2)\)-multipole frequency.
We set the luminosity distance of the sources to the fiducial value of $d_\text{L}=500\,$Mpc.

If the system of interest is precessing then the detector antenna patterns of
Eq.~\eqref{eq:strain} become time-dependent. Consequently, the waveform strain at the
detector becomes dependent on constant and time-varying extrinsic parameters
(and of course the intrinsic parameters) which we can write as
\be
\begin{split}
h(t) &= F_{+}(\theta_s, \phi_s,\psi_s(t))\; h_{+}(\ioo, \varphi_\text{ref}, t_\text{ref}) \\
    &+  F_{\times}(\theta_s, \phi_s,\psi_s(t))\; h_{\times}(\ioo, \varphi_\text{ref}, t_\text{ref}) \label{eq:ki_eq1}.
\end{split}
\ee
%
%
By introducing an effective amplitude $\mathcal{A}$
\be
\mathcal{A}(\theta_s,\phi_s) = \sqrt{F^2_{+}(\theta_s,\phi_s,\psi_s) + F^2_{\times}(\theta_s,\phi_s,\psi_s)} \label{eq:calA}
\ee
and an effective polarizability $\kappa$ via
\be
e^{i \kappa(\theta_s,\phi_s,\psi_s)} = \left[F_{+}(\theta_s,\phi_s,\psi_s) + i F_{\times}(\theta_s,\phi_s,\psi_s)\right]/\mathcal{A}(\theta_s,\phi_s) \, , \label{eq:kappa}
\ee
we can rewrite Eq.~\eqref{eq:ki_eq1} as
\be
\begin{split}
h(t) =  \mathcal{A}(\theta_s,\phi_s)\{&\cos\left[\kappa (\theta_s,\phi_s,\psi_s)\right] h_{+}(\ioo, \varphi_\text{ref}, t_\text{ref})  \\
      &+ \sin\left[\kappa(\theta_s,\phi_s,\psi_s)\right] h_{\times}(\ioo, \varphi_\text{ref}, t_\text{ref}) \} \, \label{eq:ki_eq2},
\end{split}
\ee
where we suppressed the $t$ dependence in the right-hand side.

We can now define the sky-maximized (optimized) faithfulness (match) between the target strain and the waveform template as
\begin{equation}
\M_\text{opt}  = \max_{t'_\text{ref}, \varphi'_\text{ref}, \kappa',\phi'}  \f{\langle h_0 | h'\rangle}{\sqrt{\langle h_0 | h_0 \rangle \langle h' | h'\rangle}}, \label{eq:opt_match}
\end{equation}
where $t'_\text{ref}, \varphi'_\text{ref}, \kappa'$ are template parameters to be optimized over.
The details of the optimization can be found in Refs.~\cite{Harry:2016ijz, Gamba:2021ydi}.
In particular, the $\kappa'$ optimization is performed analytically,
while $t'_\text{ref}$ is maximized via the inverse fast Fourier transform.
The maximization over $\varphi'_\text{ref}$ is performed numerically using
a dual annealing algorithm \cite{Pratten:2020ceb, Gamba:2021ydi}.
$\phi'$ represents the final degree of freedom to be maximized over,
i.e., the freedom to
shift $\phi_1, \phi_2$ by a constant amount which leaves $\Delta\phi$ unchanged.
We perform this maximization using another dual annealing algorithm \cite{Gamba:2021ydi}.
It is evident from Eq.~\eqref{eq:ki_eq2} that the match is also a function of the inclination angle $\ioo$,
but since we fix the inclinations in our comparisons at a given reference frequency, we do not need to optimize over them.

Note that $\M_\text{opt}$ is a function of $\varphi_\text{ref}$ and $\kappa$ so,
as a final step, we compute its average over an
evenly spaced grid for $\{\varphi_\text{ref}, \kappa\} \in [0, 2\pi)\times [0, \pi/2)$ with
$6\times 7=42 $ elements to obtain
\be
\M_\text{opt,av}:=\f{1}{42}\sum_{i=1}^{42}\M_{\text{opt}}(\varphi_{\text{ref},i},\kappa_i)\label{eq:M_opt_av}.
\ee
This is done to marginalize over any dependence of the match on the sky position
and obtain values which depend exclusively on the intrinsic parameters of the source.
This quantity is similar to the sky-and-polarization averaged faithfulness given by Eq.~(35)
of Ref.~\cite{Ramos-Buades:2023ehm}, but we do not average over the inclination.

For the remainder of this article, we employ the sky-averaged, optimized waveform mismatch
\be
\bar{\M}_\text{opt} := 1-\M_\text{opt,av} \label{eq:mismatch_opt}
\ee
as our faithfulness gauge.
The optimized mismatches that we quote henceforth will always be this average which we may occasionally
refer to as the ``full mismatch''.
As a check, we also store the minimum and the maximum values of $\MMo$ over the grid as well as the standard deviation all which we present
in the data provided in our \git{} repository.

To further disentangle possible causes of waveform mismodelling,
we additionally compute mismatches where we truncate
the mismatch integral before the transition to plunge,
specifically at $f_{f}={0.6} f_\text{peak}$ with $f_\text{peak}:= f\left(t=t_\text{peak}\right)$
denoting the GW frequency at the maximum (peak) amplitude of the co-precessing $\ell=2$ strain
of \NRsurP, i.e., the maximum of
\be
\mathcal{A}_\text{coprec}(t):=\sqrt{\sum_{m=-2}^{m=2}h_{2m}^\text{coprec}(t)}
\label{eq:Amp_coprec_max}
\ee
%
%
%
with $t_0 < t_\text{peak}$ \cite{Schmidt:2017btt}.
We have checked that ${0.6} f_\text{peak}\ge f_\text{MECO}$,
the minimum energy circular orbit (MECO) frequency \cite{Pratten:2020fqn}.
Accordingly, we present results for not only $\MMo$, but also for the merger-ringdown
excluded (inspiral-only) mismatch
\be
\MMo^\text{\noMR}:=\MMo(f_{f}=0.6f_\text{peak}) \label{eq:mismatch_opt_noMR}.
\ee

\section{Faithfulness Survey I: Comparisons with \texttt{NRSur7dq4}}\label{Sec:faith_survey}

We begin our survey with Sec.~\ref{sec:NRsur_survey_discrete} where
we assess the faithfulness of \SEOB, \TEOB, \TPHM{} and \XPHM{}
to \NRsurP{} using a discrete grid for the intrinsic parameters
at zero inclination, by which we mean $\ioo:=\io(f_0)=0$.
Since the direction of $\Lhat$ evolves in time, so does $\io$.
Thus, the best we can do is specify its value at some reference frequency.
In principle, we could have employed the [approximately] fixed inclination with respect to the
total angular momentum, \(\theta_\text{JN}\). However, the $\phi'$ optimization of
Eq.~\eqref{eq:opt_match} becomes non-trivial in this case.

We also include in Sec.~\ref{sec:NRsur_survey_discrete} a comparison in the
extrapolation region of \NRsurP{}, where the model is stated to be robust \cite{Varma:2019csw}.
In Sec.~\ref{sec:discrete_inc90}, we consider
the same discrete set at $\ioo=\pi/2$ thus changing the multipole content of
the strain in Eq.~\eqref{eq:GW_polarizations}.
We perform a similar analysis in Sec.~\ref{sec:NRsur_survey_random}
over a random-uniformly filled intrinsic parameter space and compare our findings with those
of Secs.~\ref{sec:NRsur_survey_discrete} and \ref{sec:discrete_inc90}.
We should add that the most recent version of \SEOB{} (\texttt{v5}) was released near
the completion of this work so we had initially conducted our survey on \texttt{SEOBNRv4PHM}.
As a useful sidenote, we present a brief comparison of \texttt{v5} and \texttt{v4} in App.~\ref{app:SEOBv4_vs_v5}.

%
%
%
\begin{table}[t]
\begin{ruledtabular}
\begin{tabular}{ll}
Total mass &$M\in \{37.5\;M_{\odot},150\;M_{\odot}\}$ \\
Mass ratio & $Q\in\{\Qone,2,4,6\}$\\
Initial frequency & $f_0=\begin{cases}
                     \{30,31,35,38.5\}\text{Hz}\ \, \text{for}\ M=37.5\Msun,\\
                     \{7.5,8,9,10\}\text{Hz}\qquad \text{for}\ M=150\Msun\\
                    \end{cases}
                    $
\\ \B\T
Spin magnitudes &$ \chi_1=\chi_2=0.8$\B\\
\T Tilt angles & $\cos\theta_{1,2} =\{ -\tfrac{\sqrt{3}}{2}+k \tfrac{\sqrt{3}}{4}\}\cup \{-1\}, k=0,\ldots, 4$  \\
& if $\theta_1=\pi,\theta_2\neq\pi$ \\
Azimuthal angles &\begin{tabular}{@{}c@{}}$\phi_1=0, \phi_2= k \pi/4, k = 1, \ldots, 8 $\end{tabular} \\
\end{tabular}
\caption{Relevant parameters for the binary systems which we use in our faithfulness survey
in Sec.~\ref{sec:NRsur_survey_discrete}.
$f_0$ denotes the initial frequency from which we start the binary evolution.
We fix $\phi_1=0$ at $f_0$ without loss of generality.
}
\label{tab:params}
\end{ruledtabular}
\end{table}
%
%

%
%

\begin{figure*}[t]
    \centering
    \includegraphics[scale=0.40]{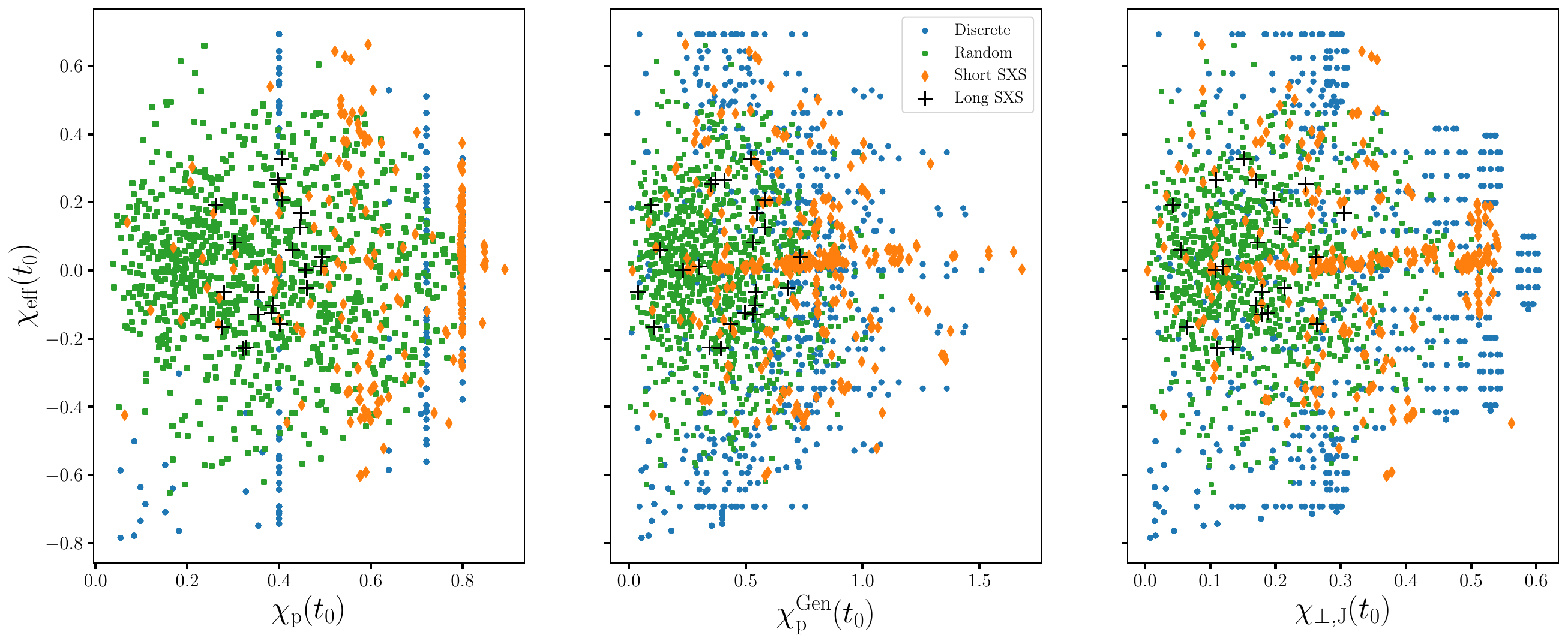}
\caption{The coverage of the spin space for all our parameter sets
shown using the parallel and perpendicular effective spin scalars computed at the initial time $t_0$.
In each panel, $\chieff(t_0)$ of Eq.~\eqref{eq:chi_eff} is plotted in the vertical axis.
From left to right, $\chip(t_0)$ [Eq.\eqref{eq:chi_p}], $\chipGen(t_0)$ [Eq.\eqref{eq:chipgen}]
and $\chiperp(t_0)$ [Eq.\eqref{eq:chiperp}]
are plotted in the horizontal axes, respectively.
The blue disks correspond to the values of these quantities coming from our discrete parameter
set of Sec.~\ref{sec:NRsur_survey_discrete} (see Table~\ref{tab:params}).
The green squares, orange diamonds and the black crosses represent the same quantities
obtained from our random-uniformly filled set (Sec.~\ref{sec:NRsur_survey_random}),
the short and the long \SXS{} sets (Secs.\ref{sec:Short_SXS} and \ref{sec:Long_SXS}), respectively.
As $\chip$ has no dependence on the azimuthal components of the spins, the parameters
of the discrete and the short \SXS{} sets yield degenerate values, clustered at $\chip\approx 0.4,0.72,0.8$, whereas both $\chipGen$ and $\chiperp$ cover their respective ranges better since they are $\phi_i$-dependent.}
\label{fig:spin_space}
\end{figure*}
%

\subsection{Discrete Parameter Set with zero Inclination}\label{sec:NRsur_survey_discrete}
As both LIGO and Virgo noise curves are frequency, and therefore mass, dependent,
we consider a set of light and heavy binaries for our survey with
$M = 37.5 \Msun$ and $150 \Msun$. We divide each set into four equal-size subsets
separated by mass ratio values of $Q=\Qone, 2, 4, 6$.
We have chosen $Q=\Qone \,(q=0.9)$ to break the $Q=1$ symmetry where the nonprecessing $(2,1)$ multipole equals zero, i.e., $h_{21}^\text{AS}=0$.
The subset with $Q=6$ is in the so-called extrapolation region of \NRsurP{} so we delegate
the comparison pertaining to it to Sec.~\ref{sec:q6_comparison}.

A lighter binary mostly accumulates SNR during its inspiral whereas a heavy enough binary
could have as much SNR accumulated during the merger-ringdown stage as the inspiral
assuming signals enter the detector band at $\gtrsim 20\,$Hz.
We illustrate this in Fig.~\ref{fig:sample_waveforms} with three light and three heavy BBH
waveforms with $Q=\Qone,2,4$ and the same spin vectors.
The merger-ringdown SNR for the light binaries ranges
from a quarter to $\lesssim$ half of its inspiral counterpart
(computed from 38\,Hz, see $f_0$'s in Table~\ref{tab:params})
while the same SNR for the heavy systems
ranges from 70\% to 90\% of its inspiral counterpart (from 12\,Hz).
Therefore, by considering light and heavy systems, we are effectively dividing our
survey sample into two halves whereby merger-ringdown modelling is much more important
for waveform faithfulness in one half (heavy BBHs) than the other (light BBHs).

Any comparison with \NRsurP{} is ultimately limited by the fact
the surrogate waveforms have a maximum inspiral time length of $4300M$.
For this reason, we have chosen $f_0=\{30,31,35\}\text{Hz}$ for
the $Q=\Qone,2,4$ subsets of the $M=37.5\Msun$ set and
$f_0=\{7.5,8,9\}\text{Hz}$ for the same subsets of the $M= 150\Msun$ set.
These yield between {27 and 47} GW cycles for light BBHs and
between {25 and 45} GW cycles for the heavy BBHs.

Each $Q$ subset consists of a grid of {spin angles} $\{\theta_1,\theta_2,\Delta\phi:=\phi_2-\phi_1\}$
where the tilt angles are evenly spaced in $\cos\theta_i$
from $-\sqrt{3}/2$ to $\sqrt{3}/2$ in steps of $\sqrt{3}/4$.
We also included the grid points with $\cos\theta_i=-1$ ($i=1$ \emph{or} 2) with the
intention to create a small subset of cases with near-transitional precession \cite{Apostolatos:1994mx}
to test the models' robustness.
Specifically, the eight cases with $Q=4,\theta_1=\pi,\theta_2=\pi/6$ yield
$\Lhat(t_0)\cdot\J_{\text{N},0}\approx -0.02$ and
$|\J_{\text{N},0}|\approx 0.025 \approx 0.033|\L_{\text{N},0}|$ (in units of $M^2$).
The addition of these extra grid points creates a somewhat skewed coverage of the parameter
space as can be seen by the blue dots in Fig.~\ref{fig:spin_space}.

$\Delta\phi$ runs from $\pi/4$ to $2\pi$ in steps of $\pi/4$.
We fix the dimensionless Kerr spin parameter $\chi$ to 0.8 for all cases
resulting in $\chip \le 0.8, \chi_{\perp,\text{J}} < 0.55, \chipGen < 1.45 $ and $|\chi_\text{eff}|<0.7$.
We summarize these parameter choices in Table~\ref{tab:params},
and show the coverage of the dimensionless intrinsic parameter space
in terms of the parallel and perpendicular spin projections
in Fig.~\ref{fig:spin_space} as the blue disks.
The figure also shows the same quantities plotted for the uniformly filled parameter set of
Sec.~\ref{sec:NRsur_survey_random} and the \SXS{} sets of Sec.~\ref{Sec:SXS_comparison}.

\begin{figure*}[t!]
    \centering
    \includegraphics[width=1.0\textwidth]{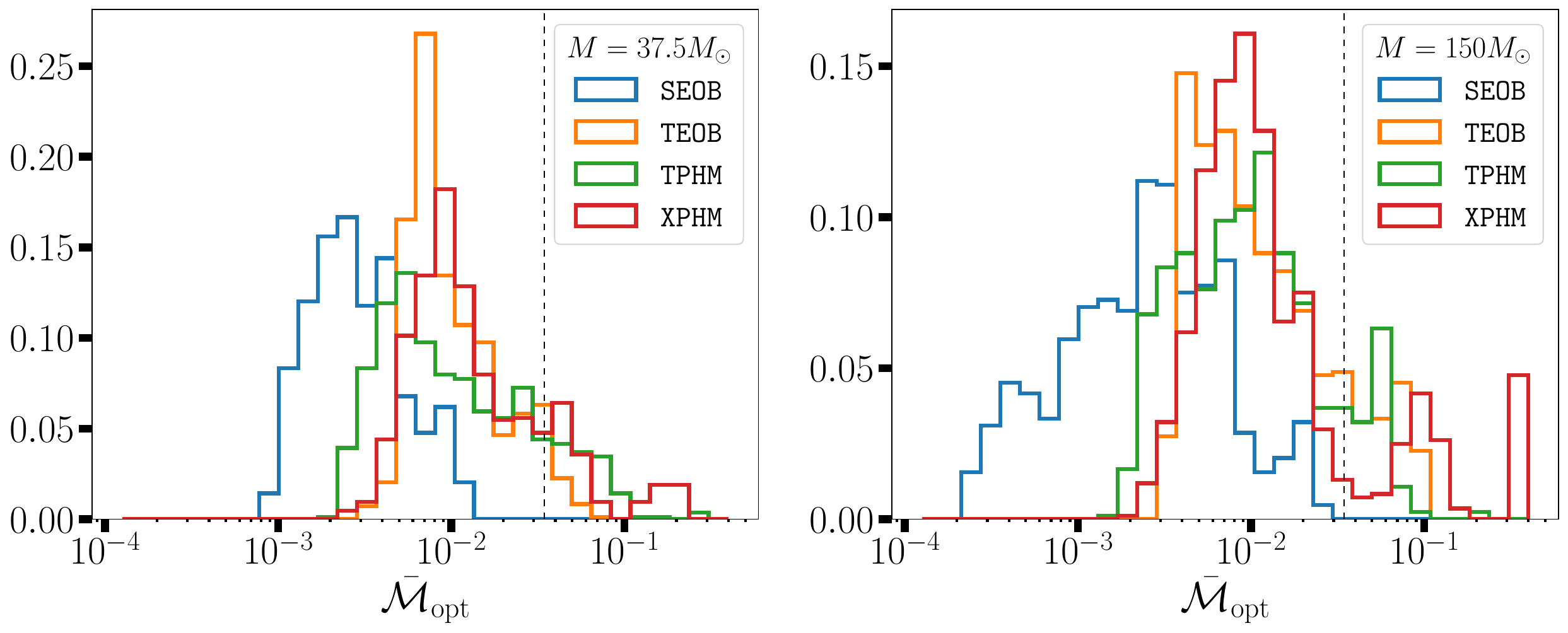}
\caption{The main result of this section: the distributions of the sky-optimized mismatch,
$\MMo$, given by Eq.~\eqref{eq:mismatch_opt} between the models listed in the legend and
\NRsurP{} which we take as a proxy for numerical relativity.
The parameters for the waveforms are given in Table~\ref{tab:params}
with the $Q=6$ subset excluded from the figure and we consider only the $\ell=2$ strain
for the mismatches.
The left (right) panels display the results for low (high) mass, $M=37.5\Msun$ ($M=150\Msun$),
systems, respectively.
\SEOB{} (\texttt{v5}), being the most recent of the four models, shows improvement, especially for heavier systems.
The dashed vertical line in both panels marks mismatch of
$1-0.9^{1/3}\approx 0.035$ translating to an event loss rate of 10\% \cite{Owen:1995tm, Flanagan:1997sx}.
}
\label{fig:main_mismatches}
\end{figure*}
%

The three-angle $\{\theta_1, \theta_2,\Delta\phi\}$ grid has in total $(6^2-1)\times8=280$ elements.
This then yields $280\times3=840$ separate binaries to consider per total mass separated
by $Q=\Qone, 2,4$.
Overall, we have $840\times 2=1680$ systems for each of which we compare the waveforms generated by $\{\SEOB,\TEOB,\TPHM,\XPHM\}$
to \NR{} via the averaged, optimized mismatch $\MMo$ given by Eqs.~(\ref{eq:opt_match},
\ref{eq:mismatch_opt}) for which the last quantity to consider is the lower limit of the match integral, i.e., $f_i$.
For the $M=37.5\Msun$ set, we pick $f_i = f_0+3\,\text{Hz}$
which captures approximately {25 to 42} GW cycles depending on the intrinsic parameters.
This allows for plenty of SNR accumulation in the inspiral stage before transitioning
to plunge. On the other hand, for the $M=150\Msun$ set, we deliberately fix
$f_i =11\,\text{Hz}$ leading to fewer inspiral cycles captured in the match
integration, thus increasing the relative contribution of the merger-ringdown stage
to the overall SNR.

In Fig.~\ref{fig:main_mismatches}, we show the main result of this section
where we plot the distributions of $\MMo$ between \NRsurP{} and
\SEOB, \TEOB, \TPHM, \XPHM{}.
In the left (right) panel, we present the mismatches for the
light (heavy), i.e., $M = 37.5\Msun\; (150\Msun)$, systems for all
$Q=\Qone,2,4$.
An apparent feature in the figure is the superior faithfulness of \SEOB{}
exhibiting no cases of $\MMo >0.035$ and having $\approx {22}\%$ of the heavy-binary mismatches below $10^{-3}$.
We also note comparable performances among
\TEOB, \TPHM{} and \XPHM.
For the heavy systems, the distributions of $\MMo$ for these approximants become wider
with more cases of $\MMo >0.035$.

Perhaps the most curious feature in Fig.~\ref{fig:main_mismatches} is the little ``bump''
in the heavy \XPHM{} histogram at $\MMo \sim 0.3$ with a similar smaller ``island'' for
the light-binary \XPHM{} mismatches at $0.1\le\MMo \le 0.3$.
For both light and heavy systems, we find that these are the same 40 cases
which have $Q=4, \theta_1=\pi$ with the worst mismatches coming from the $\theta_2 > \pi/2$
subset. For these configurations, which have the most negative $z$ projections of spins, i.e.,
$\chi_\text{eff}\lesssim -0.5$,
the default MSA prescription for the precession dynamics
of \XPHM{} breaks down \cite{Chatziioannou:2017tdw}
and the model defaults to the single-spin,
next-to-next-to-leading order (NNLO) prescription for the spin dynamics
with a 3PN approximation to $|\L|$.
This is a known shortcoming of \XPHM{} and is expected to be
fixed in the next version of the model \cite{Colleoni:2023}.


\begin{figure*}[t!]
    \includegraphics[width=1.\textwidth]{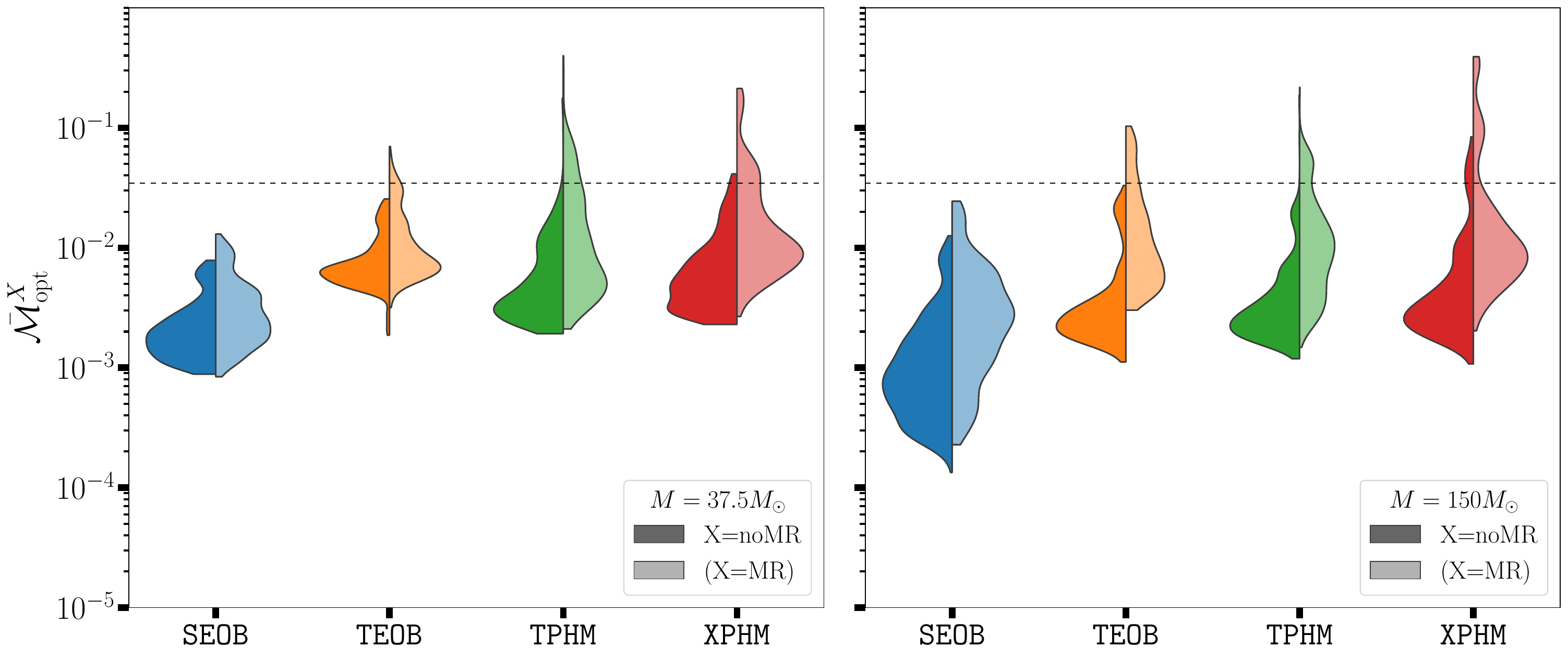}
\caption{
Merger-ringdown truncated mismatches versus full-length waveform mismatches.
The dark-shaded half violins are the distributions of $\MMno$ [Eq.~\eqref{eq:mismatch_opt_noMR}]
between the numerical relavity surrogate \NRsurP{} and the four approximants listed along
the bottom horizontal axis of each panel.
Similarly, the light-shaded half violins are the distributions of $\MMo$ [Eq.~\eqref{eq:mismatch_opt}].
which are shown in the legend as $\MMo^{X=\text{MR}}$.
These are the same quantities plotted in Fig.~\ref{fig:main_mismatches} as histograms.
The left (right) panel contains the data from the light (heavy), $M=37.5\Msun (150\Msun)$,
binaries.
We used the default Gaussian kernel density estimator of the \texttt{seaborn} library \cite{Waskom2021} to smooth the histograms for the violin plots.
The horizontal dashed line marks the mismatch value of $1-0.9^{1/3}\approx 0.035$.
}
\label{fig:violin_MR_noMR}
\end{figure*}

%
%

\subsubsection{Effects of the Merger-Ringdown Portion of GWs on Faithfulness}
\label{sec:discrete_MR_vs_noMR}
The increased amplitude/phase disagreement between NR and waveform models
in the plunge-merger-ringdown stages of the binary evolution is a well-known shortcoming,
usually illustrated in terms of time-domain plots of the waveforms.
For this reason, we investigated how the mismatches plotted in Fig.~\ref{fig:main_mismatches} change when
neglecting the {merger} ringdown portions of the waveforms.
We quantify this in terms of the merger-ringdown truncated (i.e., inspiral only) mismatch given in Eq.~\eqref{eq:mismatch_opt_noMR}.
There are two reasons for which
we \textit{a priori} expect the MR-truncated waveforms to agree better with \NR, i.e.,
we expect the $\MMno$ distributions to occupy lower values
than their $\MMo$ counterparts.
The first reason pertains to signal morphology: the inspiral
is much smoother than the MR and therefore easier to optimize over in the match computation.
The second reason has to do with the modelling: the MR stages are more difficult to model
analytically, especially without input from NR simulations. As such, the waveform models
employ NR-informed fits for the MR regime.

Our second expectation is a larger shift
between the $\MMno$ and $\MMo$ distributions for the $M=150\Msun$ sample
since the MR part of the signal is
much more important for these heavier binaries.

These expectations are confirmed in Fig.~\ref{fig:violin_MR_noMR}
where we present the distributions of $\MMno$ and $\MMo$ as [half] violin plots
for the four approximants for both the light and the heavy BBH samples.
We can clearly see that the distributions of the MR-excluded mismatches ($\MMno$),
i.e., the darker-shaded half violins, are all
shifted to lower values than the lighter-shaded half violins representing the distributions of $\MMo$.
This downward shift is more pronounced for the heavier cases ($M=150\Msun$)
with the median of the $\MMno$ distributions roughly half an order of magnitude
lower than the $\MMo$ distributions (less so for \TEOB).
The shift is less prominent for the light BBHs ($M=37.5\Msun$)
where the MR portion of the signal is much less important than the inspiral.

One can also discern a smaller secondary peak in most of the $\MMno$ distributions
which comes from the mismatches of the $Q=4$ subset.
This peak is somewhat obscured in Fig.~\ref{fig:violin_MR_noMR} where we plot
the entire emsemble of $Q=\Qone,2,4$ subsets.
We investigate this in more detail in the next section where we separate the mismatch data
by mass ratio.

One can further ask whether or not there are regions in the spin space where
there is a larger gap between $\MMno$ and $\MMo$ than other regions.
To this end, we define the following ratio $\cal{R}:=\MMo/\MMno$ and look at how
it is distributed in the $\{\chi_\perp\text{'s}, \chieff,Q\}$ space, where $\chi_\perp\text{'s}$
represents one element of the set $\{\chip, \chipGen,\chiperp\}$ at a time.

For the light BBHs, we observe $\mathcal{R}\lesssim \{2,3.5,6\}$ for
$\{\SEOB, \TEOB, \TPHM\}$ regardless of $Q$. For \XPHM,
$\mathcal{R}\lesssim \{4.6,6,9\}$ for $Q=\{\Qone,2,4\}$.
Moreover, $\mathcal{R}$ seems to increase monotonically for \TEOB{} and \TPHM{}
as the spin vectors become more planar, i.e., as $\theta_{1,2}\to \pi/2$, whereas
for \XPHM, the large $\R$ region is mostly situated in the $\theta_{1,2}\to 5\pi/6$
region. As for \SEOB, since $1\le \R \lesssim 2$,
no specific region stands out.

For the heavy BBHs, $\R$ increases significantly for all models.
\SEOB{} still yields the smallest values with $\R\lesssim 6$ regardless of the mass ratio
with a small region of $\R\approx 8$ in the $Q=4$ subset.
The values of $\R$ output by the other three models show more variation with mass ratio as
$\R\lesssim \{12,18,14\}$ for \TEOB, $\R\lesssim \{6,10,12\}$ for \TPHM,
and $\R\lesssim \{6,10,17\}$ for \XPHM{} corresponding to $Q=\{\Qone,2,4\}$.
Unlike the light set, we now observe larger-$\R$ regions in the $\chieff >0$ half of the
parameter space for \SEOB, \TEOB{} and \TPHM{} with the opposite relation applying to \XPHM.

%
%
%

\begin{figure*}[t!]
    \centering
    \includegraphics[width=1\textwidth]{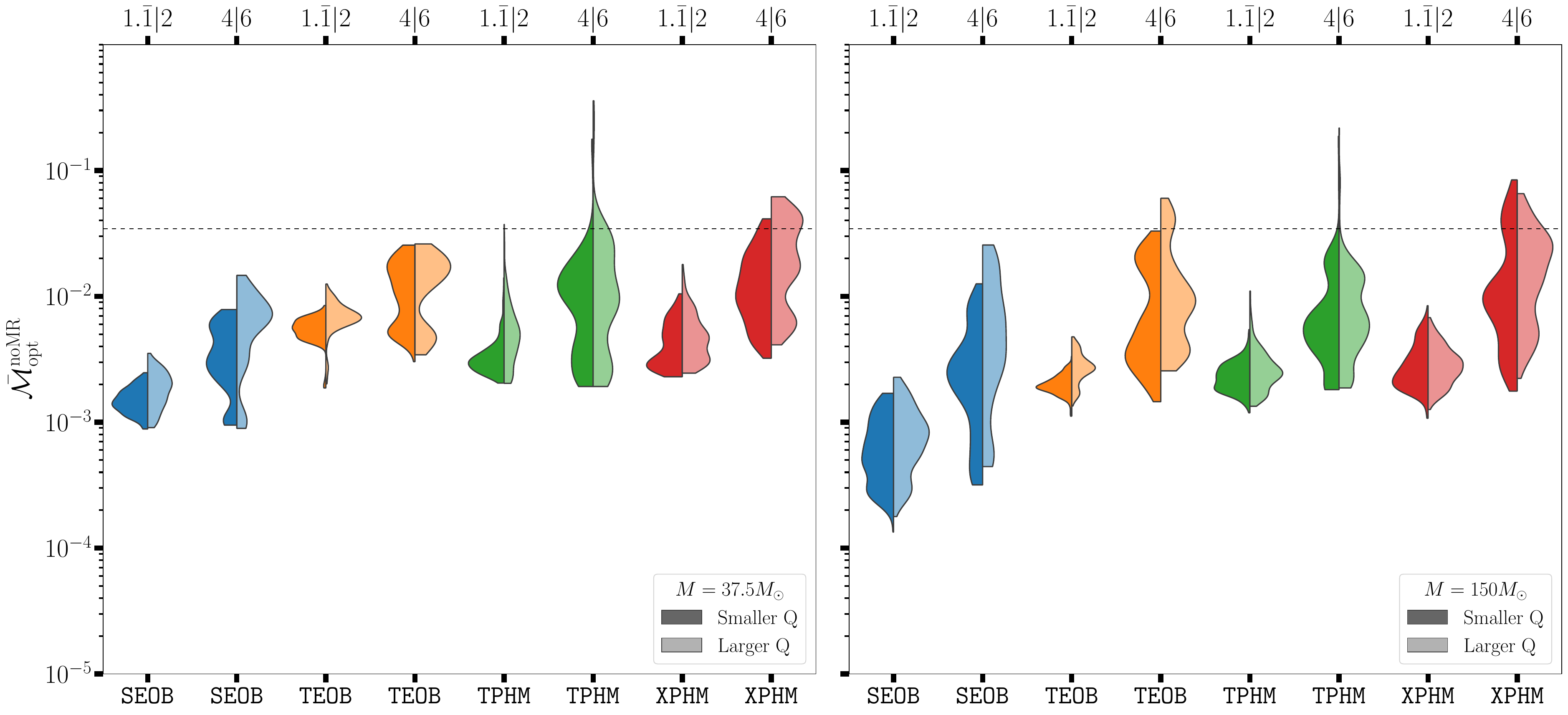}
\caption{Mass ratio separated, merger-ringdown truncated mismatches, $\MMno$, between \NRsurP{} and
the approximants $\{\SEOB,\TEOB,\TPHM,\XPHM\}$ labeled along the bottom
axes and plotted in blue, orange, green, red, respectively.
For each approximant, we present four half violins, representing, from left to right,
the $\ioo=0$ mismatches for the subsets with mass ratios $Q=\Qone,2,4,6$ which are shown along the top axes.
The $Q=6$ subset is discussed separately in Sec.~\ref{sec:q6_comparison} as it involves
comparisons with \NRsurP{} in its extrapolation region.
The horizontal dashed line marks the mismatch value of $1-0.9^{1/3}\approx 0.035$.}
\label{fig:violin_q_separated}
\end{figure*}

%
%

In order to better determine how much of the increased mismatches are genuinely due to mismodelling
of the merger-ringdown regimes,
one could construct hybrid waveforms where the inspiral-only part generated by the four models
is attached to the plunge-merger-ringdown part of the corresponding \NRsurP{} waveform.
One could then compute $\MMo$ between the hybrid waveforms and the full \NRsurP{} waveforms,
and compare these values to the ones obtained here.
Though this strategy is, in principle, straightforward, its implementation is highly non-trivial with precession.
We leave this for future work.

\subsubsection{The Dependence of Waveform Unfaithfulness on the Mass Ratio}\label{sec:Q_1_2_4cases}

Another known degradation of waveform faithfulness occurs for systems with
more component mass asymmetry, i.e., large (small) values of $Q\ (q)$.
In terms of our sample, this should translate to increasing mismatches
with increasing $Q$ (or decreasing $q$).
In order to better reduce the potential ``contamination'' of the mismatches due to
merger-ringdown mismodelling, we present only $\MMno$, separated by $Q$
values here. Accordingly, we should expect to see
$ \MMno(Q=\Qone) \preceq \MMno(Q=2) \preceq \MMno(Q=4)\preceq \MMno(Q=6)$, where we are
appropriating the $\preceq$ symbol to mean that the distributions
have higher occupancy per bin at lower mismatch values.
The results of this breakdown by mass ratio are shown in Fig.~\ref{fig:violin_q_separated},
where we plot the distributions of $\MMno$ as separate half violins for the
$Q=\Qone,2,4, 6$ subsets, but delegate our discussion of the last
subset (\(Q=6\)) to Sec.~\ref{sec:q6_comparison}.

Let us recall that increasing $Q$ increases the number of cycles in the range $[f_i,f_f]$,
so part of the increased mismatch may be simply due to having longer waveforms.
However, this is not the only factor. Given a value of $Q$, the spin configurations that yield
the longest waveforms are those with $\theta_i=\pi/6$ and the shortest waveforms are the ones
with $\theta_1=5\pi/6\;(150^\circ),\theta_2=\pi$. It is true that the cases with
$\{Q,\theta_1,\theta_2\}=\{6,\pi/6,\pi/6\}$ have higher mismatches than, e.g., the cases with
$\{\Qone,\pi/6,\pi/6\}$, where the former cases have nearly twice as many cycles in the $[f_i,f_f]$ interval.
On the other hand, the mismatches for $\{Q,\theta_1,\theta_2\}=\{6,\pi/6,\pi/6\}$
are lower than the cases with
$\{6, \pi/2\lesssim\theta_1\lesssim 5\pi/6, \pi/2\lesssim\theta_2\lesssim 5\pi/6\}$,
where the latter cases have 10 to 20 less cycles in band.
Therefore, waveform length alone can not explain the increased mismatches.

Fig.~\ref{fig:violin_q_separated} essentially corroborates our expectations.
We find that for both
light and heavy systems, the best agreement with \NR{} for each approximant is always by the
$Q=\Qone$ subset with the $ Q=2$ subset yielding slightly worse mismatches.
What is additionally apparent from the figure is the relative upward shift of the mismatch
distributions for the $Q=4,6$ subsets, i.e., a clear deterioration of waveform
faithfulness for more mass asymmetric systems.
We also observe an emergence of multimodalities in the distributions
of the $Q=4,6$ mismatches.

In more detail, we observe that the worst mismatches from the $Q=4$ subset
predominantly come from the cases with $\chip\gtrsim 0.7$
($0.6\lesssim \chipGen \le 1 \text{ and } \chi_{\perp,\text{J}}\gtrsim 0.4$),
where we quote the values at $f_0$.
This is somewhat expected as these cases precess more strongly.
However, for the heavy-mass subset,
some of the worst mismatches show up by $\chip\gtrsim0.4$ indicating additional
mismatch dependence on the parallel projection of the spins.
This might be connected with some of the observed multimodalities.
For example, we discovered that the two distinct peaks in \TEOB's
$Q=4$ mismatch distributions can be mapped to two separate regions in the parallel-perpendicular
spin projection space. Specifically, we find that for $M=150\Msun$,
the mismatches of the upper peak ($\MMno > 10^{-2}$) solely come from cases with
$\chi_\text{eff}<0$. 
Light \SEOB{} exhibits a less prominent bimodality separated at $\MMno \approx 0.004$
with the mismatches coming from the upper peak corresponding to cases with
$\chi_p \gtrsim 0.7$ and mostly $\chi_\text{eff}<0$.
Light \TPHM{} also exhibits a bimodality separated at $\MMno \approx 0.004$ corresponding to
$\chi_p \gtrsim 0.7$ regardless of the value of $\chieff$.
Heavy \XPHM{} distribution separates at $\MMno \approx 0.02$ with mismatches of
the upper peak \emph{all} coming from $\chi_\text{eff}<0$ cases.

%
%
\begin{figure}[t]
\centering
 \includegraphics[width=0.48\textwidth]{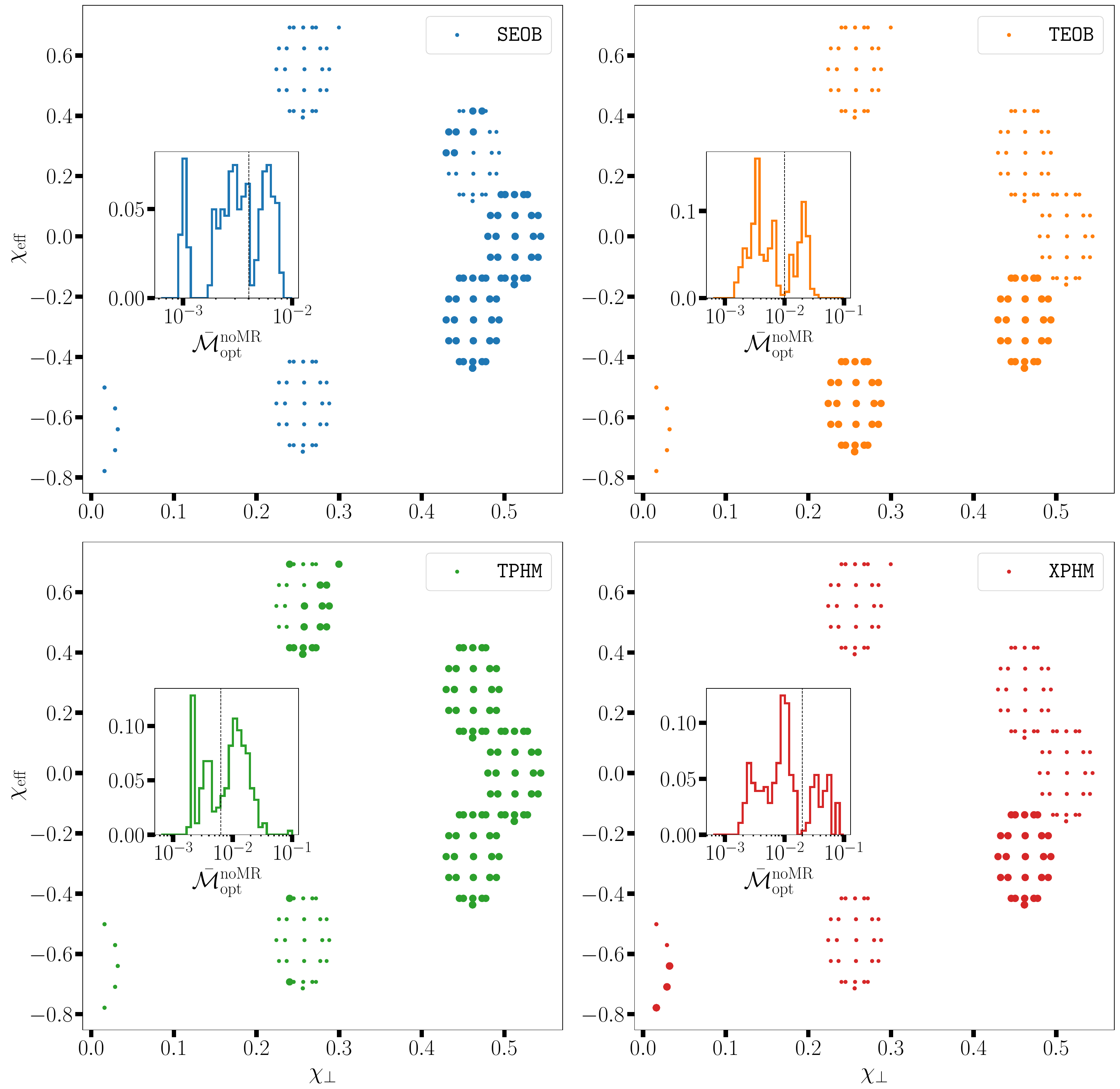}
 \caption{The association of certain peaks in the mismatch
 distributions with specific regions in the spin parameter space.
 In each of the four panels, we scatter-plot the $\{\chiperp,\chieff\}$ values of the
 $Q=4$ subset of the discrete parameter space.
 The larger disks correspond to points which yield
 values of $\MMno$ [Eq.~\eqref{eq:mismatch_opt_noMR}] in the right-most peaks of the histograms plotted in the insets where the thick,
 vertical dashed lines separate the multiple modes (peaks) of each histogram.
 The smaller squares mark the complementary $Q=4$ cases.
 The \SEOB, \TPHM{} histograms in the first and third insets correspond to the $Q=4$ half violins
 from the left panel of Fig.~\ref{fig:violin_q_separated}, while the rest from the right panel.
 }
 \label{fig:mismatch_chiPerp_chiEff}
\end{figure}

We show how the multimodalities of these four cases map to $\{{\chiperp},\chi_\text{eff}\}$
space in Fig.~\ref{fig:mismatch_chiPerp_chiEff} for the $Q=4$ subset.
For this figure, we opted to use $\chiperp$ instead of $\chip$ (or $\chipGen$) for visual clarity.
From the figure, we observe that for $\chiperp \gtrsim 0.4$ ($\chip\gtrsim 0.5$) cases,
the $\chi_\text{eff}<0 $ corner of the parameter space seems to be
more challenging than the $\chi_\text{eff}>0 $ corner for all approximants except for \TPHM.
The aforementioned breakdown of \XPHM's MSA prescription is also exhibited in the
$\chi_\text{eff} \lesssim -0.5, \chiperp \lesssim 0.05\;(\chip \lesssim 0.2)$ corner of the parameter space.

Finally, let us finish this section with a brief breakdown of the full mismatches ($\MMo$)
in terms of mass ratio so as to complement Fig.~\ref{fig:violin_q_separated}.
For \SEOB,  we find no cases of $\MMo >0.035$ for $Q\le 4$ as is clear from
Figs.~\ref{fig:main_mismatches} and \ref{fig:violin_MR_noMR}.
$\{\TEOB,\TPHM,\XPHM\}$ are also robust for $Q\le 2$ with very few cases of
$\MMo >0.035$ for either light or heavy BBHs, worst being 3\% of the
$M=150\Msun \ (37.5\Msun)$ cases for \TEOB{} (\TPHM), but their faithfulness
degrades for higher values of $Q$.
For example, at $Q=4$, $\{14\%, 40\%, 50\%\}$ of the light-mass cases yield $\MMo >0.035$ which becomes $\{40\%, 36\%, 49\%\}$ for the heavy mass set for $\{\TEOB,\TPHM,\XPHM\}$, respectively.
We can also contrast these percentages with their $\MMno >0.035$ counterparts
which are $\{0\%,2\%,4\%\}$ for the light and $\{0\%,1\%,2\%\}$
for the heavy sets.
In short, there is a noticeable degradation of model performance in going from
$Q\lesssim2$ to $Q\gtrsim 4$ for all models including \SEOB{} though overall it is always
better than 0.965 faithful for $Q\le 4$.

\subsubsection{The Effect of AS/Co-precessing multipoles on Faithfulness}\label{sec:ASmode_MMs}
Since each precessing multipole of the four models is constructed using Eq.~\eqref{eq:hlm_Twist1}
or some variant of it (e.g., using co-precessing multipoles instead of AS multipoles), there are two main modelling
systematics at interplay here:
\begin{inparaenum}[(i)]
\item systematics coming from the modelling of precession dynamics that are manifest
in the Euler angles used in the frame rotation,
\item systematics in the AS (or co-precessing) multipoles that are being Euler rotated.
\end{inparaenum}
We focus on the latter  in this section.

Even if the Euler angles of $\{\SEOB,\TEOB,\TPHM,\XPHM\}$
equal those of \NR{} (not actually the case), strain mismatches can still be large
if the Euler-rotated multipoles poorly match \NR's co-precessing multipoles.
Moreover, in the case of unequal Euler angles, the AS/co-precessing multipole mismatch may
still end up being the dominant systematic.
We investigate this here by computing AS/co-precessing multipole mismatches between the four
models and \NR's co-precessing multipoles. 

%
\begin{figure*}[t]
\centering
 \includegraphics[width=0.99\textwidth]{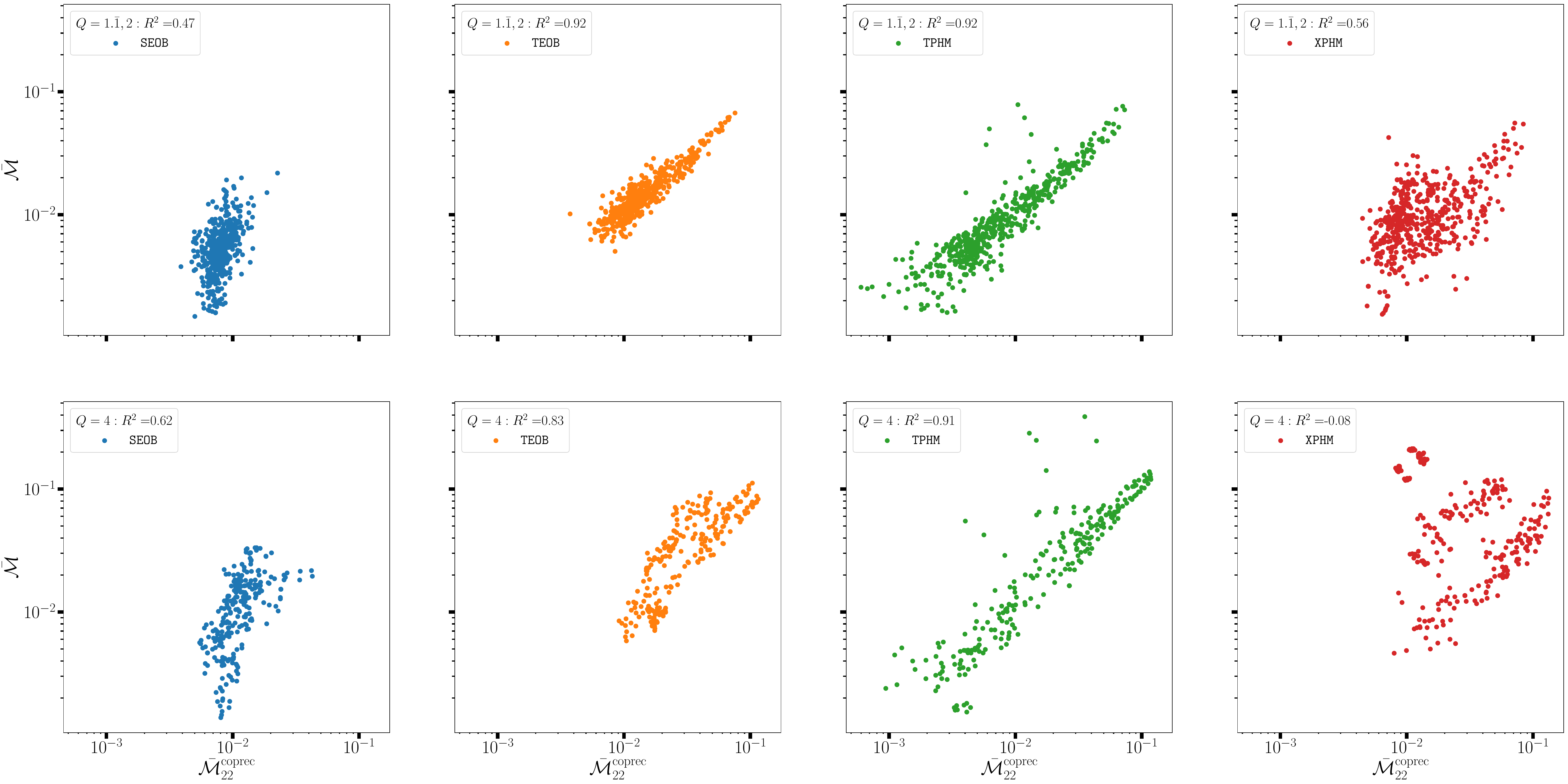}
 \caption{The correlation between $\MM_{22,\text{AS}}$ and $\MM(\ioo=0)$ for the $M=37.5\Msun$
 cases, where $\MM_{22,\text{AS}}$ is the mismatch [via Eq.~\eqref{eq:standard_match}] between \NR's co-precessing $(2,2)$ multipole and each AS $(2,2)$ multipole of the four models labelled in the legends
 where we also show the square of the Pearson correlation coefficient.
 The upper panels show the mismatches from the combined $Q=\Qone$ and $Q=2$ subsets, while the
 lower panels show the results from the $Q=4$ subsets.
 The cluster of red dots in the upper left corner of the $Q=4$ \XPHM{} panel are the cases for which the
 aforementioned MSA prescription of \XPHM's precession dynamics breaks down.
 }
 \label{fig:Coprec_22_mismatches}
\end{figure*}


First, let us add a few details.
To our knowledge, among the four models, \TEOB{} is the only one that rotates actual AS multipoles.
\SEOB{} twists AS multipoles with the constant spin parameters $\chi_i$ replaced by the time varying
$\bm{\chi}_i(t)\cdot \Lhat(t)$ evolved via \SEOB-specific dynamics.
\TPHM{} and \XPHM{} rotate AS multipoles with modified remnant properties \cite{Estelles:2021gvs, Pratten:2020ceb}.
Though these multipoles may be referred to as co-precessing, they are all approximations and thus,
not exactly equal to $h^\text{coprec}_{\ell m}$ of Eq.~\eqref{eq:coprec_modes}.
Therefore, let us denote these approximate co-precessing multipoles by
${h}^\text{AS}_{\ell m}$ as well.
These still satisfy the $m\leftrightarrow -m$ multipole symmetry.
Though exact for non-precessing systems, this is an approximation for the true co-precessing
multipoles which are known to violate this symmetry (see, e.g., Fig.~2 of Ref.~\cite{Ramos-Buades:2020noq}).

We can now ask:
how faithful are AS multipoles to the true co-precessing multipoles?
And what is the penalty in using AS multipoles to construct precessing waveforms?
Ref.~\cite{Ramos-Buades:2020noq} provides the first detailed answer to this question
using comparisons to 72 NR simulations though only six cases have $Q\ge 4$.
They find that the AS $(2,\pm2)$ multipoles are faithful representations of
their co-precessing counterparts with only one (five) out of the 72 simulations
resulting in mismatches larger than 0.03 (0.01) (see their Fig.~2 and Table~III),
but that the AS $(2,\pm1)$ multipoles may not be considered to be so (ibid.).
Especially relevant here is their specific comparison with a short BAM \cite{Husa:2007hp, Brugmann:2008zz}
simulation (ID 28) with $Q=3,\chip=0.8$, where
they identify the $m$-multipole asymmetry as the cause of the high mismatch.

In the following, we conduct a study similar to Ref.~\cite{Ramos-Buades:2020noq}'s.
Specifically, we compute mismatches between the co-precessing $(2,\pm2)$ multipoles of \NRsurP{}
and those of $\{\SEOB,\TEOB,\TPHM,\XPHM\}$. We do this by modifying the source code
for the $\texttt{gwsurrogate}$ package. We focus on $h_{22}^\text{AS}$ which is dominant in Eq.~\eqref{eq:hlm_Twist1},
which we also treat as a proxy for results pertaining to $h_{2,-2}^\text{AS}$.
We proceed by first computing a new quantity: $\MM_{22,\text{AS}}$,
the standard $\{t_\text{c},\varphi_\text{c}\}$-optimized mismatch
between \NR's co-precessing $(2,2)$ multipole and each AS $(2,2)$ multipole of the four models.
We then look for correlations between these mismatches and the
strain mismatches $\MM(\ioo=0), \MMo(\ioo=0)$.
Indeed, as Fig.~\ref{fig:Coprec_22_mismatches} highlights,
there is a very prominent correlation between $\log  \MM_{22,\text{AS}}$
and $\log \MM$ for \TPHM{} and \TEOB, with \XPHM{} showing a weaker
trend and \SEOB{} hardly showing any.
The linear correlation of the logs implies a powerlaw relation between these mismatches.
Additionally, we observe that

\begin{enumerate}[label=(\roman*)]
 \item the correlations are stronger between $\MM_{22,\text{AS}}$ and $\MM$
 than $\MMo$. This means that the optimization of the mismatch over extrinsic parameters
 partially ``smears out'' this relation;
 \item the correlations are much more prominent for the light BBHs, where more inspiral
 cycles are included in the match integrals;
 \item the correlations persist in all mass ratio subsets, but are stronger for $Q=\Qone$ and 2.
\end{enumerate}

For \TPHM{} and \TEOB, the $Q=\{\Qone,2\}$ subset correlation coefficients are larger than 0.9.
Moreover, we observe that for the cases that yield $\MM_{22,\text{AS}}\gtrsim 0.05$,
both \TEOB's and \TPHM's $\MM$ can be fit by lines parametrized by $\MM_{22,\text{AS}}$
with slopes $\gtrsim 1$, not just their logarithms.
This is an indication that the AS multipole unfaithfulness, when large enough, becomes the dominant
systematic in waveform faithfulness for these two models.
This linearity is less obvious for \SEOB{} and \XPHM. The former model produces the most
faithful AS multipoles: $\MM_{22,\text{AS}}<0.01 $ for almost every case,
whereas the latter model has the most outliers indicating other systematics contaminating this
relation such as {the already mentioned MSA-related breakdown}.
Despite these exceptions, when $\MM_{22,\text{AS}}$ is large enough,
the observations above also hold for these two models.

This systematic becomes more severe with increasing mass ratio as we determined by
comparing the slopes of the linear fits to the mismatch data for \TEOB{} and \TPHM{}
between the $Q=\Qone,2$, $Q=4$ and the $Q=6$ subsets.
In each comparison, the slope of the larger $Q$ subset was greater than the slope of the smaller $Q$ subset,
though we observed this to be more severe for \TEOB{} than \TPHM.

When we looked at the intrinsic parameters of the cases that yield the highest values for $\MM_{22,\text{AS}}$ for each approximant, we found that the worst
mismatches come from the $65^\circ\lesssim\theta_{1,2}\lesssim 115^\circ$ region,
resulting in the largest planar spin projections. 
We have already identified this region as the most challenging one consistent with the literature.
The fact that the AS multipole mismatches are also the highest in this region tells
us that the true co-precessing multipoles differ much more than the AS multipoles for cases with
mostly planar spins.
We believe this to be mostly due to the imposed $m\leftrightarrow -m$ symmetry on the AS multipoles,
which we confirmed to be more violated for the strongly precessing cases.
We did this by comparing $\MM_{22,\text{AS}}$ with $\MM^\text{sym}_{22,\text{AS}}$, i.e.,  the AS multipole mismatches computed with respect to the $m$-symmetrized co-precesing \NRsurP{} multipoles, and found that the quantity $\MM_{22,\text{AS}}- \MM^\text{sym}_{22,\text{AS}}$ peaks
for $65^\circ \lesssim\theta_{1,2} \lesssim 115^\circ$, i.e., for the cases with mostly in-plane spins.

When we extended the above analysis of the AS $(2,2)$ multipole to the AS $(2,1)$ multipole,
we encountered some unexpectedly high values of $\MM_{21,\text{AS}}$ for $Q=\Qone$
for all four models. 
This contradicts the thus-far observed trend of increasing mismatches with increasing $Q$.
Upon closer examination, we discovered that the problem was due to \NRsurP:
since for $Q=1$, $h_{21}^\text{AS}=h_{21}^\text{coprec}=0$ exactly,
the model generates a co-precessing (2,1) multipole that is barely above numerical error for $Q=\Qone$.
For this reason,  we choose to disregard all $Q=\Qone$ values of $\MM_{21,\text{AS}}$.

Focusing on the $Q\ge 2$ part of the discrete set, we observe that the previously observed
tight correlation for the AS (2,2) multipole becomes much less strong for the AS (2,1) mode.
In fact, the linear trend between $\log\MM_{21,\text{AS}}$ and $\log\MM$
is only discernible  for \TPHM, with \SEOB{} and \TEOB{} showing some positive
correlation at the high mismatch end as well. As was the case with the AS (2,2) multipole,
the trends are more visible for the $M=37.5\Msun$ BBH set.
As for the cases that yield the worst AS (2,1) mismatches,
we find more of a spread in the $\{\theta_1,\theta_2\}$ space, but with the worst
common mismatches coming from the $\theta_1\le \pi/2,\theta_2\ge\pi/2$ corner,
while the cases with maximum
or near maximum $|\chieff|$ values consistently yield the lowest mismatches.
We should keep in mind that even for the worst cases where this correlation is tight,
it need not necessarily affect the overall unfaithfulness $\MMo$ significantly
as the AS (2,1) multipole amplitude is roughly an order of magnitude smaller than its (2,2) counterpart.
And as we discuss further below, the AS $(2,\pm 1)$ multipoles have a rather weak effect on the overall
strain faithfulness.

The AS $(2,\pm 1)$ multipoles' importance for faithful precessing waveform construction was investigated by
Ref.~\cite{Ramos-Buades:2020noq} where they showed that while the inclusion of the
AS $(2,\pm 1)$ multipoles degrades the precessing $(2,\pm 1)$ multipole faithfulness
(see their Table~III and Fig.~4),
it marginally improves the precessing $(2,\pm 2)$ multipoles.
How these changes affect the faithfulness of the waveform strain  is illustrated for a $Q=5$ case
in their work, where it is shown
that the inclusion of the AS $(2,\pm1)$ multipoles
lowers the $\ell=2$ strain unfaithfulness by 0.01 in the highest mismatch regions
as can be gathered by comparing the middle left with the middle right panel in their Fig.~5.
In the words of Ref.~\cite{Ramos-Buades:2020noq}, this indicates that
the improvement in the precessing $(2,\pm2)$ multipoles,
due to the inclusion of the $(2,\pm1)$ AS multipoles, compensates for
the degradation of the precessing $(2,\pm1)$ multipoles.

We extend their work here by turning ``off'' $h_{2\pm1}^\text{coprec}$ for
\NRsurP{} and looking at the unfaithfulness of the resulting strain with respect to
the full-multipole content \NRsurP{} strain.
Specifically, we compute $\MMno$ and $\MMo$ for the $\ell=2$ strain at inclinations of zero and $\pi/2$. Though we use the same multipole content as Ref.~\cite{Ramos-Buades:2020noq},
they considered all inclinations in the one example that they present whereas we suffice with
two inclinations here, but repeat the computation for all 1120 cases of the discrete set.

For $\ioo=0$, we find that $\MMo$ ranges from $\sim~10^{-8}$ to $10^{-2}$ with
mismatches increasing with $Q$ and the highest values coming from $\theta_{1,2}=\pi/2$ cases of the $Q=4,6$ subsets. For $\ioo=\pi/2$, we observe $10^{-5}\lesssim \MMo < 0.035$,
with the dependence on $Q$ becoming less important.
In other words, when the co-precessing $(2,\pm 1)$ multipoles are omitted,
we observe increased strain unfaithfulness at higher inclinations.
Accordingly, the co-precessing $(2,\pm1)$ multipoles improve
the strain faithfulness more for higher inclinations, consistent with the findings of Ref.~\cite{Ramos-Buades:2020noq}.
Interestingly, the largest values of $\MMo(\ioo=\pi/2)$ come from cases with $\theta_{1,2}\ge 5\pi/6$,
not $\pi/2$ as was the case for $\ioo=0$.


\subsubsection{Mismatches in the Extrapolation Region of \NRsurP}\label{sec:q6_comparison}
Since the approximants exhibit a degradation in faithfulness as $Q$ increases,
we extended our investigation into the extrapolation regime of
\NRsurP{} by setting $Q=6$ for the same
discrete grid as before, resulting in two additional subsets of light
and heavy BBHs with 280 cases each.
We have opted to present this comparison separately from the $Q\le 4$ cases
of the previous subsections since, as Ref.~\cite{Varma:2019csw} advises,
we must exercise caution when using \NRsurP{} in its extrapolation region, i.e., $4<Q<6, 0.8<\chi_i < 1$.
Nonetheless, comparisons with 100 \texttt{SXS} simulations
with $Q=6$ and $\chi_i \le 0.8$ have yielded 
mismatches ranging from $\sim 10^{-4}$ to $\sim 10^{-2}$
with the median at $\sim 10^{-3}$.  
Overall, this is roughly an order of magnitude worse
than \NRsurP's performance in its training region, $Q\le 4, \chi_i\le0.8$,
but good enough for our purposes here.
A more direct approach would be to compare the four models with the
aforementioned 100 \texttt{SXS} simulations, but we were unable to retrieve these
from the \SXS{} database.
We present comparisons using different sets of \SXS{}
simulations in Sec.~\ref{Sec:SXS_comparison}.

Returning again to Fig.~\ref{fig:violin_q_separated}, where the $Q=6$ $\MMno$
distributions are plotted as the rightmost half violins for each color,
we see that the worsening trends observed for the ($Q$\,=\,4)-case mismatches
persist for the $Q=6$ cases.
As before, the mismatch distributions exhibit bi or trimodalities with the
larger-mismatch peaks of $\{\SEOB,\TEOB,\XPHM\}$'s distributions coming from
cases with either large planar spins, i.e.,
$\chip \gtrsim 0.7$ or with moderate $\chip$ values combined with negative $\chi_\text{eff}$ values.
\TPHM{} once again exhibits mismatches that are more symmetric with respect to $\chieff$.
Overall, the $Q=6$ $\MMno$ values for the four models are slightly worse than the $Q=4$ values.
Interestingly,  the $\chi_\text{eff}\lesssim -0.5,
\chip\lesssim 0.2$ corner of the parameter space that previously yielded
high mismatches for \XPHM{} has shrunk to a single point that has the most negative $\chieff$
value of the entire $Q=6$ set.
This minor improvement in \XPHM's performance
is partly due to the fact that we no longer have the near-transitional precession cases encountered
for $Q=4$ where $|\J_{\text{N},0}| \ll |\L_{\text{N},0}|$.
which were causing the MSA prescription to breakdown forcing the model to default to the NNLO prescription for precession dynamics.

Taking $\MMno$ as a gauge of inspiral-only model faithfulness,
we find that $\{\SEOB,\TEOB,\TPHM,\XPHM\}$ are reasonably faithful at this mass ratio with
the percentage of cases of $\MMno>0.035$ being $\{0\%,0\%,5\%,28\%\}$ for
the light and $\{0\%,14\%,2\%,11\%\}$ for the heavy-mass sets.
As for the full mismatch $\MMo$,
the percentage of cases greater than $0.035$ is, respectively, $\{0\%,34\%,37\%,75\%\}$ for the light
and $\{16\%,41\%,46\%,77\%\}$ for the heavy-mass set.

\subsection{Behavior of Model Unfaithfulness under changing Inclination}\label{sec:discrete_inc90}

Our aim in this section is to investigate how the inclusion of the precesssing
$(2,1)$ multipole, i.e., $h_{21}$, affects the faithfulness.
Eq.~\eqref{eq:GW_polarizations} tells us that
the contribution of $h_{21}$ is maximal for $\io=\pi/2$ yielding
$|{}_{-2}Y^{21}|/|{}_{-2}Y^{22}|=2 $. 
Though $|h_{21}^\text{AS}| <|h_{22}^\text{AS}|$ always holds true,
for precessing multipoles, one can observe that $|h_{21}| \sim |h_{22}|$
(see, e.g., Fig.~1 of Ref.~\cite{Varma:2019csw}).
This is due to the power from the AS $(2,\pm2)$ multipoles rotating into $h_{21}$
which is time (frequency) dependent, as is $\io$.
Accordingly, the ${}_{-2}Y^{21} h_{21}$ term in the strain \eqref{eq:GW_polarizations}
can be dominant.
Therefore, comparisons of strain mismatches for cases with large inclinations
offer an indirect way to assess
the faithfulness of ``subdominant'' multipoles such as $h_{21}$ which, as we just explained,
may not be subdominant at all for certain configurations.
The analysis almost equally applies to $h_{2,-1}$ due to the precessing multipoles approximately
inheriting the $m\leftrightarrow-m$ symmetry
of the AS multipoles.
For a more detailed investigation of the mismodelling of
$h_{21}$ and higher multipoles $(\ell>2)$, see Ref.~\cite{Ramos-Buades:2020noq}.

This analysis also indirectly gauges the performance of the Euler angles used in the frame rotations.
Since, as shown, the faithfulness of the co-precessing $(2,\pm 2)$ multipoles matters much
more than that of the co-precessing $(2,\pm 1)$ multipoles, the cases for which the former are 
very faithful, but the resulting $\MMo$ values are high may tell us something about the mismodeling
of the Euler angles.

%
%
%
\begin{figure}[t!]
    \centering
    \includegraphics[width=0.49\textwidth]{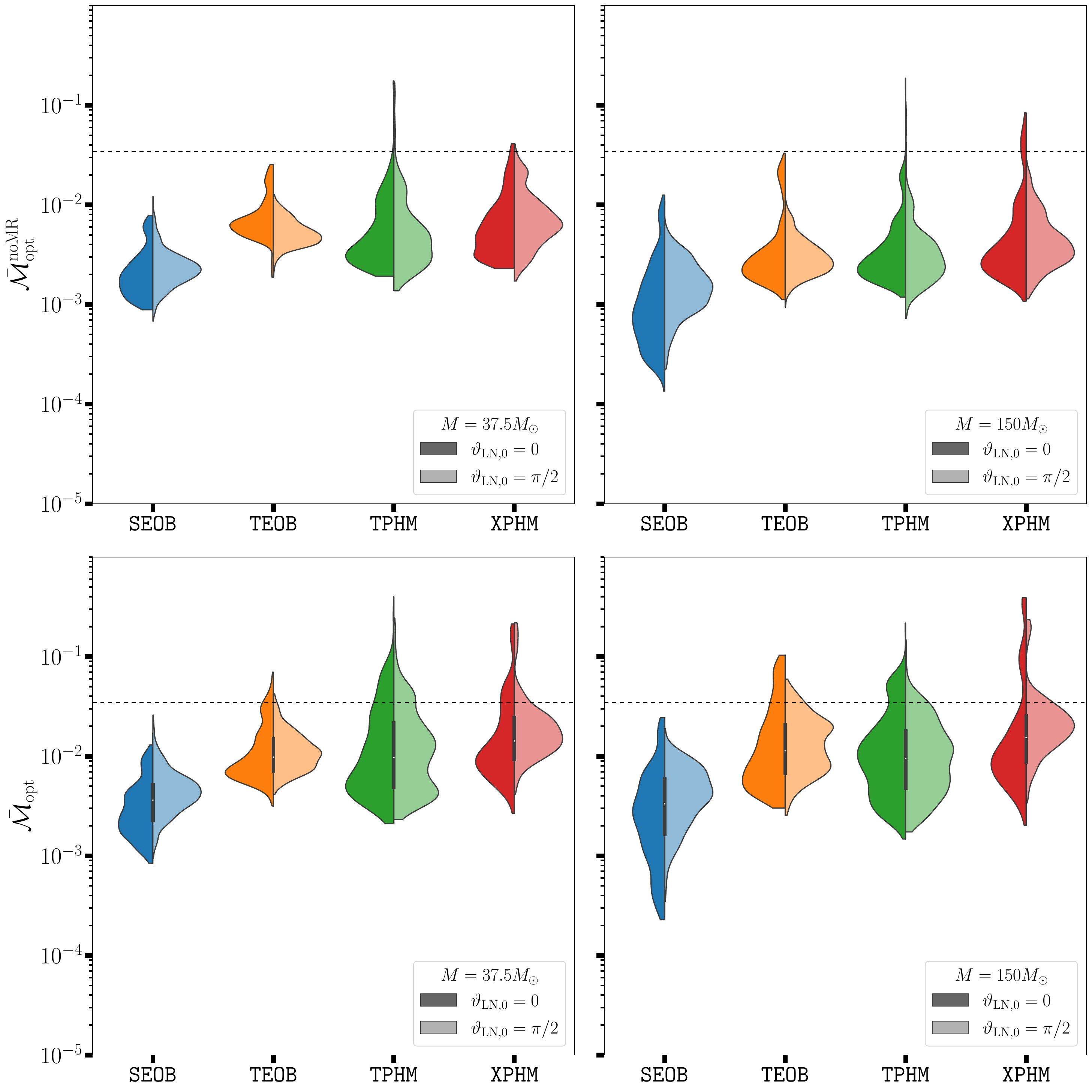}
\caption{Comparison of the mismatch distributions between face-on ($\ioo=0$)
and edge-on ($\ioo=\pi/2$) configurations.
In the top row, we display violin plots of $\MMno(\ioo=0)$ vs. $\MMno(\ioo=\pi/2)$
for the $M=37.5\Msun$ ($M=150\Msun$) sets in the left (right) panels.
Likewise, the bottom row shows $\MMo(\ioo=0)$ vs. $\MMo(\ioo=\pi/2)$ for the same sets.
Note that we exclude the $Q=6$ subset here, but we include it in Fig.~\ref{fig:violin_inc_0vs90_byQs}.}
\label{fig:violin_inc_0vs90}
\end{figure}
%
%
%
%

%
%
\begin{figure*}[t!]
    \centering
    \includegraphics[width=1\textwidth]{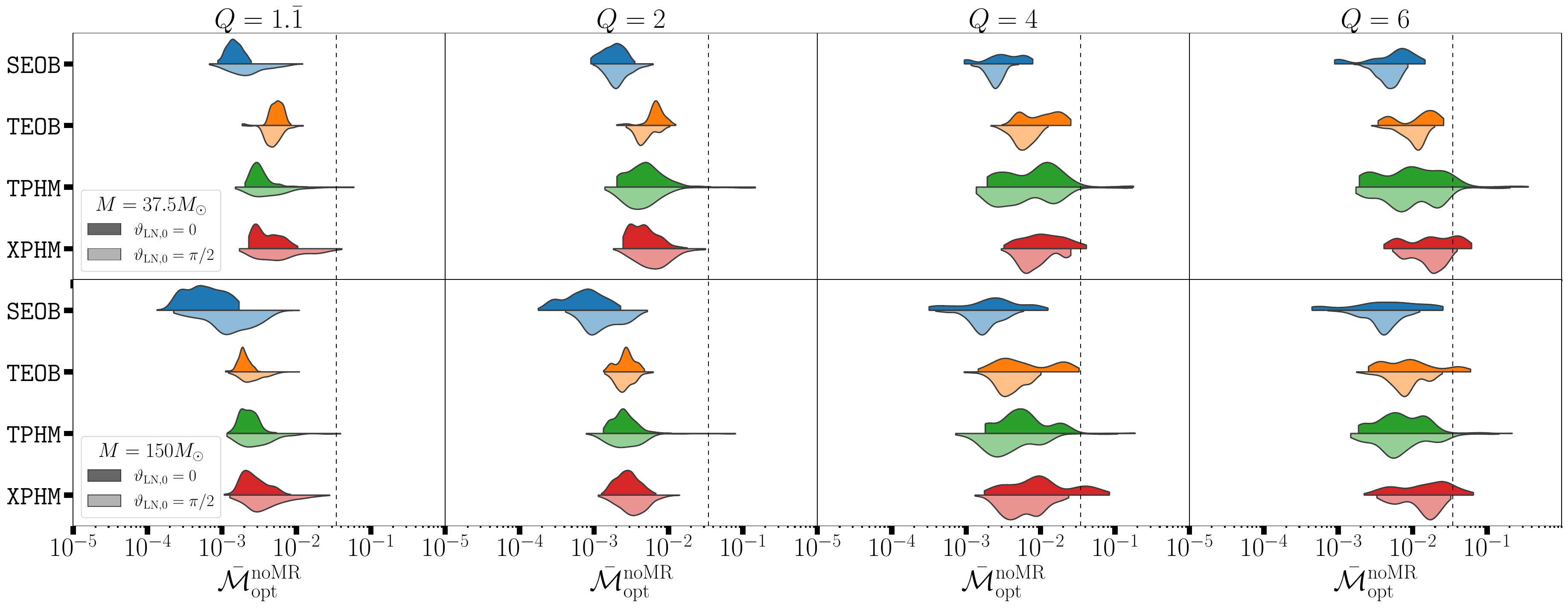}
\caption{Comparison of the mismatch distributions between ``face-on'' ($\ioo=0$)
and ``edge-on'' ($\ioo=\pi/2$) configurations separated by mass ratio.
In particular, we present results here only for the MR-truncated, sky-angle
optimized mismatch $\MMno$. The top row displays results for the light ($M=37.5\Msun$)
BBH set, and the bottom row the heavy ($M=150\Msun$) BBHs.}
\label{fig:violin_inc_0vs90_byQs}
\end{figure*}
%
%

We proceed by re-employing the discrete parameter set of Sec.~\ref{sec:NRsur_survey_discrete}
with $\ioo=\pi/2$,
and compute the mismatches $\MMno,\MMo$ as before.
Our results are presented in Fig.~\ref{fig:violin_inc_0vs90}
where we show the distributions of the mismatches between the $\ioo=0$ and
$\ioo=\pi/2$ sets for the $Q\le 4$ portion for both light and heavy binaries.
Several features stand out in the figure which we highlight next.

First, as in the $\ioo=0$ case,
the distributions of $\MMo$ are in general broader than $\MMno$ distributions and
shifted to higher values. This shift is larger for the heavy masses.
Second, the inspiral-only faithfulness of all approximants
remains mostly unchanged between $\ioo=0$ and $\ioo=\pi/2$,
with \TEOB{} even showing a slight improvement for $\ioo=\pi/2$.
There is also an overall trend of the high-mismatch tails being reduced.

In Fig.~\ref{fig:violin_inc_0vs90_byQs}, we show a decomposition of $\MMno(\ioo=0)$ and
$ \MMno(\ioo=\pi/2)$ in terms of mass ratio, where the upper
(lower) panels correspond to the $M=37.5\Msun \ (M=150\Msun)$ set.
We note several observations.
First, the $\ioo=\pi/2$ distributions exhibit less or no modality
compared with the $\ioo=0$ distributions.
Second, $Q\ge 4$ $\ioo=\pi/2$ mismatches have narrower distributions and mostly
occupy lower values than their $\ioo=0$ counterparts.
The converse holds true for the $Q\le 2$ mismatches with the exception of \TEOB.
Third, the medians of the $\ioo=\pi/2$ distributions shift to lower values
than the medians of the $\ioo=0$ distributions as
$Q$ increases, especially for the \textsc{EOB} models.
The especially remarkable feature is the improved inspiral-only faithfulness for all models for $Q=4$
at $\ioo=\pi/2$ as compared with $\ioo=0$.

For $Q=6$, there is mostly a reduction of the bi/trimodalities which we consider to be is an improvement.
However, the situation is somewhat more complicated for $Q=6$ as
the values of $\MMno(\ioo=0)$ of the left-most peaks
are {mostly} lower than the smallest values of $\MMno(\ioo=\pi/2)$ whereas
the values of $\MMno(\ioo=0)$ of the right-most peaks are {mostly} higher than the largest values of
$\MMno(\ioo=\pi/2)$. So, the question of whether or not the models show improved
inspiral-only faithfulness for $Q=6$ may have a subjective, rather than an objective answer.
Therefore, let us focus on the $Q=4$ cases next.

Given the above observations, an obvious question is: what is the cause of
the improved faithfulness for $Q=4$ in going from $\ioo=0$ to $\ioo=\pi/2$?
We hypothesize that this may be due to the $h_{2,\pm1}$ multipoles improving the
strain when they are more faithful than $h_{2,\pm2}$.
Given that at $\ioo=\pi/2$, we have $|{}_{-2}Y^{2,\pm1}|/|{}_{-2}Y^{2,\pm2}|=2$,
this improvement more likely matters only for cases where $|h_{2,\pm1}| \gtrsim |h_{2,\pm2}|$.
This may then result in $\MMno(\ioo=\pi/2)$ being less than $ \MMno(\ioo=0)$.
Henceforth, we will simply mention the (2,1) and (2,2) multipoles with the same results applying
to the $(2,-1), (2,-2)$ multipoles.

We test our hypothesis on the \TEOB{} mismatches first since it shows the most $Q=4$ improvement for $\ioo=\pi/2$
as well as showing improvements for $Q=\Qone, 2$.
We begin by computing two new quantities: $\MM_{22}^\text{\noMR},\MM_{21}^\text{\noMR}$, i.e.,
the inspiral-only $\{t_c,\varphi_c\}$-maximized mismatches [see Eq.~\eqref{eq:standard_match}] of 
the precessing $(2,2)$ and $(2,1)$ multipoles between \NRsurP{} and \TEOB. 
We also compute the ratio
\be
R:= \f{2}{N}\sum_{i=0}^N\f{|h_{21,i}|}{|h_{22,i}|} \label{eq:mode_ratio},
\ee
%
where we have $N+1$ samples in the time domain labelled by $i$.
$R$ is the Riemann sum of the following integral
\be
\int \text{d}t \frac{\left| {}_{-2}Y^{21}(\pi/2,0)\,h_{21}(t)\right|}{\left| {}_{-2}Y^{22}(\pi/2,0)\,h_{22}(t)\right|}
\label{eq:ratio_integral}
\ee
$R> 1$ tells us that the $(2,\pm1)$ terms have larger
magnitudes than their $(2,\pm2)$ counterparts in the waveform strain.
Thus when $R>1$ and $ \MM_{21}^\text{\noMR}<\MM_{22}^\text{\noMR}$, we expect the more
faithful $h_{21}$ to result in improved faithfulness
for $\ioo=\pi/2$ as compared to $\ioo=0$, i.e.,
$\MMno(\ioo=\pi/2)\preceq \MMno(\ioo=0)$ in the respective distributions.

We proceed by counting the cases satisfying the condition:
$[R> 1] \land [\MM_{21}^\text{\noMR}<\MM_{22}^\text{\noMR}] $ and 
comparing the resulting distributions for $\MMno(\ioo=\pi/2)$ with those of $\MMno(\ioo=0)$.
The expectation is that we should see the distributions of the affirmative cases satisfying 
$\MMno(\ioo=\pi/2)\preceq \MMno(\ioo=0)$ and that there are enough such cases to affect the overall distributions as seen in Fig.~\ref{fig:violin_inc_0vs90_byQs}.
We find our hypothesis validated by the $\Mtot=\Ml, Q=4$ subsets
of \SEOB{} and \TEOB{} as well as the $\Mtot=\Mh, Q=4$ subset of \TPHM.
We chose these subsets because it is evident from Fig.~\ref{fig:violin_inc_0vs90_byQs}
that this improvement is most prominent for them. We also confirmed our hypothesis for
the entire $\Mtot=\Ml,Q\le 4$ \TEOB{} set. We did not use \XPHM{} because 
it is non-trivial to obtain its precessing multipoles in the time domain as it is a frequency
domain approximant.

The MR-included version of Fig.~\ref{fig:violin_inc_0vs90_byQs} (which we do not include)
shows less modality in the $\MMo(\ioo=\pi/2)$ distributions. However, there are fewer
$Q$ subsets where $\ioo=\pi/2$ mismatches are lower than the $\ioo=0$ mismatches.
The exceptions to this relation are \SEOB's light and heavy $Q=4,6$ mismatches.
For these, we verified that the above arguments held.

The fixed angle $\vartheta_{\text{JN}}$ is also at play here:
precession-induced modulations to the waveform decrease as $\vartheta_{\text{JN}}\to 0$ \cite{Apostolatos:1994mx}. Using the \texttt{LALSimulation} function \texttt{SimInspiralTransformPrecessingWvf2PE}, 
we computed $\vartheta_{\text{JN}}$ and observe that as the $Q=4$ subset's spin tilt angles go from pointing ``northward''
to southward, $\left.\vartheta_{\text{JN}}\right|_{\ioo=\pi/2}$ drops below 
$\left.\vartheta_{\text{JN}}\right|_{\ioo=0}$.
In other words, the higher inclination cases yield smaller precession-induced modulations,
which ties in with the discussion above, but does not explain why 
the precessing $(2,\pm 1)$ multipoles are sometimes
more faithful than $(2,\pm2)$ when both have the same co-precessing (AS) multipole content dominated by $h_{2,\pm2}^\text{coprec}$ ($h_{2,\pm2}^\text{AS}$).

To answer this question, one would have to dismantle the individual co-precessing (AS) multipole
contribution to the twist expression \eqref{eq:hlm_Twist1} and compare the interplay between
various multipoles. 
We do not pursue this matter any further here. 
Ref.~\cite{Ramos-Buades:2020noq} present additional analysis in this direction, where they
compare waveform strain and individual precessing multipole mismatches for cases in which
all [useful] AS multipole content has been included and for cases where some AS multipoles, e.g., $(2,\pm1)$
have been excluded.

The final observation relevant to this section is that just as in the $\ioo=0$ case,
the distributions for $\MMo(\pi/2)$ are shifted upward, i.e., to worse values, as compared to
$\MMno(\ioo=\pi/2)$ distributions, especially for $Q=4,6$.
The fact that $\MMno \le \MMo$ is not surprising since the merger-ringdown part of the
signal is more complex as we already mentioned.
We can nonetheless ask whether or not there
might be regions in the spin space where there is a larger gap between $\MMno$ and $\MMo$ than
other regions. To this end, we again looked at how the ratio $\cal{R}=\MMo/\MMno$ is distributed in the
spin space. The approximants show a lot variability with respect to
each other and across $Q$ values. However, one nearly common feature is this ratio
reaching $\ord(10)$ for $\theta_{1,2}= \pi/2$ with \XPHM{} as the exception.
Additionally, \SEOB{} has the smallest maxima for the ratio while \XPHM{} has the largest.
\XPHM{} has more occurences for large ratios in the $\chieff<0$ half whereas the other three
models yield more large ratios in the $\chieff>0$ half.
\XPHM{} also behaves differently for the $Q=4$ subset with the ratio showing very little
variation in the spin space and tending to the smallest values at $\theta_{1,2}= \pi/2$.


\subsection{Comparisons over a uniformly filled Parameter Space}\label{sec:NRsur_survey_random}
%
%

We now consider a uniformly
populated parameter space consisting of 1000 BBHs
with $M\in [{35}, {225}]\Msun, Q\in[1,4], \chi_i \in [0.1,0.8]$.
We choose to make the spin tilt angles uniform in their cosine, i.e.,
$\cos\theta_i\in[-1,1]$ with $\phi_1=0,\phi_2\in[0,2\pi)$.
To ensure once again that the resulting inspirals are not longer than 4300M,
we set $f_0$ equal to a linearly decreasing function
of $M$ from {35}\,Hz to $5.75$\,Hz as $M$ increases from {35}$\Msun$ to ${225}\, M_\odot$.
We fix $f_i=f_0+{3}$\,Hz ($11$\,Hz) for $M\lesssim 210\Msun \, (\gtrsim 210\Msun)$
in the integral \eqref{eq:inner_prod}.
In the following, we present the mismatches similarly to what we have done
for the discrete set of Sec.~\ref{sec:NRsur_survey_discrete} so that
we can illustrate the differences in the approximant performance
between a more traditional parameter space (uniformly filled) and one for which
certain parameters are tailored to potentially better expose precession related modelling systematics.

\begin{figure}[t!]
    \centering
    \includegraphics[width=0.5\textwidth]{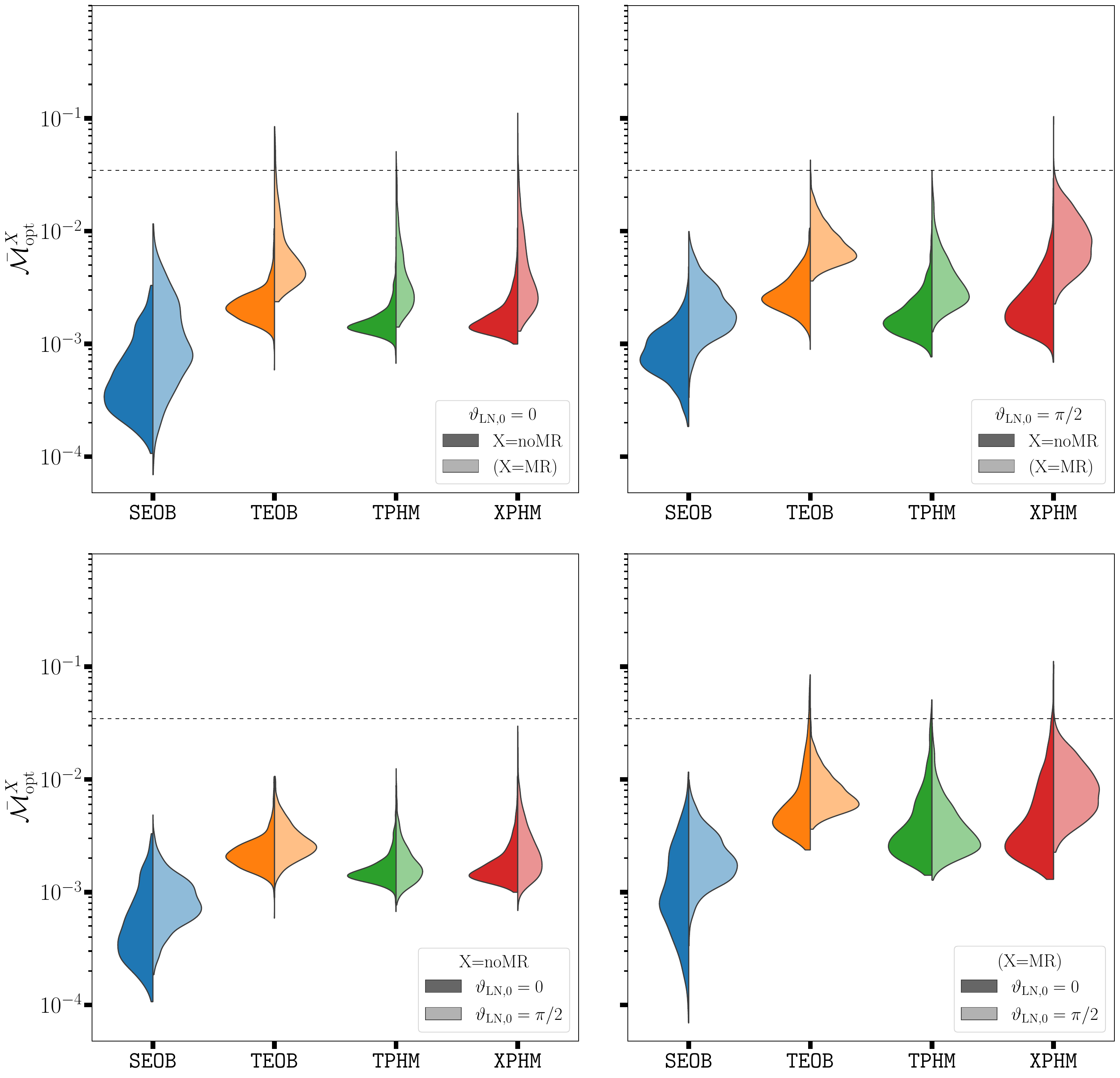}
\caption{Mismatches between \NRsurP{} and \SEOB, \TEOB, \TPHM, \XPHM{} for the uniformly
filled parameter space of 1000 cases. Upper panels compare $\MMno$ with $\MMo$ at inclinations
of $\ioo=0$ (left) and $\pi/2$ (right) and the lower panels compare $\MMno(\ioo=0)$
with $\MMno(\ioo=\pi/2)$ (left) and $\MMo(\ioo=0)$ with $\MMo(\ioo=\pi/2)$ (right).}
\label{fig:violin_random_Set}
\end{figure}

We summarize our main results in Fig.~\ref{fig:violin_random_Set},
where in the top row we compare $\MMno$ with $\MMo$ at inclinations of $\ioo=0$ ($\pi/2$)
in the left (right) panels.
As expected, the relation $\MMno \preceq \MMo$ holds likely due to the combination
of MR modelling systematics and simpler signal morphology for the inspiral.
The shifts between the peaks of $\MMno$ and $\MMo$ distributions are larger for $\ioo=\pi/2$
than $\ioo=0 $ as can be discerned from the upper left and right panels of the figure.
This was also the case for the discrete set though this is somewhat hard to extract from Fig.~\ref{fig:violin_inc_0vs90}.
This is a reaffirmation of the fact that the modelling of the plunge-merger-ringdown stages
of the precessing $(2,\pm1)$ multipoles remains challenging.

In the bottom panels of Fig.~\ref{fig:violin_random_Set}, we show how
the mismatches shift upward more for \SEOB{} and \XPHM{} than for \TEOB{} and \TPHM{}
as the inclination goes from 0 to $\pi/2$, more consistent with the $M=150\Msun $(right) panels of
Fig.~\ref{fig:violin_inc_0vs90} than $M=37.5\Msun$ (left) ones.
This somewhat makes sense as $150\Msun$ roughly equals the median of $M\in[35,225]\Msun$,
whereas $37.5\Msun$ is at the lower end.

%
%
%
%

\begin{figure}[t]
    \centering
    \includegraphics[width=0.5\textwidth]{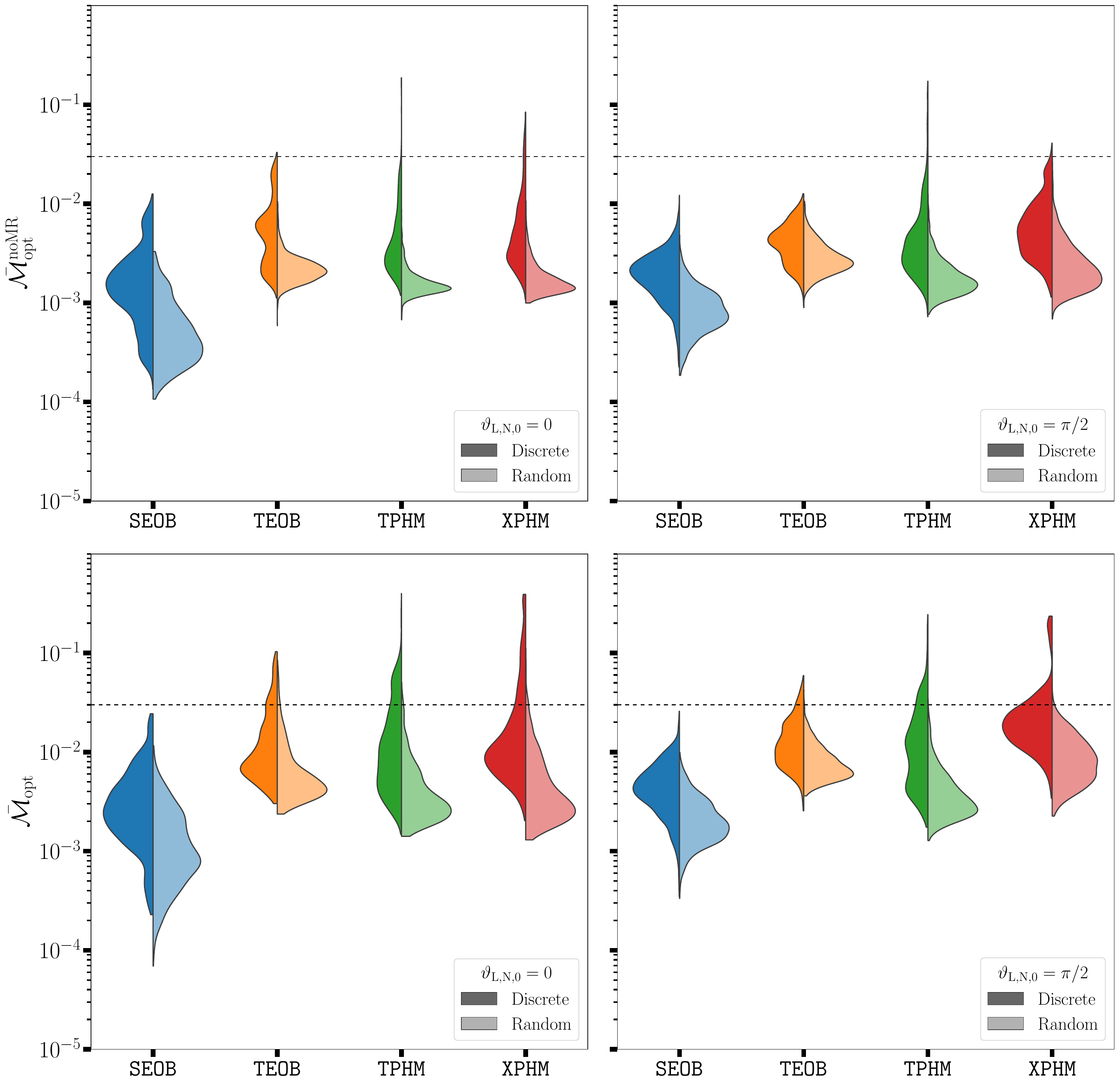}
\caption{Comparison of model mismatches to \NRsurP{} between the discrete parameter set and
the random-uniformly filled one.
The left half violins are the mismatch distributions from the entire
$M=\{37.5\Msun, 150\Msun\}, Q=\{\Qone,2,4\}$ discrete set while the right ones
are the mismatch distributions from the 1000-case random-uniformly filled set.
In the top (bottom) row, we plot $\MMno$ ($\MMo$) at inclinations of 0 (left panels) and $\pi/2$ (right panels).
}
\label{fig:compare_discrete_vs_uniform}
\end{figure}

Fig.~\ref{fig:compare_discrete_vs_uniform} better highlights the differences in
model faithfulness resulting from using a uniformly filled set
vs. our purpose-built discrete set of Secs.~\ref{sec:NRsur_survey_discrete}, \ref{sec:discrete_inc90}.
In this figure, the left half violins (darker shaded) are the mismatch distributions from the entire
$M=\{37.5\Msun, 150\Msun\}, Q=\{\Qone,2,4\}$ discrete set, while the lighter-shaded right violins
correspond to the mismatch distributions from the 1000-case random-uniformly filled set.
One immediately notices that the distributions for the latter (uniform) set
have smaller minima, means and maxima than the former (discrete) set.
Though the shifts between the respective minima are not so large,
the offset between the means and the maxima are quite prominent.
The distributions for the uniform set are also smoother, narrower and unimodal
owing to the nature of the uniformly-filled intrinsic parameter space.

%
%
%
%

The main conclusion of Fig.~\ref{fig:compare_discrete_vs_uniform} is that the model
faithfulness is better for the uniform set than our purpose-built discrete set.
This is, of course, not surprising since intrinsic parameters of the discrete set are such that
it contains many more
cases with strong precession than the uniform set, where $\chi_{\text{p}}(t_0)\ge 0.6$
for 60\% of the cases for the former vs. 13\% for the latter.
We already documented this contrasting coverage of the spin space in Fig.~\ref{fig:spin_space}.
Additionally, one third of the discrete set is made up of $Q=4$ cases
whereas the uniform set has $\sim 17\%$ of its cases with $3.5\le Q<4$.
Recall that these high-$Q$ cases appear to be the most challenging for the models
that we consider here, as can be gathered from Figs.~\ref{fig:violin_q_separated}
and \ref{fig:violin_inc_0vs90_byQs}.

Perhaps the difference between the two sets is best illustrated by considering
the ``extremal corner'' of the five-dimensional $\{Q,\chi_1,\chi_2,\theta_1,\theta_2\}$
parameter space. Let us as, an example,  assign to this region the following bounds:
$Q\in[3.5,4],$ $\chi_i\in[0.7,0.8],$ and $\cos(\theta_i)\in[-0.1,0.1]$.
A straightforward computation shows that the uniform set needs to consist of 20379 points
in order to have a $50\%$ chance that one case will be inside this extremal region,
whereas the discrete parameter set
has 8 out of 840 cases in this extremal corner (see Table~\ref{tab:params}).

Let us elaborate why the differences shown in Fig.~\ref{fig:compare_discrete_vs_uniform} matter.
Our concern is for strongly precessing events like GW200129\_065458
for which the PE routines should mostly draw samples from the large $\chip$ region of
the spin space. This in itself is not a problem. However, if one employs a model whose faithfulness,
to a target model such as \NRsurP,
has been documented over a uniformly filled parameter space,
then one may overestimate the faitfulness of that model.
For example, a figure akin to our Fig.~\ref{fig:violin_random_Set} may lead
to the conclusion that model ``X'' is better than 0.035 faithful in the entire parameter space,
but it may turn out that this model X actually has a large mismatch tail extending to $\sim 0.1$ in the relevant
corner of the parameter space as illustrated in our Fig.~\ref{fig:compare_discrete_vs_uniform}.
This would be quite problematic for reliable parameter estimation from strongly precessing events.
In short, uniformly sampled intrinsic parameter spaces may not contain the best set of parameters
to properly ``stress-test'' precessing approximants.

Let us conclude this section by providing additional metrics.
As done in the previous section, we once again look at the ratio $\cal{R}=\MMo/\MMno$. 
For $\ioo=0$, we observe that
$\mathcal{R} \lesssim 5 $ everywhere in the parameter space for all four models.
For $\ioo=\pi/2$, we record that $\mathcal{R}\lesssim 10 $.
For both inclinations, these upper bounds are roughly half of the corresponding bounds
of $\mathcal{R}$ for the discrete parameter set. As before \XPHM{} has the largest values of $\cal{R}$
and \SEOB{} the smallest with \TPHM{} yielding very similar magnitudes for $\cal{R}$
in the parameter space. One major difference with respect to the discrete set is that
as we uniformly fill the mass ratio space from $Q=1$ to 4, the clearcut regions
of large $\cal{R}$ observed for the discrete set at, e.g., $Q=4$, are no longer there.
Overall, the values of $\cal{R}$ are more uniformly spread in the spin space and show less
variation.

Finally, we count the percentage of cases with $\MMno>0.035$ and $ \MMo>0.035$
for both inclinations. As is clear from Fig.~\ref{fig:violin_random_Set},
this is 0\% for all \SEOB{} mismatches and 0\% for all other $\MMno$ values
regardless of inclination.
As for $\MMo$ percentages, we record $\{2\%, 0.4\%, 0.8\%\}$ for $\MMo(\ioo=0)>0.035$,
and $\{0.1\%, 0\%, 0.2\%\}$ for $\MMo(\ioo=\pi/2)>0.035$, respectively
for $\{\TEOB,\TPHM,\XPHM\}$.
These can be compared with the ones from the discrete set provided at the end of Sec.~\ref{sec:Q_1_2_4cases}.

\section{Faithfulness Survey II: Comparisons with Numerical Relativity Waveforms}\label{Sec:SXS_comparison}
In the second part of our survey, we assess the faithfulness of
$\{\SEOB,\TEOB,\TPHM,\XPHM\}$ to numerical relativity waveforms from
the \SXS~\cite{sxs_tools} catalog. Specifically, we compare against 317 short \SXS{} waveforms
and {23} long \SXS{} waveforms of the \textsc{LVCNR} catalog \cite{Schmidt:2017btt},
where short (long) means that the total number of GW cycles
in the simulations is $\lesssim 70\, (\gtrsim 125)$.
As such, the short \SXS{} waveforms are more comparable in length to our discrete and uniform
set \NRsurP{} waveforms.
For consistency with the previous sections,
we set $M=37.5\Msun, 150\Msun$ once again for the short waveform comparisons.
For the long waveforms, we consider several other values for $M$ as there are only
23 cases to compare.

\subsection{Short \SXS{} Waveforms}\label{sec:Short_SXS}
%
%
\begin{figure}[t]
     \centering
        \includegraphics[width=0.5\textwidth]{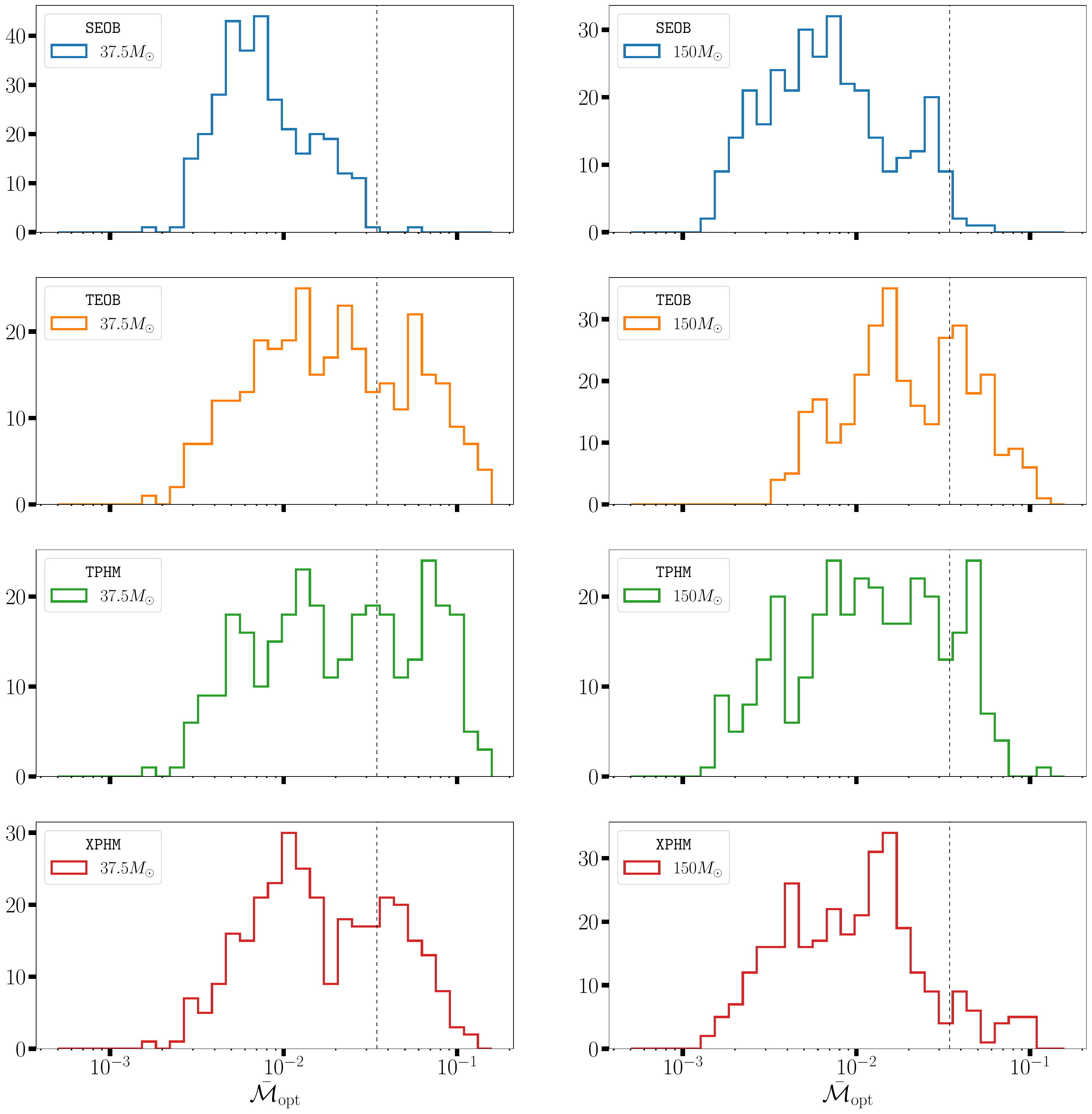}
    \caption{Mismatches defined by Eq.~\eqref{eq:mismatch_opt}
        between the four approximants and the 317 short \SXS{} waveforms shown at an inclination of $\ioo=0$
        for a direct comparison with Fig.~\ref{fig:main_mismatches}.
        The left and right columns of panels correspond to
        mismatches from the light ($M=37.5\Msun$) and heavy ($M=150\Msun$) binaries.
        }
        \label{fig:SXS_mismatches}
\end{figure}
%
%

\begin{figure*}[t]
     \centering
     \begin{subfigure}[b]{0.48\textwidth}
         \centering
         \includegraphics[width=\textwidth]{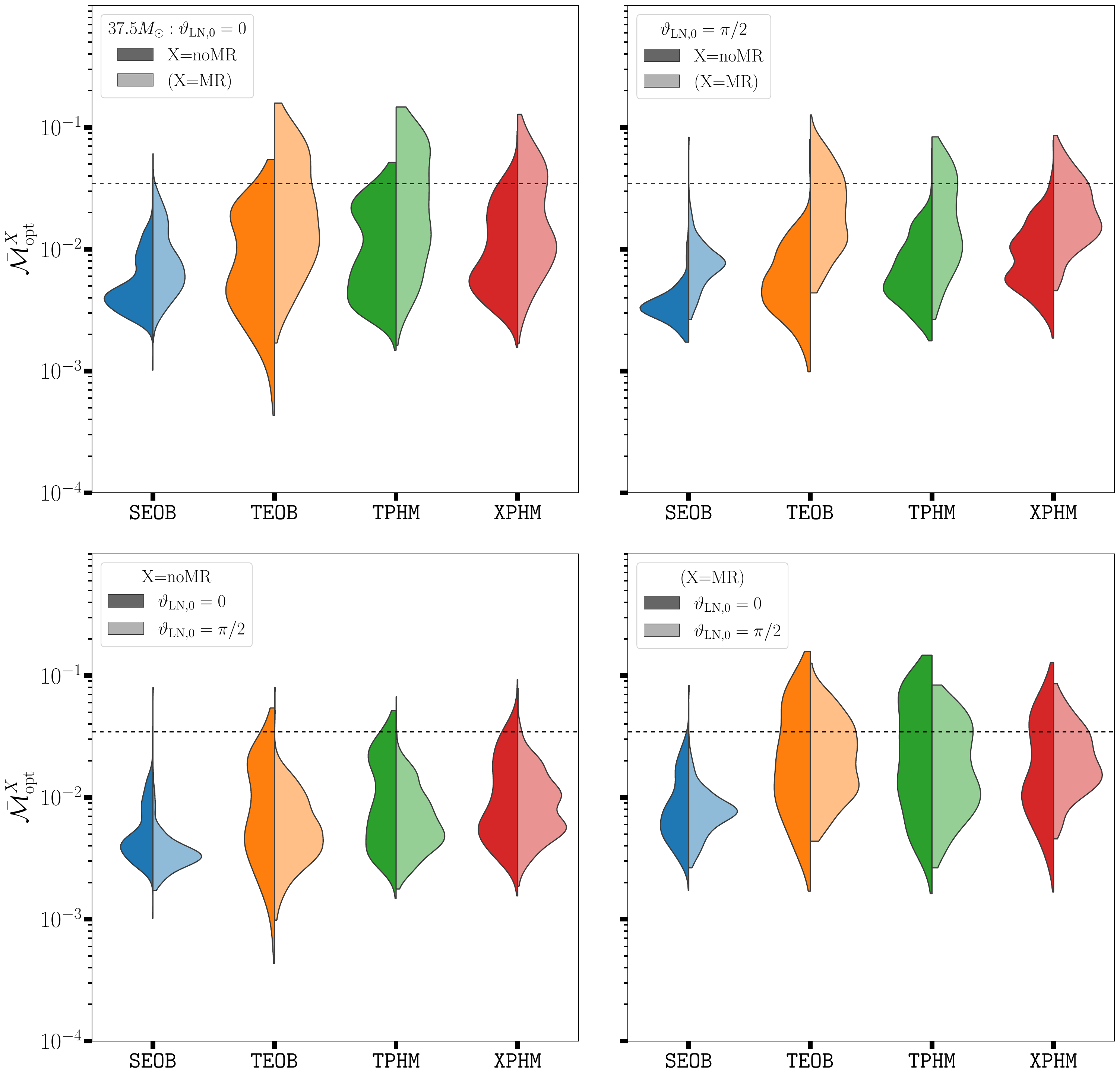}
     \end{subfigure}
     \hfill
     \begin{subfigure}[b]{0.48\textwidth}
         \centering
         \includegraphics[width=\textwidth]{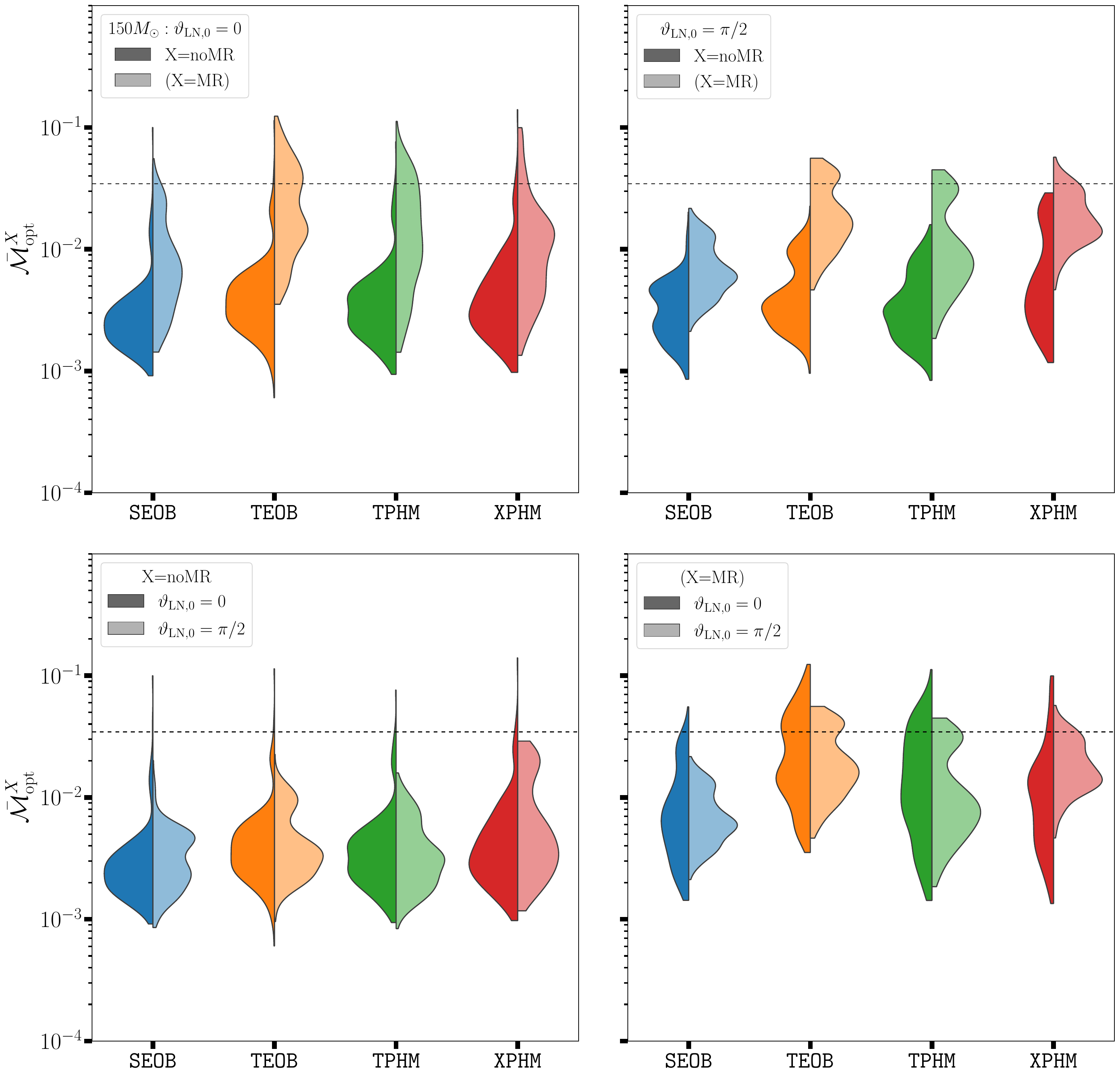}
         \end{subfigure}
        \caption{Merger-ringdown truncated and full mismatches
        between the four approximants and the 317 short \SXS{} waveforms shown for inclinations of 0 and $\pi/2$. The left (right) $2\times2$ blocks of figures correspond to
        light, $M=37.5\Msun$, (heavy, $M=150\Msun$) binaries.
        The top panels compare $\MMno$ with $\MMo$ for $\ioo=0$ and $\pi/2$ and the bottom panels compare $\MMno(\ioo=0)$ with $\MMno(\ioo=\pi/2)$ and $\MMo(\ioo=0)$ with $\MMo(\ioo=\pi/2)$.}
        \label{fig:SXS_short_violin}
\end{figure*}
We employ a specific data set containing 317 short \SXS{}
waveforms, with $\chip \lesssim {0.89}$, $\chieff \in [{-0.6, 0.66}]$
and $Q \lesssim 6$. These have between 13 and 70 GW cycles.
Though there are roughly 1500 \SXS{} BBH simulations from which we can extract waveforms,
we purposefully formed a specific subset similar to that of the discrete set of Secs.~\ref{sec:NRsur_survey_discrete}, \ref{sec:discrete_inc90}.
Specifically, our short \SXS{} set has 98 cases with $Q<1.3$, 114 cases with $1.9< Q<2.1$ and
101 cases with $3.99<Q<4.01$, thus having nearly the same mass ratio split as our discrete set.
As there are only three $Q\approx5$ and one $Q\approx6$ simulations in our set, we discuss
them separately than the 313 $Q\lesssim4$ cases.
The relevant spin space parameters of these simulations are plotted in Fig.~\ref{fig:spin_space},
where it can be seen that there are many cases with $\chip \approx 0.8$ and $|\chieff|\ll 1$.
We obtained the parameters for each simulation using \texttt{LALSimulation}
\cite{lalsuite}\footnote{We employ the tools at \url{https://git.ligo.org/waveforms/lvcnrpy/-/tree/master} with public SXS data} which requires the recasting of \SXS{} data into the \textsc{LVCNR} catalog format for which we employed the tools of the \texttt{sxs} package \cite{Boyle_The_sxs_package_2023}.
We provide the list of the 317 \SXS{} simulations that we use in the linked \git{} repository.

For the match computation, we retain the same values of $d_\text{L}=500\,$Mpc,
$f_{f}=1024\,$Hz
and the same grid over the $\{\kappa,\phi_\text{ref}\}$ space.
As \SXS{} simulations start from differing frequencies, we do not employ a fixed
value for $f_0$, but use instead the \texttt{f\_lower\_at\_1MSUN} attribute \cite{Schmidt:2017btt} of each simulation's data
to obtain the frequency in Hertz for $1\Msun$ then rescale for $M=37.5\Msun, 150\Msun$.
As before, we set $f_i = f_0+3\,$Hz ($11\,$Hz) in the match integral for light (heavy)
systems.

The main results for zero inclination ($\ioo=0$) are shown in Fig.~\ref{fig:SXS_mismatches}
where we plot the distributions of $\MMo$.
What stands out the most is the superior performance of \SEOB{} as compared with the
other three models. It yields only a handful of cases of $\MMo>0.035$ whereas \TEOB, \TPHM{}
and \XPHM{} have many mismatches above 0.035, especially for the light mass BBHs,
with a few cases of even $\MMo>0.1$, which have
$\{Q,\chip\}=\{\Qone,0.89\},\{2,0.85\},\{4, 0.8\},\{6,0.77\}$.
A quick count yields that $\{0.3,30,35,27\}\% $ $(\{1,31,17,9\}\%)$ of the 317 cases have $\MMo>0.035$ for
$\{\SEOB,\TEOB,\TPHM,\XPHM\}$ for the light (heavy) binaries.

\subsubsection{Effects of Merger-Ringdown on Faithfulness}
\label{sec:MR_short_SXS}
\begin{figure}[t!]
     \centering
        \includegraphics[width=0.5\textwidth]{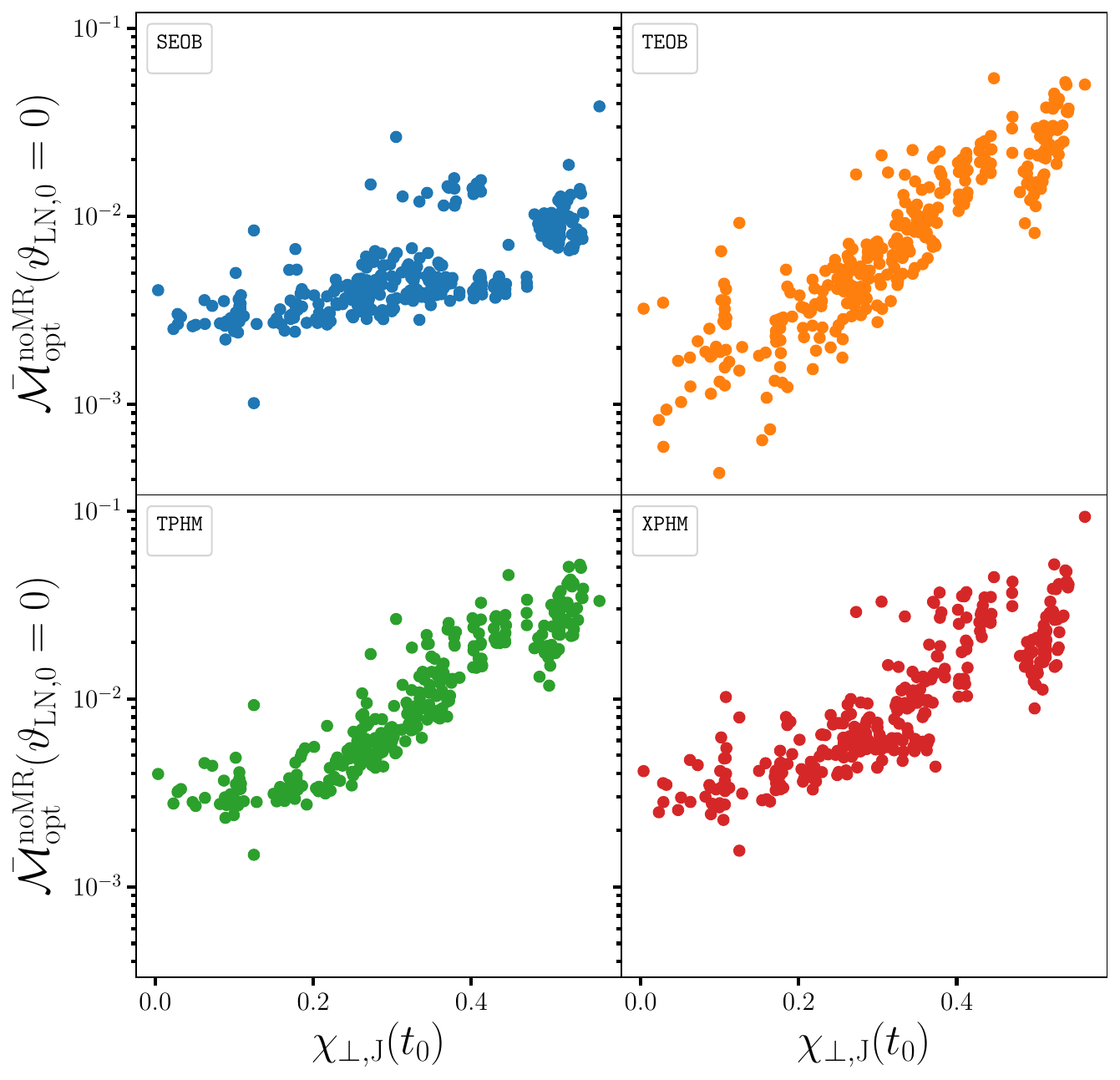}
    \caption{Inspiral-only mismatches at $\ioo=0$, i.e., $\MMno(\ioo=0)$ between our entire $Q\lesssim 4$
    short \SXS{} set of waveforms and \SEOB, \TEOB, \TPHM, \XPHM{} plotted as a function of $\chiperp$
    [Eq.~\eqref{eq:chiperp}]. Since $\chiperp$ is maximized for $\theta_{1,2}\to \pi/2$, we
    observe increasing mismatches as the spins become more planar, i.e., for stronger precession.
    }
        \label{fig:mismatches_noMR_inc0_misc}
\end{figure}

As we did in Sec.~\ref{sec:discrete_MR_vs_noMR},
we compare $\MMno$ with $\MMo$ for all four models in order to
investigate how the model faithfulness is affected by the inclusion of the MR regime.
When considering the inspiral-only mismatch $\MMno$,
the percentages above drop significantly, to $\{0.3,4,4,5\}\% $ $(\{0.6,1,0.6,1\}\%)$.
This drop is shown across the top row of Fig.~\ref{fig:SXS_short_violin} where
we display the distributions for both $\MMno$ and $\MMo$
at $\ioo=0$ and $\pi/2$ for $M=37.5\Msun, 150\Msun$.
As was the case with our discrete set, we observe an increased shift between the $\MMno$ and $\MMo$
distributions as the binary mass increases from $37.5\Msun$ to $150\Msun$.
The upward shift also increases more when the inclination goes from 0 to $\pi/2$ though less
clearly so for the heavy binaries.
We also observe that all four models behave extremely similarly for \emph{inspiral}-only
heavy BBHs, for which we show the distributions of $\MMno(\ioo=0)$ in the upper left panel of
Fig.~\ref{fig:mismatches_noMR_inc0_misc}.
This is not the case for the longer-inspiral light-mass binaries except for \SEOB{}
for which we observe very similar $\MMno$ distributions at both masses.
We discuss our results for $\MMno$ further below.

Looking, as before, at the spread of the ratio $\cal{R}=\MMo/\MMno$ over spin space, we see that
$\cal{R}\lesssim \text{4}$ for light \SEOB, \TPHM{} and \XPHM, with \TEOB's range up to 7.
For heavy BBHs, \SEOB, \XPHM{} yield $\cal{R}\lesssim \text{7}$, with $\cal{R}\lesssim \text{14}$ for \TPHM{} and $\cal{R}\lesssim \text{22}$ for \TEOB.
The near-maximal and maximal values cluster around the $\theta_{1,2}\approx\pi/2$ region of the parameter space.
Moreover, the $\theta_1>\pi/2$ half of the parameter space yields lower values for $\cal{R}$
than the $\theta_1<\pi/2$ half though with plenty of variability in the $\{\theta_1,\theta_2\}$ space. The range of $\cal{R}$ shows a weaker dependence on the mass ratio here than was observed in
Sec.~\ref{sec:discrete_MR_vs_noMR}. This is due to the fact that we had previously used the same spin
parameters at all three mass ratios, whereas in this section, the spin
parameters and the mass ratios are dictated by the \SXS{} simulations.

The mismatch distributions also exhibit some multimodality regardless of histogram binning, which
is more prominent for \TEOB, \TPHM{} and \XPHM{} for $M=37.5\Msun$ as evident in Fig.~\ref{fig:SXS_mismatches}. This was not the case for the $Q\le 4$ discrete
set as can be gathered from Fig.~\ref{fig:main_mismatches} though multimodalities emerged for its $Q=4$ subset shown in Figs.~\ref{fig:violin_q_separated},
\ref{fig:violin_inc_0vs90_byQs}.
More multimodalities are seen in Fig.~\ref{fig:SXS_short_violin} in the various distributions such as
\TEOB{} and \TPHM's $\MMo(\ioo=\pi/2)$ for $\Mtot=150\Msun$ (lower right panel).
Some of the modes seem to be due entirely to the MR part of the mismatch.
For example, neither the $M=37.5\Msun$, nor the $M=150\Msun$ distributions for \SEOB's
$\MMno(\ioo=0)$ show any of the secondary modes present in the $\MMo(\ioo=0)$ distributions as seen in Fig.~\ref{fig:SXS_mismatches}, which stand out more clearly using different histogram binning.
We observe similar, MR-related modes appearing in all the distributions for $\MMo$ for all models.
Therefore, we narrow our focus to $\MMno$ in the next few paragraphs.

%
%
\subsubsection{Dependence of Inspiral-only Faithfulness on Intrinsic Parameters}
%
%
\begin{figure*}[t]
     \centering
        \includegraphics[width=1\textwidth]{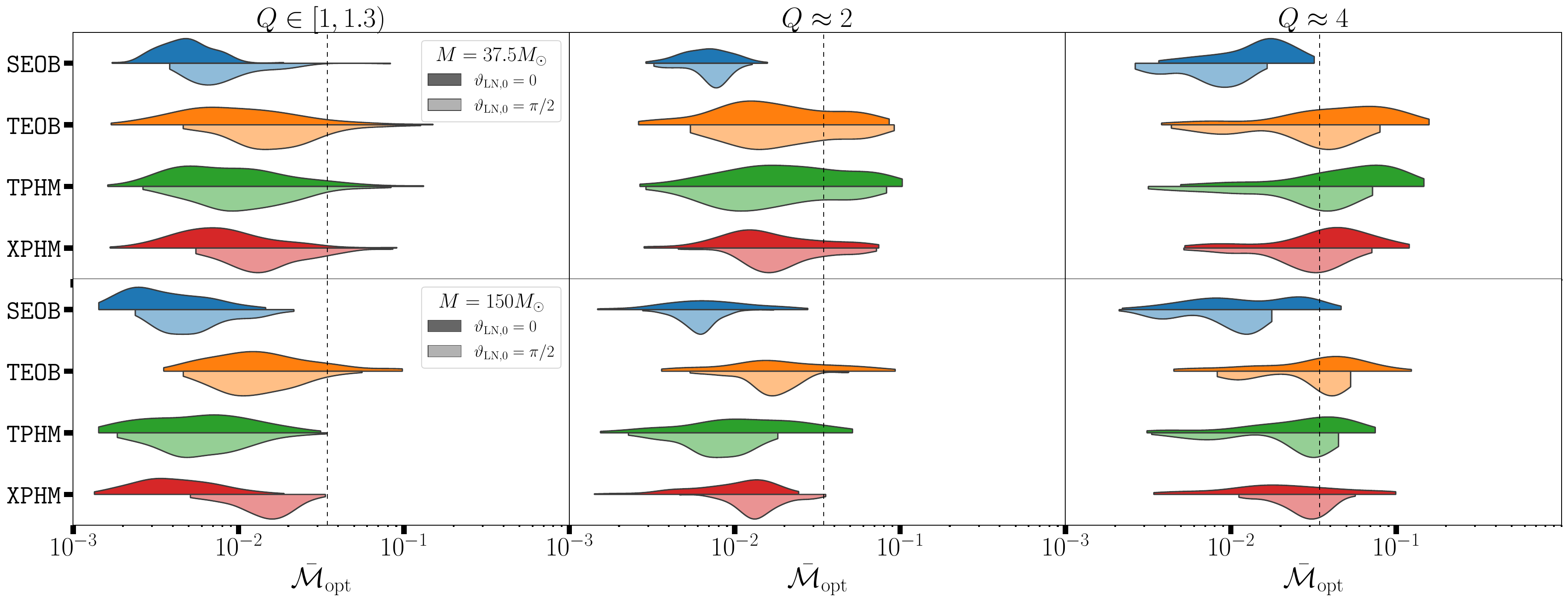}
    \caption{Sky-optimized mismatches [Eq.~\eqref{eq:mismatch_opt}] between
    $\{\SEOB,\TEOB,\TPHM,\XPHM\}$ and the 317 short SXS
waveforms shown for inclinations of $\ioo=0$ and $\pi/2$, separated into three subsets in terms of the
mass ratio $Q$. The top (bottom) panels show the mismatch distributions for binaries with total mass
$\Mtot=\Ml\,(\Mh)$.
        }
        \label{fig:SXS_mismatches_by_Q_and_inc}
\end{figure*}

As done in Sec.~\ref{sec:NRsur_survey_discrete}, we investigate whether or not specific modes
in the $\MMno$ distributions can be attributed specific regions in the $\{Q,\theta_1,\theta_2\}$ space.
As briefly mentioned, we observe unimodal, positively skewed distributions for the heavy-mass $\MMno(\ioo=0)$
for all models as well as for the light-mass $\MMno(\ioo=0)$ \SEOB{} distribution.
However, the same distributions for light-mass $\{\TEOB,\TPHM,\XPHM\}$
exhibit a clear secondary mode peaking between $\MMno= 0.02$ and 0.03, which emerges at $Q\approx2$,
and becomes dominant at $Q\approx 4$.
This behavior is consistent for $\{\TEOB,\TPHM,\XPHM\}$ indicating very 
similar inspiral performance for these models.

The tilt angles corresponding to the $Q\approx2$ secondary mode all come from
cases with $\theta_{1,2} \approx \pi/2, \chi_{1,2}\approx 0.8$ similar to our results in
Sec.~\ref{sec:Q_1_2_4cases}.
For $Q\approx4$, it is mostly the cases with $\theta_{1} \approx \pi/2$ and $\chi_1\approx 0.8$
that yield the mismatches in the now-dominant mode 
of the $\MMno$ distributions for $\{\TEOB,\TPHM,\XPHM\}$. 
There is no strong $\theta_2$ dependence as $|\Sb|\ll|\Sa|$ for $Q\ge4$. There is however a second smaller region of high mismatches clustered
around $\theta_1\approx 3\pi/4$.
The common theme in the above analysis is that strongly precessing cases remain challenging
for the models. This is even true for comparable mass ratios, where we found that the highest
$\MMno(\ioo=0)$ values come from cases with largest planar spin projections.
We demonstrate this in Fig.~\ref{fig:mismatches_noMR_inc0_misc},
where we plot $\MMno(\ioo=0)$ vs. $\chiperp$ for the entire $\Mtot=\Ml,Q\lesssim 4$ short \SXS{} set.
We opt for $\chiperp$ as $\chip$ yields many degenerate values and $\chipGen$ correlates less clearly.
Recall that $\chiperp$ becomes maximum for $\theta_{1,2}\to \pi/2$, thus the figure tells us that
the highest mismatches come from cases with $\theta_{1,2}\approx \pi/2$.

\subsubsection{Model Faithfulness under increasing Inclination}
\label{sec:short_SXS_inc}
Returning to Fig.~\ref{fig:SXS_short_violin}, we first focus on the bottom row where
we compare $\ioo=0$ mismatches with the $\ioo=\pi/2$ ones.
For light BBHs, we observe once again that the distributions for $\ioo=\pi/2$ are narrower,
consistent with the left panels of Fig.~\ref{fig:violin_inc_0vs90}.
Overall the inspiral-only performance of the models is comparable modulo \SEOB's superior performance
for light BBHs. \TEOB{} and \TPHM{} yield very similar $\MMno(\ioo=\pi/2)$ distributions.
For light BBHs, \XPHM's distribution is also very similar to these two.
For heavy BBHs, there is less similarity in the distributions.
Nonetheless, the ranges of the mismatches for all models are very comparable modulo
a secondary mode appearing for \XPHM{} seen in the upper right panel of Fig.~\ref{fig:violin_inc_0vs90}.
A quick count yields $\{0.3,0.6,0.6,1\}\% $ of the light mass, and $\{0, 0,0,0\}\%$ of the heavy
mass cases resulting in $\MMno(\ioo=\pi/2)>0.035$. These percentages are lower than their
$\ioo=0$ counterparts. A similar count returns $\{0.3,28,23,20\}\%\; (\{0,22,4,8\}\%) $
for the percentage of light (heavy) cases yielding $\MMo(\ioo=\pi/2)>0.035$, which are also lower than the corresponding
$\ioo=0$ percentages.

The large offsets between the $\MMno(\ioo=\pi/2)$ and $\MMo(\ioo=\pi/2)$ distributions
for both light and heavy masses shown in Fig.~\ref{fig:SXS_short_violin} is consistent with our previous results of Sec.~\ref{sec:discrete_inc90}.
As before, we also observe that changing inclination has a subdominant effect on model faithfulness
compared to the MR portion of the waveforms.
As for the ratio $\cal{R}$, we observe that
$\cal{R}\lesssim \{\text{3, 8, 5, 5}\}$ for the light and
$\lesssim \{\text{5,10,9,14}\}$ for the heavy $\{\SEOB, \TEOB, \TPHM, \XPHM\}$ mismatches.
Comparing these with the $\ioo=0$ ratios of Sec.~\ref{sec:MR_short_SXS} tells us that
$\cal{R}$'s range has barely changed for the light binaries.
In the case of the heavy binaries, $\cal{R}$'s maximal values have decreased for \SEOB, \TEOB{} and \TPHM, while it has increased for \XPHM.
The large changes in the maximal values of $\cal{R}$ are due to a few cases, each of which might be
challenging to a particular model, while not to others, and vice versa.

The $\theta_{1,2}\approx\pi/2$ region still results in larger values of $\cal{R}$.
And the $\theta_1>\pi/2$ half of the parameter space still yields lower values for $\cal{R}$
than the $\theta_1<\pi/2$ half though this is less distinct than it was for $\ioo=0$, and with
\XPHM{} violating this trend with the region $\theta_{1,2}\gtrsim 2\pi/3$ also returning
large values of $\cal{R}$. Consistently with our previous findings, the maximal values of $\cal{R}$
do not vary much with changing mass ratio.

A striking feature in Fig.~\ref{fig:SXS_short_violin} is the
prominent bimodalities in the $\ioo=\pi/2$ mismatch distributions for $\Mtot=150\Msun$,
which are not present for $\Mtot=37.5\Msun$.
This bimodality is most prominent for the heavy-mass \TEOB{} and \TPHM{} mismatches,
and is seen, albeit less strongly, for the heavy \SEOB{} and the light \TEOB, \TPHM{} mismatches
(referring specifically to $\MMo$).
Decomposing these mismatches further into their mass ratio subsets as shown in
Fig.~\ref{fig:SXS_mismatches_by_Q_and_inc} clarifies that the bimodalities are due
to the mismatches clustering into two separate regions for $Q\lesssim 2$ vs. $Q\approx 4$.
We had already observed a similar behavior for the $\ioo=0$ mismatches as $Q$ increased from 1 to 4,
which we replot in Fig.~\ref{fig:SXS_mismatches_by_Q_and_inc}, mirroring our $\ioo=\pi/2$
results. The cases causing these large mismatches have $\theta_1\in [1.45,1.6]$ and $\chip\approx 0.8$, which, for $Q\approx4$, result in values of $\MMno(\ioo=\pi/2$) much larger than the ones
coming from the $Q\lesssim 2$ cases. These inspiral-only mismatches further degrade by up to an
order of magnitude when we consider the full mismatches $\MMo$.
Note that the values of $\theta_1,\theta_2$ do not alone explain the increased mismatches for
$Q\approx 4$ as the $Q\approx 2$ subset has even more cases with $\theta_{1,2}\approx \pi/2$.
The number of cycles do not provide a satisfactory explanation either as our chosen $Q\approx 4$
simulations are comparable in length to the $Q\approx 2$ subset.

Another interesting feature of Fig.~\ref{fig:SXS_mismatches_by_Q_and_inc} is the fact that
for $Q\approx 4$ the $\ioo=\pi/2$ mismatches are mostly lower than $\ioo=0$ ones.
Specifically, for all light BBHs and the heavy \SEOB{} and \TPHM{}, we clearly have
$\MMo(\ioo=\pi/2) \preceq \MMo(\ioo=0)$. Focusing on the $Q\approx4$ \SEOB{} subsets alone,
we confirmed that the improved faithfulness at $\ioo=\pi/2$ is due to the arguments made in Sec.~\ref{sec:discrete_inc90}, i.e.,
for cases with $|h_{21}|\gtrsim |h_{22}|$
and $\MM_{21}\lesssim \MM_{22}$, we see $\MMo(\ioo=\pi/2) \preceq \MMo(\ioo=0)$
for the distributions.
In short, as we had observed in Sec.~\ref{sec:discrete_inc90}, when the $h_{2,\pm 1}$
multipoles are more NR faithful than $h_{2,\pm 2}$ and their amplitudes are comparable
to $|h_{2,\pm 2}|$, the resulting high-inclination strain matches are mostly better than zero-inclination
matches.

\subsubsection{Effects of Co-precessing Multipoles on Faithfulness}
\label{sec:coprec_SXS_multipoles}
%
\begin{figure}[t!]
    \centering
    \includegraphics[width=0.48\textwidth]{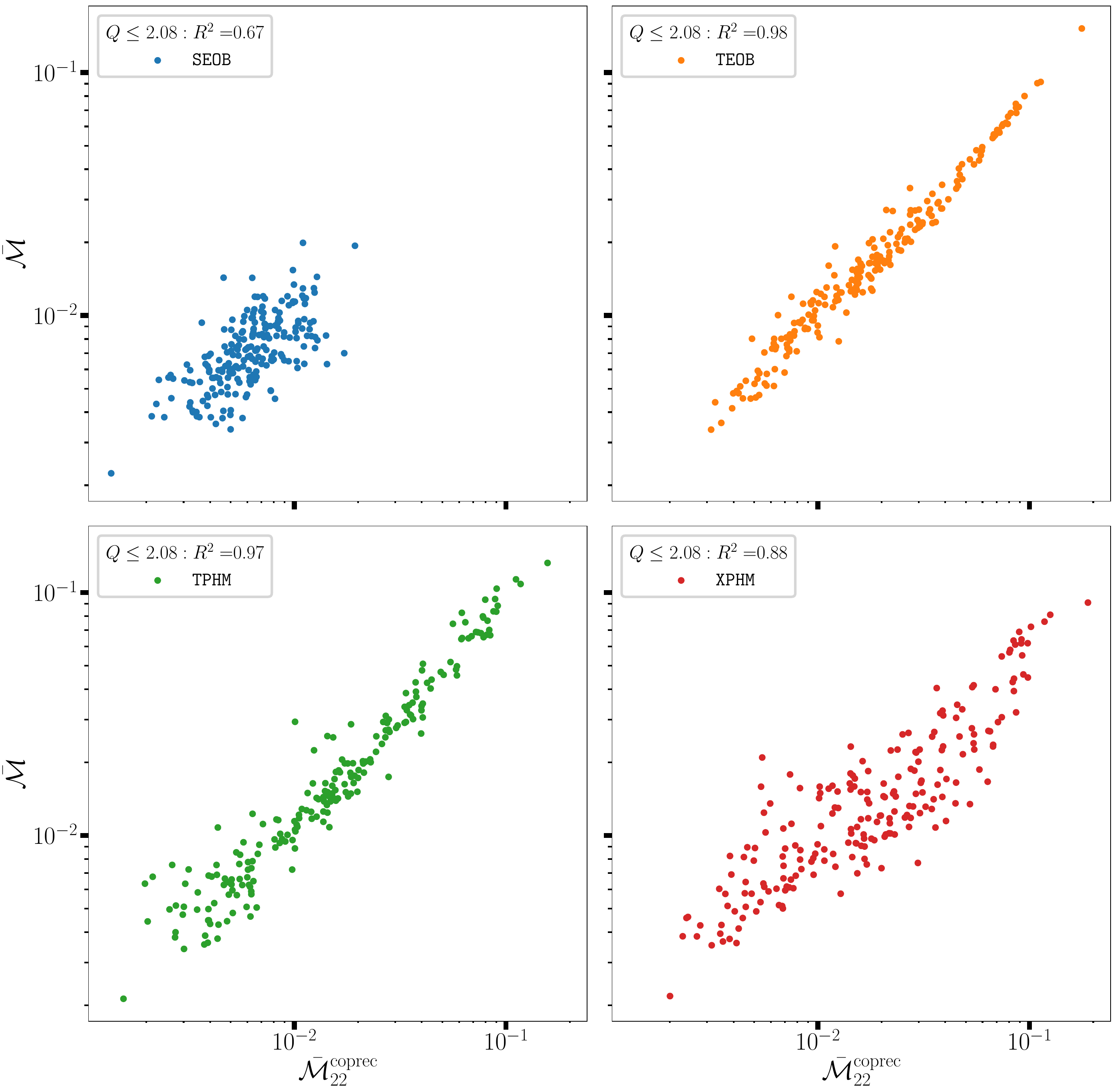}
\caption{The correlation between $\MM_{22,\text{AS}}$ and $\MM(\ioo=0)$ for the
$M=37.5\Msun, Q\lesssim 2$ cases of our short \SXS{} set.
$\MM_{22,\text{AS}}$ is the mismatch [Eq.~\eqref{eq:standard_match}] between the \SXS{}
co-precessing $(2,2)$ multipole and each AS $(2,2)$ multipole of the four models.
Each legend also gives the $R^2$ value for the $\log$-$\log$ data sets.
This figure is akin to the top panel of Fig.~\ref{fig:Coprec_22_mismatches}.}
\label{fig:SXS_coprec_mismatches}
\end{figure}

As we had done in Sec.~\ref{sec:ASmode_MMs}, we investigate once again whether or not
the unfaithfulness of the AS (co-precessing) multipoles have any effect on model performance.
As before, we find a correlation between $\log(\MM_{22,\text{AS}})$ and $\log(\MM(\ioo=0))$,
where $\MM_{22,\text{AS}}$ is the mismatch between the \SXS{} co-precessing $(2,2)$ multipole\footnote{
We employ the \texttt{Scri} package \cite{Boyle:2013nka, Boyle:2014ioa, Boyle:2015nqa, mike_boyle_2020_4041972} to extract the \SXS{} co-precessing multipoles.}
and each model's AS $(2,2)$ multipole.
The correlation for our short \SXS{} set is stronger than the discrete set of Sec.~\ref{sec:NRsur_survey_discrete} and persists whether or not we consider $\log\MM$ or $\log\MMo$
or inspiral-only vs. full mismatches. It is also seen for the $\Mtot=150\Msun$ waveforms,
which for the \SXS{} set are in general longer than the discrete set, which were limited by \NRsurP's
length restrictions. As before, the correlation is strongest for the $\Mtot=\Ml$ \TEOB{} and \TPHM{}
mismatches, but also persists for \SEOB{} and \XPHM{} as can be seen in Fig.~\ref{fig:SXS_coprec_mismatches}, where we plot only the data from the $Q\lesssim 2$ subset.
Thus, this figure can be directly compared with the top row of Fig.~\ref{fig:Coprec_22_mismatches}.
In Fig.~\ref{fig:SXS_coprec_mismatches}, we also provide the $R^2$ values for the
$\log$-$\log$ scatter plots, which for \TEOB{} and \TPHM{} are near unity, and equals almost 0.9 for \XPHM.

Following Sec.~\ref{sec:ASmode_MMs}, we additionally compute the slopes of the linear fits
to $\MM(\ioo=0)$ as a function of $\MM_{22,\text{AS}}$ for \TEOB, \TPHM{} and \XPHM.
We find that the entire $Q\lesssim 4 $ set of $\MM(\ioo=0)$ (313 out 317 cases)
for \TEOB{} and \TPHM{} can be fit by lines with slopes of approximately 0.9, and 0.5 for \XPHM.
This is consistent with our findings in Sec.~\ref{sec:ASmode_MMs} that once the AS
$(2,\pm 2)$ multipole unfaithfulness exceeds a certain value, it becomes the dominant source
of unfaithfulness for the strain mismatch.

Since the $(2,\pm 2)$ AS/co-precessing multipoles also make up the dominant contribution to the precessing
$(2,\pm 1)$ multipoles, we expect the correlation to persist for $\MM(\ioo=\pi/2)$, albeit
less strongly. Indeed, the $R^2$ values provided in Fig.~\ref{fig:SXS_coprec_mismatches}
drop to $\{0.2,0.77,0.85,0.6\}$, respectively for four models.
And the slopes of the linear fits are comparable.
Finally, as we have already seen in Sec.~\ref{sec:ASmode_MMs}
the AS $(2,\pm1)$ multipoles do not yield a similar correlation:
their magnitudes are too small.
As done in Sec.~\ref{sec:ASmode_MMs}, we also checked how much these modes matter
for strain faithfulness. Specifically, we re-constructed the \SXS{} strains without
the $h_{2,\pm1}^\text{coprec}$ content and computed the mismatch to the unaltered
\SXS{} strains. As before, we found that the mismatch is at most 0.015 (0.07) at $\ioo=0 \;(\ioo=\pi/2)$.
At both inclinations, it is only few cases that yield such mismatches with the majority of the values
below 0.001 (0.01).

\begin{table}
 \begin{tabular}{|c|c|cc|cc|}
 \hline
\multicolumn{2}{|c|}{} & \multicolumn{2}{c|}{$\MMo(\ioo=0)$}& \multicolumn{2}{c|}{$\MMo(\ioo=\pi/2)$}\T\B\\
 \cline{3-6}
   \hline
\SXS{} ID &	Model &	$ 37.5\Msun$ &	$ 150\Msun$ &	$37.5\Msun$ &	$ 150\Msun$ \T \B \\
\hline
\multirow{4}{*}{\texttt{0057}} &	\SEOB &	0.027 &	0.032 &	0.011 &	0.0076 \\
 &	\TEOB &	0.024 &	0.038 &	0.015 &	0.010 \\
 &	\TPHM &	0.029 &	0.032 &	0.012 &	0.0081 \\
 &	\XPHM &	0.036 &	0.038 &	0.015 &	0.0098 \\
\hline
\multirow{4}{*}{\texttt{0058}} &	\SEOB &	0.0069 &	0.0041 &	0.0055 &	0.0050 \\
 &	\TEOB &	0.022 &	0.014 &	0.018 &	0.019 \\
 &	\TPHM &	0.026 &	0.0087 &	0.013 &	0.0086 \\
 &	\XPHM &	0.012 &	0.010 &	0.013 &	0.018 \\
\hline
\multirow{4}{*}{\texttt{0062}} &	\SEOB &	0.014 &	0.019 &	0.0039 &	0.0026 \\
 &	\TEOB &	0.020 &	0.038 &	0.0079 &	0.0076 \\
 &	\TPHM &	0.016 &	0.023 &	0.0051 &	0.0045 \\
 &	\XPHM &	0.025 &	0.047 &	0.0062 &	0.0077 \\
\hline
\multirow{4}{*}{\texttt{0165}} &	\SEOB &	0.061 &	0.055 &	0.017 &	0.016 \\
 &	\TEOB &	0.088 &	0.076 &	0.028 &	0.023 \\
 &	\TPHM &	0.063 &	0.11 &	0.028 &	0.027 \\
 &	\XPHM &	0.13 &	0.093 &	0.080 &	0.057 \\
 \hline
 \end{tabular}
 \caption{The sky-optimized mismatches between the four $Q>4$ \SXS{} waveforms and
 the four models that we consider in this work.}
 \label{tab:Q_ge_5_SXS_MMs}
 \end{table}

\subsubsection{$Q\ge 5$ Comparisons}
As already noted, we only have four cases with $Q\ge5$ in our short \SXS{} set.
Three of these, namely \SXS:\texttt{0057,0058,0062}, have
$Q\approx 5, -0.18\lesssim \chieff\lesssim 0$ and $ 0.45\lesssim \chip\lesssim 0.5$, and one,
\SXS:\texttt{0165}, has $Q\approx 6, \chieff \approx -0.45, \chip\approx 0.77$.
We should caution that we detected ``V''-shaped kinks in the phases of both the precessing and co-precessing
$(2,2),(2,1)$ multipoles.
The first kinks appear at $\approx 100M$ before the peak of the multipole amplitude.
Additionally, \texttt{SXS:0057} is very short, i.e., its inspiral lasts
$\approx 800M$.
For these reasons, we present only the full wavelength mismatch, $\MMo$ for these cases.

We list the values of $\MMo$ for $\Mtot=\Ml,\Mh$ and $\ioo=0,\pi/2$ for each model in
Table~\ref{tab:Q_ge_5_SXS_MMs}.
One interesting finding is that  $\MMo(\ioo=\pi/2) < \MMo(\ioo=0)$
for all models except for the $\Mtot=\Mh$ mismatches for \SXS:\texttt{0058}.
Most of the improved faithfulness at $\ioo=\pi/2$ is due to a more faithful
$h_{21}$ multipole satisfying the relation $|h_{21}|\gtrsim |h_{22}|$ as we have previously
explored in detail.
As we have only four cases with $Q\gtrsim 5$, we move on to comparisons of longer waveforms.


\begin{figure*}[t!]
    \centering
    \includegraphics[width=1\textwidth]{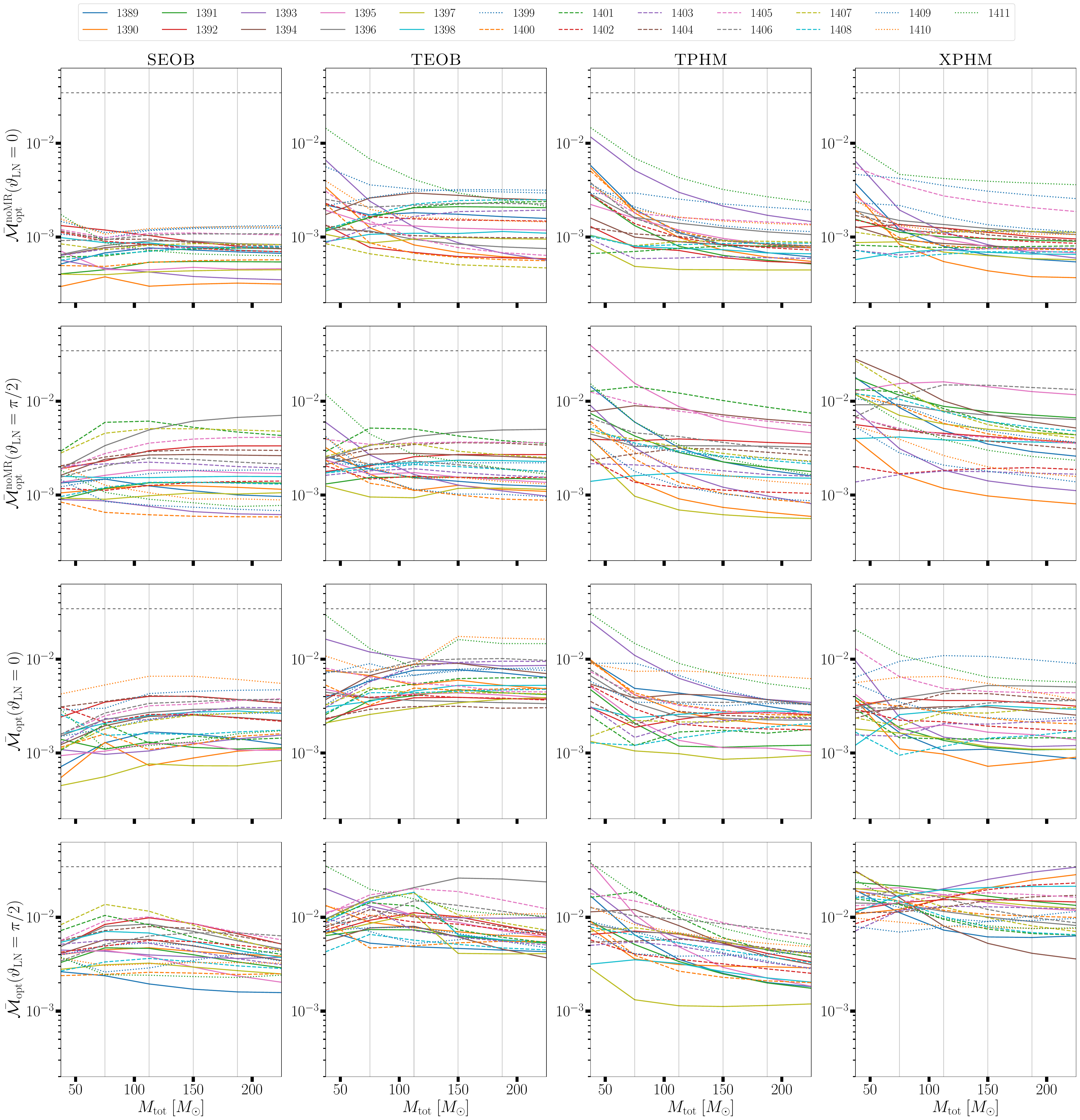}
\caption{
Trace plots of the mismatches between $\{\SEOB,\TEOB,\TPHM,\XPHM\}$
and numerical relativity waveforms from 23 long \texttt{SXS} simulations as a function of the total binary mass.
We computed the mismatches at the values of $\Mtot=\{37.5,75,112.5,150,187.5,225\}\Msun$
(marked by the faint vertical lines) and linearly connected the data points. 
The top (bottom) two rows show $\MMno$ ($\MMo$)
with odd (even) rows corresponding to $\ioo=0$ ($\ioo=\pi/2$) results.
The four columns represent, from left to right, the results for \SEOB, \TEOB, \TPHM, \XPHM.
}
\label{fig:spaghetti}
\end{figure*}

\subsection{Long \SXS{} Waveforms}\label{sec:Long_SXS}

For this assessment, we employ 23 long \SXS{} waveforms, namely simulations
\texttt{1389}-\texttt{1411} \cite{Boyle:2019kee}.
The total number of GW cycles in these simulations varies approximately between $ 128$ and $ 305$,
and the relevant parameters have the following ranges:
$Q\le 4, \chip \lesssim 0.49, \chi_{\rm eff} \in [-0.2, 0.3]$
though only \texttt{1409}-\texttt{1411} have $Q=4$ with the rest having $Q< 2$.
With only 23 cases as compared with $\ord(10^3)$ in Sec.~\ref{Sec:faith_survey}
and $\ord(10^2)$ in Sec.~\ref{sec:Short_SXS},
we could afford the computational time to expand our total mass sample to the following values:
$M=\{37.5,75,112.5,150,187.5,225\}\Msun$,
with $f_i=f_0+3\,$Hz for all match integrals.
Note that because the ASD drops steeply from $\ord(10)$\,Hz to 20\,Hz
(see Fig.~\ref{fig:sample_waveforms}), only 1 to 2 inspiral cycles contribute to the match integrals
for the $\Mtot=225\Msun$ cases even though $f_i\in [4,5]\,$Hz for these.
Similarly, we record 2.5 to 4 inspiral cycles for $\Mtot=187.5\Msun$ and
4.5 to 7 cycles for $\Mtot=150\Msun$.
Accordingly, we observe that $\text{(MR-only SNR)}\gtrsim \text{(inspiral-only SNR)}$ for $\Mtot \gtrsim 175\Msun$.

We present the resulting mismatches as trace plots in Fig.~\ref{fig:spaghetti}.
In the top two rows of the figure,
we display $\MMno$ at $\ioo=0$ and $\pi/2$ then repeat this for $\MMo$ in the
bottom two rows with the four columns corresponding to the four approximants in the usual order.
First, we note that with the exception of one \TPHM{} $\ioo=\pi/2$ case (corresponding to \texttt{SXS:1395}),
all mismatches are less than 0.035.
Moreover, we also observe that $\MMno(\ioo=0)<0.01$ holds nearly universally and that
this inequality still holds at $\ioo=\pi/2$ for the \textsc{EOB} models.
Our finding of $\MMno \preceq \MMo$ from the previous sections persists here as well,
which can be gathered by comparing row one with row three, and row two with row four.
The dependence of the mismatches on inclination also seems to be consistent with our previous
results as the relation $\MMo(\ioo=0)\preceq \MMo(\ioo=\pi/2)$ is retained as seen by
comparing row one with row two, and row three with row four of the same figure.

As for the dependence of the mismatches on the total mass, we observe mostly a flattening of each
mismatch curve as $\Mtot$ exceeds a certain threshold, more strongly so for the $\ioo=0$ cases.
This flattening is consistent with the results of, e.g., Refs.~\cite{Pratten:2020fqn, Estelles:2021gvs, Gamba:2021ydi, Ramos-Buades:2023ehm}.
Interestingly we observe it for both $\MMno$ and $\MMo$. For the flattening of the former,
our reasoning is as follows: as the binary gets heavier, the signal shifts toward lower frequencies:
e.g., $f_\text{peak}\lesssim 50\,$Hz for $\Mtot=150\Msun$ and $f_\text{peak}\lesssim 30\,$Hz for $\Mtot=225\Msun$.
In this regime, the detectors are less sensitive (see Fig.~\ref{fig:sample_waveforms}),
therefore the mismatch expression is also less sensitive to the differences between waveforms.
So the dependence on the total mass becomes irrelevant for $\MMno$ beyond a certain threshold.

The flattening of the $\MMo$ curves are partly due to this reason combined with the fact that
for the cases with only a few inspiral cycles, i.e., $\Mtot \gtrsim 150\Msun$, the MR part of the
signal becomes the dominant contribution to the mismatch.
In other words, the contribution to the mismatch from $\lesssim 5$ inspiral cycles does not change the overall
$\MMo$ significantly enough. However, the MR part of some of the heavier cases falls in the $\ord(100)\,$Hz
region, where the detectors are more sensitive, as is, accordingly, our full-band mismatch $\MMo$.
This is why we observe flatter curves for $\MMno(\ioo=0)$ than for $\MMo(\ioo=0)$ in Fig.~\ref{fig:spaghetti}. Also note that some curves do not flatten out at all. We are unable to provide
a single universal, ``one size fits all'' explanation for these as the shape and mass dependence of each curve
varies from one model to another.

One final prominent feature of the figure is the change in \XPHM's faithfulness, $\MMo$, in going from
$\ioo=0$ to $\pi/2$, for which we see that the values
of the mismatches shift from mostly being below 0.01 to mostly being above it,
while still remaining below 0.035.
This behavior might be indicative of more severe MR-related issues for \XPHM's precessing
$(2,\pm1)$ multipoles than for other models.
We re-computed $\MMo(\ioo=\pi/2)$ using the updated \textsc{SpinTaylor} version of \XPHM{} \cite{Colleoni:2023} which showed improvement:
the range of $\MMo(\ioo=\pi/2)$ dropped from $[0.004,0.03]$ to $[0.003,0.02]$.
More noticeably, the $\Mtot \gtrsim 100\Msun$ values of $\MMo(\ioo=\pi/2)$ obtained with the \textsc{SpinTaylor} version
are all $\lesssim 0.02$ in contrast to many cases of $0.02 <\MMo(\ioo=\pi/2)\lesssim 0.035$ 
seen in the lower right panel of Fig.~\ref{fig:spaghetti}. 


\section{Injection/Recovery Study}
\label{Sec:PE}

As the final part of our survey, we investigate the performance of the precessing approximants
in an injection-recovery PE study. Specifically, we inject two different \SXS{} ($\ell=2$)-only strains
into LIGO noise and recover the parameters of the injected waveforms with \SEOB, \TPHM{} and \XPHM.
We are greatful to Charlie Hoy and Lorenzo Pompili for assisting us with the \SEOB{} runs.
Our omission of the \TEOB{} results is due to:
\begin{inparaenum}[(i)]
\item the precessing model can only be used for PE with the \bajes{} library \cite{Breschi:2021wzr}
which is external to the computational infrastructure that we employed for the PE runs with
$\{{\SEOB},\TPHM,\XPHM\}$;
\item \bajes{} runs crashed with waveform errors when the upper bounds of the $\chi_{1,2}$ priors were above $0.8$ despite \TEOB{} generating reasonable waveforms in that regime. This error first arose when attempting to analyse GW200129\_065458 with \TEOB{} and persisted when injection/recovery runs were carried out.
We hope to collaborate with \bajes{} developers in the future so that these issues can be resolved.
\end{inparaenum}

Returning to our study here. We performed ``zero-noise'' injections of the NR waveforms
\texttt{SXS:0050} (henceforth \texttt{0050}) and \texttt{SXS:0628} (henceforth \texttt{0628})
while using an estimated O4a PSD \cite{aLIGO_O4_high_psd} for our likelihood computations.
These simulations have key parameters
$\{Q,\chieff,\text{max}(\chip)\}\approx \{3,0.001,0.5\},\{2,-0.174,0.84\}$ respectively.
The former has a higher component mass asymmetry and moderate spins, only in the orbital plane, whereas
the latter has a moderate mass ratio, but  significant precession.
We perform each injection twice: for a low mass and a high mass BBH
given in the detector frame, specifically,
$\Mdet=56.5\Msun, 150\Msun$ for \texttt{0050} and
$\Mdet=67\Msun,150\Msun$ for \texttt{0628}.
The values for the lighter masses are chosen so that the \SXS{} reference frequency corresponds to
$f^\text{det}_0\approx20\;\mathrm{Hz}$ for both simulations.
The values for the heavier masses
are chosen to put more emphasis on the plunge-merger-ringdown portions of the waveforms,
consistent with the rest of this article.
The source-frame masses are smaller by a factor of $(1+z)$ where $z$ is the redshift of the
sources at the injected luminosity distance which we set to $\{\{308, 649\}, \{555, 1020\} \}$Mpc,
respectively for light/heavy \texttt{0050} and light/heavy \texttt{0628}.

\begin{figure*}[t!]
    \centering
    \includegraphics[width=0.99\textwidth]{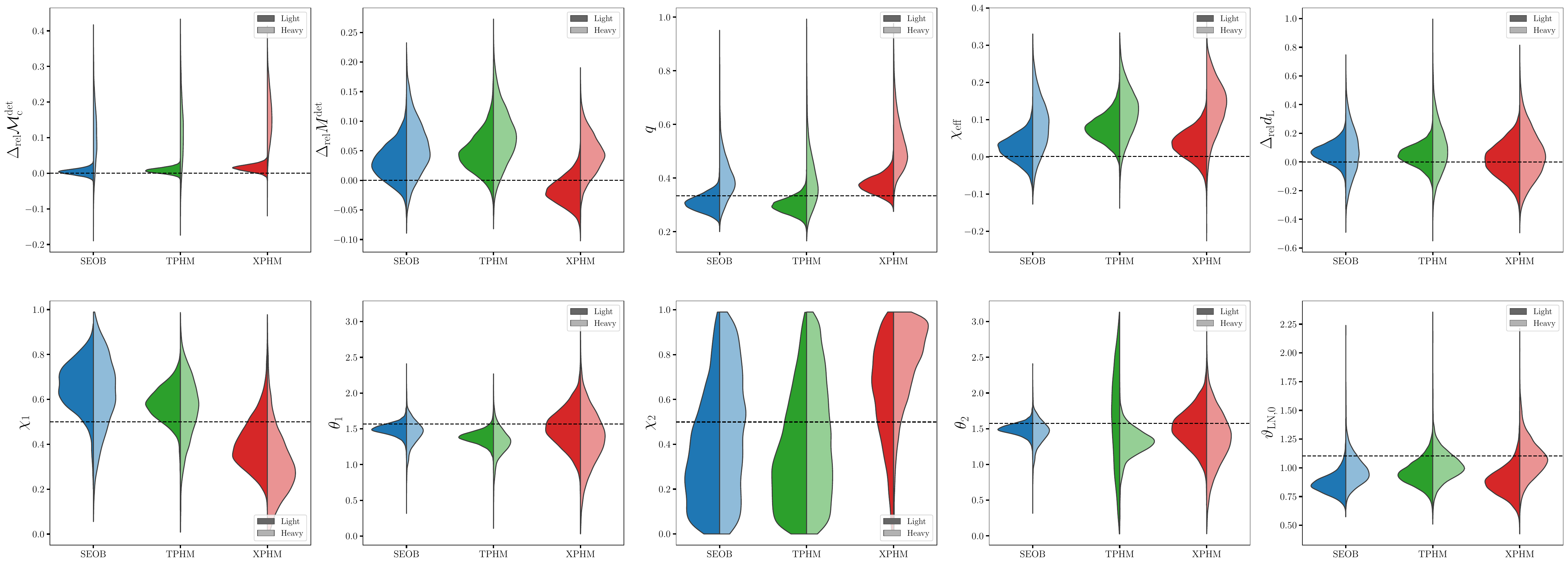}
\caption{
Recovery performances of \SEOB, \TPHM{} and \XPHM{} for a synthetic
signal generated by a ``zero-noise'' injection of the \SXS{} simulation \fifty{} into the O4a LIGO Hanford-Livingston network. In each panel, we present the results from both the light and the heavy mass runs
as the darker and lighter halves of each violin, respectively.
Shown are the recovered posteriors for
$\{\M_\text{c}^\text{det}, \Mdet,q,\chieff, d_\text{L},\chi_1,\theta_1,\chi_2,\theta_2,{\ioo}\}$
with the dashed horizontal lines marking the injected values.
For $\{\M_\text{c}^\text{det}, \Mdet, d_\text{L}\}$, we present posteriors for
the relative difference between the injected and the recovered values, i.e.,
$\Delta_\text{rel}X:=(X_\text{r}-X_\text{inj})/X_\text{r}$.
The first panel, i.e., $\Delta_\text{rel}\M^\text{det}_\text{c}$ is somewhat hard to read
because the models produced very narrow posteriors for the light mass run,
but much wider ones for the heavy one.}
\label{fig:0050_violins}
\end{figure*}

The reason why the value of $d_\text{L}$ for the $150\Msun$ \texttt{0050} run differs from that of
\ste{} is due to our choice of extrinsic parameters.
Specifically, we fine-tuned $\{\ioo,\theta_s,\phi_s,\psi_s,d_\text{L}\}$ such that the total network SNR, 
$\rho_\text{tot}$, accumulated by the H1-L1 network
is 40 for each case resulting in a precession SNR of $\rho_\text{p}=\{\{11,6.4\},\{15,10\}\}$,
respectively for light/heavy \fifty{} and light/heavy \ste.
This quantity was first defined in Ref.~\cite{Fairhurst:2019vut} and represents the
contribution to the SNR coming from the second most significant precessing harmonic.
It is proportional to the total SNR, thus becomes larger for louder events, and has to be greater than
2.1 for the effects of precession on the signal to not be attributed to noise alone \cite{Fairhurst:2019vut}.
Let us add that we opted here for $\rho_\text{tot}=40$ as our trial runs with $\rho_\text{tot}=20$
resulted in the posteriors of precession-related quantities, such as $\chip$,
being prior dominated, hence uninformative.
The specific values for all our injected parameters can be found in the \texttt{config} files
uploaded to our \git{} repository.

We employ the \texttt{bilby} \cite{Ashton:2018jfp, Romero-Shaw:2020owr}
and \texttt{bilby\_pipe} \cite{bilby_pipe} modules with the  \texttt{dynesty} sampler \cite{Skilling:2004, Skilling:2006gxv, Speagle:2019ivv,sergey_koposov_2023_8408702}.
We set the number of live points, \verb|nlive|, to 1000, and \verb|naccept| to 60.
Other \texttt{dynesty} attributes are set to their default values.
When employing the \textsc{IMRPhenom} models, we run two parallel chains and combine the results at the end. For \SEOB, we run only one chain per injection.

The priors for the parameters are as follows. For the lighter mass runs,
we use the \texttt{bilby} function \verb|UniformInComponentsChirpMass| with
the range $\M_\text{c} \in [10,35]\Msun$, the \verb|UniformInComponentsMassRatio|
function with range $q \in [0.083,1]$, and
constraints of $m_i \in [1,1000]\Msun$.
The luminosity distance prior range is given by $d_\text{L}\in [100,5000]\,$Mpc
via the function \verb|UniformSourceFrame|  using $\Lambda$CDM cosmology
parameterized by Planck 2015 data \cite{Planck:2015fie}.
The spin magnitudes are uniform in the range $\chi_i\in[0,0.99]$ while
the spin tilt angles are uniform in their sines: $\sin\theta_i\in [0,1]$,
and the $\J$ frame inclination angle is uniform in cosine: $\cos\vartheta_\text{JN,0}\in [-1,1]$, where
$\vartheta_\text{JN,0}$ is the angle between $\J$ and $\hat{\mathbf{N}}$ at $f=f_0$.
Other angles are all uniform in their respective ranges.
For the heavier $\Mdet=150\Msun$ runs, we set the chirp mass prior range to
$\M_\text{c} \in [45,85]\Msun\, ( [40,90]\Msun)$ for \fifty{}(\ste)
with the other priors kept the same.
We find that such an injection/recovery PE ``job'' consisting of two chains of \texttt{IMRPhenomTPHM}
\emph{and} two chains of \texttt{IMRPhenomXPHM} takes roughly one day using $100$ processors. 
With the same resources, one \texttt{SEOBNRv5PHM} chain can be obtained in $\sim 1.5$ days.


\subsection{Main Results}\label{sec:PE_main_results}
\begin{figure*}[t!]
    \centering
    \includegraphics[width=0.99\textwidth]{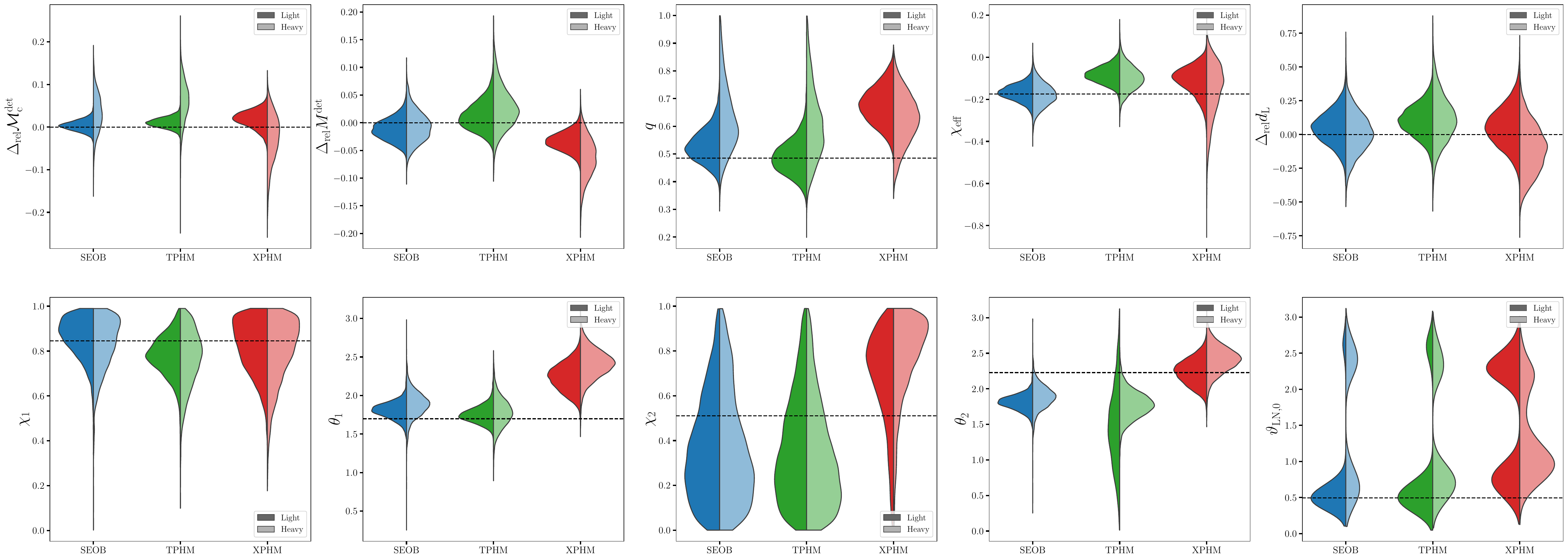}
\caption{
Same as Fig.~\ref{fig:0050_violins}, but for the \SXS{} simulation \ste.}
\label{fig:0628_violins}
\end{figure*}

Overall, we have {12} sets of posterior parameters to present resulting from
$\{\SEOB,\TPHM,\XPHM\}\times \{\fifty,\ste\}\times\{$light, heavy$\}$ runs.
For the sake of brevity, we show only a subset of posteriors for important intrinsic and extrinsic
parameters, namely the set $\{\M_\text{c}^\text{det}, \Mdet,q,\chieff, d_\text{L},\chi_1,\theta_1,\chi_2,\theta_2,{\ioo}\}$
in Figs.~\ref{fig:0050_violins} and \ref{fig:0628_violins}, where, for $\{\M^\text{det}_\text{c},\Mdet,d_\text{L}\}$ we present the relative differences between the injected and the recovered values.
When relevant, we also discuss the posteriors for other parameters besides these.
Table~\ref{tab:PE_posteriors} presents the 5th, 50th and 95th percentiles of the recovered
posteriors for the relevant parameters.
We discuss the recovery of precession-related quantities further below in Sec.~\ref{sec:prec_recovery}.
For reference, we display in Table~\ref{tab:pe_matches}, the mismatches between the injected waveforms
and those generated by {\SEOB}, \TPHM{} and \XPHM{} using the injected parameters.

\begin{table}[h!]
 \begin{tabular}{ccccc}
 \hline
  Model &\texttt{0050}  light & \texttt{0050}  heavy & \texttt{0628} light & \texttt{0628} heavy\\
  \hline\hline
    \SEOB & $1.28\times10^{-2}$ & $1.39\times 10^{-2}$ & $1.05\times10^{-2}$ & $7.94\times 10^{-3}$\\
    \TPHM & $2.13\times10^{-2}$ & $1.00\times10^{-2}$ & $1.75\times10^{-2}$ & $9.81\times10^{-3}$   \\
    \XPHM & $2.43\times10^{-2}$ & $1.73\times10^{-2}$ & $1.67\times10^{-2}$ & $1.63\times10^{-2}$  \\
  \hline
 \end{tabular}
 \caption{Mismatches calculated between the injected \SXS{} waveforms and the recovery models evaluated
 at the same parameters as the injected waveforms.  The values presented here are
 calculated using Eq.~\eqref{eq:standard_match}. }\label{tab:pe_matches}
\end{table}

We start with Fig.~\ref{fig:0050_violins}, where we present both the
light and heavy mass posteriors from the \fifty{} recovery runs as violin plots.
A common feature seems to be that \SEOB{} and \TPHM{} posteriors
are more similar to each other than \XPHM.
In general, one model recovers a particular set of parameters better than the other two. 
For example, for the light mass case, \SEOB{} recovers
$\{\M^\text{det}_\text{c},q,\chieff\}$ better, whereas \XPHM{} recovers 
$\{\chi_1,\theta_1,\theta_2,d_\text{L}\}$ better. 
\TPHM{} recovers $\ioo$ better than the other two models, and is sometimes better than \SEOB{} or \XPHM{}
for other parameters, but not both.
$\chi_2$ is  recovered at roughly the same confidence by the models though the \XPHM{} posterior is skewed oppositely.
Looking at the median values and the 90\% confidence intervals (CIs) for $m_1,m_2$, i.e., the source-frame component masses, in Table~\ref{tab:PE_posteriors}, we see \SEOB/\TPHM{} slightly overestimating $m_1$ while underestimating
$m_2$ with \XPHM{} behaving in the opposite manner.
Additional corner plots of the posteriors can be found in our \git{} repository.

\SEOB's and \TPHM's performances are even more similar for the heavy mass case.
\SEOB/\TPHM{} recover $\{\M^\text{det}_\text{c},q,\chi_1\}$ better, whereas
\XPHM{} does so for $\{\theta_1,\theta_2,d_\text{L},\ioo\}$.
$\chi_2$ posteriors for \SEOB/\TPHM{} look nearly flat indicating the dominance of the uniform prior in the posterior,
whereas \XPHM's posterior rails against the $\chi=1$ Kerr bound.
$\{\Mdet,\chieff\}$, are recovered equally marginally by \SEOB{} and \XPHM.
The recovery posteriors for $\M^\text{det}_\text{c}$ are rather wide
with the 90\% CIs ranging roughly from $54\Msun$ to $68\Msun$ for \SEOB, 
$54\Msun$ to $70\Msun$ for \TPHM, and from $58\Msun$ to
$70\Msun$ for \XPHM{},  given an injected value of $55\Msun$.

The recovery of \ste's parameters reveals yet more differences in model performance, but with the
similarity in \SEOB{} and \TPHM's posteriors persisting.
Let us recall that \ste{} has stronger precession than \fifty, but non-zero parallel spin projections.
For the light-mass run, \SEOB{} recovers $\{\M^\text{det}_\text{c}, \chieff\}$ better
while \TPHM{} does so for $\{\Mdet,q,\theta_1 \}$ with \XPHM{} recovering $\theta_2$ better.
\SEOB, \TPHM{} recover $\ioo$ very similarly, and \SEOB, \XPHM{} do so for $d_\text{L}$.
\SEOB/\XPHM{} both rail against the $\chi_1=1$ bound, but not \TPHM.
On the other hand, \SEOB/\TPHM{} rail against the $\chi_2=0$ bound, with \XPHM{} somewhat
railing against $\chi_2=1$.
Though \XPHM{} recovers $\M^\text{det}_\text{c}$ in the 90\% CI,
its recovery of the other mass parameters is rather poor. For example, its posteriors for
both the detector and the source frame (total, primary, secondary) masses do not contain the injected parameters
within their respective 90\% CIs, whereas \SEOB's and \TPHM's posteriors do, with the injected values recovered near the
peaks of the distributions. This can also be partly gathered by looking at the recovered $\{m_1,m_2\}$
posteriors in Table~\ref{tab:PE_posteriors} under the \ste{} Light rows.
The fact \XPHM{} under(over)estimates $m_1 (m_2)$ is why it recovers $\M_\text{c}$ within its 90\% CI.

As for the heavy-mass \ste, \SEOB/\TPHM{} recover both the detector and source frame masses robustly.
\XPHM{} underestimates the injected values for $m_1,m^\text{det}_1$ by three standard deviations or more. 
Despite this, its $q$ posteriors are comparable to that of \SEOB's.
Overall, \SEOB{} recovers $\{\Mdet,\chieff,d_\text{L}\}$ better, while \TPHM{} does so for $\{q,\chi_1,\theta_1\}$, 
and with \XPHM{} doing so for $\{\M^\text{det}_\text{c},\theta_2\}$, but cannot recover $\theta_1$
with any confidence.
The various railings of the $\chi_{1,2}$ posteriors against the $\chi_{1,2}=0,1$ bounds, seen for the light-mass
injection, are exacerbated.
As was the case with \fifty, the recovered posteriors for $\M^\text{det}_\text{c}$ are wider for the heavy-mass
run with the 90\% CIs ranging roughly from $58\Msun$ to $67\Msun$ for \SEOB, $59\Msun$ to $69\Msun$ for \TPHM, and from $53\Msun$ to
$64\Msun$ for \XPHM{} for an injected value of $60.5\Msun$.
Another prominent feature to note is the bimodality in the $\ioo$ posteriors which we did
not observe for \fifty.
Given that we only inject and recover the $\ell=2$ strain, some bimodality in the inclination is expected
due to the distance-inclination degeneracy. However, when there is precession, the $(2,\pm1)$ multipoles
may contain enough power to break this, and more so for larger $Q$.
Indeed, we find that the ratio $|{}_{-2}Y_{21} h_{21}|/|{}_{-2}Y_{22}h_{22}|$ is larger for
\fifty{} than for \ste, thus explaining why we observe no bimodality in the inclination posteriors for \fifty.

An interesting result that warrants further discussion
is the consistent overestimation of $\chieff$ for the \fifty{} run where the injection has $\chieff=0.001$.
This can be seen in the upper right panel of Fig.~\ref{fig:0050_violins} (also upper left panels of Figs.~\ref{fig:0050_light_chi_hist}, \ref{fig:0050_heavy_chi_hist}) and in Table~\ref{tab:PE_posteriors}.
This is consistent with the broad findings of Ref.~\cite{Ng:2018neg}.
Given that the total SNRs are fixed to 40 with precession SNRs all exceeding 6, this is somewhat suprising.
Perhaps, this may be due to the fact that the injected value is small.
Interestingly, App.~D of  Ref.~\cite{Ng:2018neg} contradicts us, mostly finding that the posteriors slightly underestimate the injected values of $\chieff=0$. We believe the disagreement is mostly due to the facts
that (i) they set $\theta_{1,2}=10^\circ,30^\circ$ thus have to fix $\chi_{1,2}=0$ to obtain
$\chieff=0$, which is different than our situation which has $\theta_{1,2}\simeq 90^\circ$;
(ii) they employ the \textsc{IMRPhenomPv2} model which is no longer considered the state of the art.
It might be case that the recovery of small injected values of $\chieff$ with large $\chip$ is
in general more challenging. Otherwise, we can expect $\chieff$ to be well recovered by the current models
as seen in, e.g., Table~I of Ref.~\cite{Estelles:2021gvs} for \TPHM{} and
Table~II of Ref.~\cite{Pratten:2020ceb} for \XPHM.
A systematic injection/recovery campaign would certainly find a trend (if there is any).
Let us reiterate that $\chi_\text{eff}$ is an effective measure of the impact of spin \emph{in the inspiral}, which may not necessarily hold for high-mass systems, i.e., few inspiral cycles. Indeed, we observe that the overestimation is more severe for our heavy mass injection.
For the \ste{} run, \SEOB{} recovers $\chieff$ very well, but \TPHM{} and \XPHM{} overestimate it once
again.

%
\begin{table*}
 \begin{tabular}{c|lllllllllll}
 \hline
 &\quad $m_1(\Msun)$ \quad &\quad $m_2(\Msun)$\quad &\quad $\chi_1$\quad &\quad $\chi_2$\quad & 
 \quad$\chieff$\quad &\quad $\chip$\quad & \quad$\theta_1$\quad &\quad $\theta_2$\quad & \quad$d_\text{L}$(Mpc)\quad &\quad $\ioo$ \T \B \\
 \hline
 \hline
$\texttt{0050}$ Light &  &  &  &  &  &  &  &  &  &  & \B\T  \\ 
inj. & $39.74$ & $13.25$ & $0.50$ & $0.50$ & $0.001$ & $0.50$ & $1.57$ & $1.57$ & $308$ & $1.05$ \B\T  \\ 
$\texttt{SEOB}$ & $41.56^{+3.79}_{-3.21}$ & $12.75^{+1.08}_{-0.93}$ & $0.66^{+0.16}_{-0.18}$ & $0.32^{+0.48}_{-0.29}$ & $0.023^{+0.056}_{-0.059}$ & $0.66^{+0.16}_{-0.18}$ & $1.49^{+0.14}_{-0.15}$ & $1.78^{+0.92}_{-1.13}$ & $328^{+35}_{-36}$ & $0.86^{+0.17}_{-0.12}$ \B\T  \\
$\texttt{TPHM}$ & $42.57^{+4.08}_{-3.33}$ & $12.63^{+1.06}_{-0.97}$ & $0.58^{+0.12}_{-0.12}$ & $0.31^{+0.48}_{-0.28}$ & $0.077^{+0.068}_{-0.057}$ & $0.57^{+0.12}_{-0.13}$ & $1.38^{+0.16}_{-0.16}$ & $1.67^{+1.01}_{-1.15}$ & $322^{+36}_{-39}$ & $0.95^{+0.20}_{-0.16}$ \B\T  \\
$\texttt{XPHM}$ & $37.84^{+3.00}_{-2.85}$ & $14.17^{+1.09}_{-1.00}$ & $0.39^{+0.21}_{-0.15}$ & $0.64^{+0.29}_{-0.45}$ & $0.034^{+0.065}_{-0.067}$ & $0.38^{+0.22}_{-0.15}$ & $1.51^{+0.45}_{-0.44}$ & $1.44^{+0.89}_{-0.88}$ & $311^{+61}_{-62}$ & $0.88^{+0.23}_{-0.19}$ \B\T  \\
\hline
$\texttt{0050}$ Heavy &  &  &  &  &  &  &  &  &  &  & \B\T  \\ 
inj. & $99.24$ & $33.08$ & $0.50$ & $0.50$ & $0.001$ & $0.50$ & $1.57$ & $1.57$ & $649$ & $1.05$ \B\T  \\ 
$\texttt{SEOB}$ & $98.85^{+10.75}_{-9.01}$ & $39.01^{+11.26}_{-8.04}$ & $0.64^{+0.24}_{-0.29}$ & $0.49^{+0.43}_{-0.44}$ & $0.085^{+0.121}_{-0.111}$ & $0.62^{+0.24}_{-0.29}$ & $1.46^{+0.25}_{-0.36}$ & $1.36^{+1.12}_{-0.94}$ & $694^{+186}_{-174}$ & $0.97^{+0.25}_{-0.17}$ \B\T  \\
$\texttt{TPHM}$ & $100.63^{+15.24}_{-12.94}$ & $38.68^{+15.13}_{-8.97}$ & $0.57^{+0.21}_{-0.25}$ & $0.43^{+0.48}_{-0.38}$ & $0.122^{+0.105}_{-0.110}$ & $0.54^{+0.20}_{-0.24}$ & $1.32^{+0.31}_{-0.38}$ & $1.43^{+1.12}_{-1.01}$ & $705^{+198}_{-164}$ & $1.01^{+0.23}_{-0.19}$ \B\T  \\
$\texttt{XPHM}$ & $90.36^{+7.01}_{-8.13}$ & $46.31^{+13.09}_{-7.91}$ & $0.30^{+0.29}_{-0.20}$ & $0.80^{+0.17}_{-0.39}$ & $0.139^{+0.106}_{-0.123}$ & $0.36^{+0.22}_{-0.14}$ & $1.38^{+0.62}_{-0.67}$ & $1.16^{+0.70}_{-0.61}$ & $679^{+202}_{-168}$ & $1.08^{+0.31}_{-0.24}$ \B\T  \\
\hline
$\texttt{0628}$ Light &  &  &  &  &  &  &  &  &  &  & \B\T  \\ 
inj. & $40.44$ & $19.61$ & $0.85$ & $0.51$ & $-0.174$ & $0.84$ & $1.70$ & $2.23$ & $555$ & $0.52$ \B\T  \\ 
$\texttt{SEOB}$ & $38.65^{+3.54}_{-3.57}$ & $20.36^{+2.25}_{-1.88}$ & $0.87^{+0.10}_{-0.18}$ & $0.33^{+0.44}_{-0.29}$ & $-0.167^{+0.065}_{-0.066}$ & $0.84^{+0.10}_{-0.17}$ & $1.81^{+0.17}_{-0.19}$ & $1.89^{+0.82}_{-1.12}$ & $584^{+114}_{-114}$ & $0.52^{+2.05}_{-0.22}$ \B\T  \\
$\texttt{TPHM}$ & $40.44^{+4.66}_{-4.06}$ & $19.62^{+2.27}_{-1.96}$ & $0.78^{+0.14}_{-0.15}$ & $0.33^{+0.48}_{-0.28}$ & $-0.083^{+0.071}_{-0.065}$ & $0.77^{+0.13}_{-0.15}$ & $1.75^{+0.20}_{-0.16}$ & $1.46^{+1.00}_{-0.87}$ & $612^{+108}_{-110}$ & $0.55^{+2.11}_{-0.28}$ \B\T  \\
$\texttt{XPHM}$ & $34.46^{+2.85}_{-2.57}$ & $23.26^{+2.22}_{-2.51}$ & $0.83^{+0.14}_{-0.23}$ & $0.72^{+0.22}_{-0.50}$ & $-0.102^{+0.091}_{-0.131}$ & $0.63^{+0.21}_{-0.20}$ & $2.24^{+0.29}_{-0.30}$ & $0.79^{+0.56}_{-0.47}$ & $585^{+145}_{-134}$ & $2.01^{+0.57}_{-1.53}$ \B\T  \\
\hline
$\texttt{0628}$ Heavy &  &  &  &  &  &  &  &  &  &  & \B\T  \\ 
inj. & $84.07$ & $40.77$ & $0.85$ & $0.51$ & $-0.174$ & $0.84$ & $1.70$ & $2.23$ & $1020$ & $0.52$ \B\T  \\ 
$\texttt{SEOB}$ & $76.22^{+8.92}_{-8.92}$ & $46.75^{+12.28}_{-8.17}$ & $0.84^{+0.13}_{-0.24}$ & $0.32^{+0.52}_{-0.28}$ & $-0.179^{+0.094}_{-0.093}$ & $0.79^{+0.15}_{-0.24}$ & $1.89^{+0.27}_{-0.30}$ & $1.73^{+0.91}_{-1.13}$ & $1029^{+259}_{-255}$ & $0.95^{+1.74}_{-0.56}$ \B\T  \\
$\texttt{TPHM}$ & $78.37^{+13.73}_{-10.79}$ & $46.87^{+11.36}_{-8.94}$ & $0.78^{+0.17}_{-0.25}$ & $0.29^{+0.50}_{-0.26}$ & $-0.100^{+0.102}_{-0.096}$ & $0.75^{+0.19}_{-0.26}$ & $1.78^{+0.33}_{-0.31}$ & $1.62^{+0.98}_{-1.03}$ & $1130^{+323}_{-285}$ & $0.85^{+1.77}_{-0.44}$ \B\T  \\
$\texttt{XPHM}$ & $73.16^{+6.51}_{-5.59}$ & $45.84^{+8.28}_{-9.45}$ & $0.82^{+0.15}_{-0.30}$ & $0.81^{+0.16}_{-0.46}$ & $-0.139^{+0.163}_{-0.269}$ & $0.53^{+0.20}_{-0.17}$ & $2.42^{+0.31}_{-0.33}$ & $0.78^{+0.71}_{-0.50}$ & $909^{+315}_{-298}$ & $1.13^{+1.27}_{-0.53}$ \B\T  \\
\hline
 \end{tabular}
 \caption{
The results of our injection/recovery parameter estimation runs.
 For a selected subset of parameters given in the first row,
 we present the medians and $\{$5th, 95th$\}$ percentile error bars for the posteriors recovered by \SEOB, \TPHM{} and \XPHM{}
 for four injections of \SXS{} numerical relativity waveforms into LIGO Hanford-Livingston O4a sensitivity.
 The labels \fifty, \ste{} in the first column are the \SXS{} simulation numbers.
 Light and Heavy refer to the total mass of the binary black hole system.
 The row label ``inj.'' denotes the injected value. The $\{$5th, 95th$\}$ error bars are simply the following percentile differences: $\{$5th-50th, 95th-50th$\}$.
 }\label{tab:PE_posteriors} 
 \end{table*}

Overall, the least informative recovery is for $\chi_2$ as can be gathered from Figs.~\ref{fig:0050_violins},
\ref{fig:0628_violins} and Table~\ref{tab:PE_posteriors}.
Though the models capture the injected values within their 90\% CIs,
these are two to three times wider than the ones corresponding to $\chi_1$, thus hardly informative.
The only clear result is the opposite behavior between the \SEOB/\TPHM{} and \XPHM{} posteriors with the
former favoring low values and the latter high values. We should also remark that the heavy \fifty{} \SEOB/\TPHM{} posteriors are mostly prior dominated.
To better quantify this, we compute the Jensen-Shannon (JS) divergences, $D_\text{JS}$,
between the posteriors and the priors. Given two distributions $p_1(x),p_2(x)$, the JS divergence
is a measure of how similar they are, given by \cite{Lin:1991}
\be
D_\text{JS}=\f{1}{2}\left[\sum_x p_1(x) \ln\left(\f{p_1(x)}{\bar{p}(x)}\right)
+p_2(x)\ln\left(\f{p_2(x)}{\bar{p}(x)}\right)\right] \label{eq:JS_divergence},
\ee
where $\bar{p}(x)=(p_1(x)+p_2(x))/2$. $D_\text{JS}=0$ means identical distributions.
We find that for the heavy-mass \fifty{} run, both \SEOB{} and \TPHM{} $\chi_2$ posteriors yield
$D_\text{JS} <0.01$ indicating high resemblance to the flat $\chi_2$ priors as can be seen from
Fig.~\ref{fig:0050_violins}.

Similarly, for $\theta_2$, the \SEOB/\TPHM{} posteriors are mostly prior dominated.
For three \TPHM{} runs, the JS divergence between the $\theta_2$ posteriors
and the prior are all $\lesssim 0.01$, indicative of very similar distributions.
Even for the light-mass \ste{} run, \TPHM{} yields $D_\text{JS}\approx 0.019$.
\SEOB{} posteriors give $D_\text{JS}\approx 0.03$ for the same run, and $ D_\text{JS}< 0.015$ for 
the other three runs.
On the other hand, the JS divergences between \XPHM's posteriors and the priors are all greater
than 0.1 except for the light \fifty{} run which still yields $D_\text{JS}>0.02$, which is usually
taken to be the threshold for similarity between two distributions.

\subsection{Inferring Precession from the Injections}\label{sec:prec_recovery}
In the previous section, we deliberately omitted the discussion on the recovery of $\chip$ as we want to present it along with the other precession scalars, namely $\chipGen$ and $\chiperp$.
Recall that our injected signals have total SNRs of 40 each, part of which is due to the precession SNR,
specifically, $\rho_\text{p}=\{\{11,6.4\},\{15,10\}\}$, respectively for light/heavy
\fifty{} and light/heavy \ste{}.

\begin{figure}[t!]
    \centering
    \includegraphics[width=0.49\textwidth]{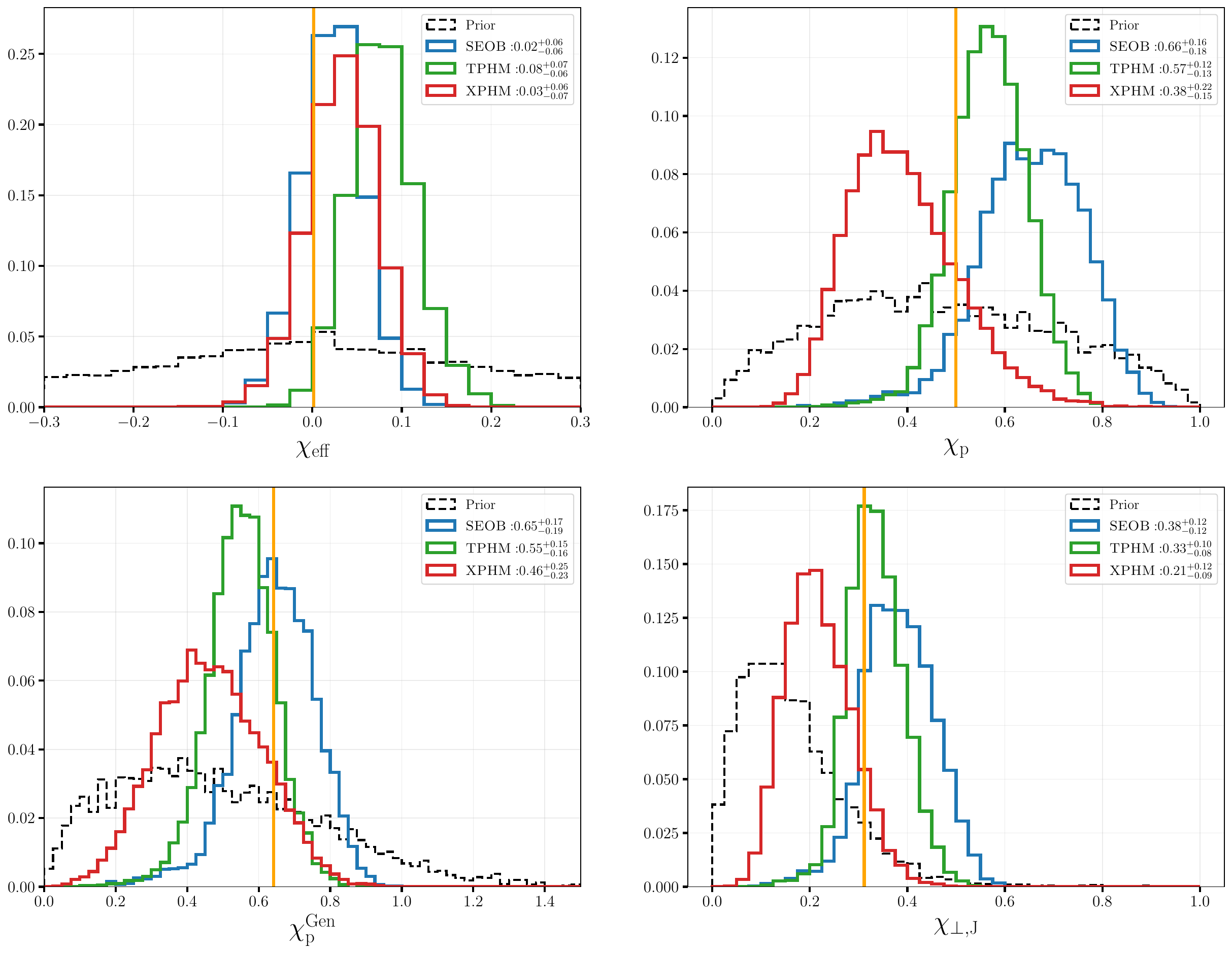}
       \caption{
       Posterior distributions recovered by \TPHM{} and \XPHM{}
       for $\{\chieff,\chip,\chipGen,\chiperp\}$ for the injection of the $\texttt{SXS}$ simulation
       \fifty{} with $\Mdet=56.5M_{\odot}$.
       The black dashed distribution is the prior for each quantity.
       The orange line marks the injected value. The color coding of each model is consistent
       with the entire article.}
\label{fig:0050_light_chi_hist}
\end{figure}

\begin{figure}[t!]
    \centering
    \includegraphics[width=0.49\textwidth]{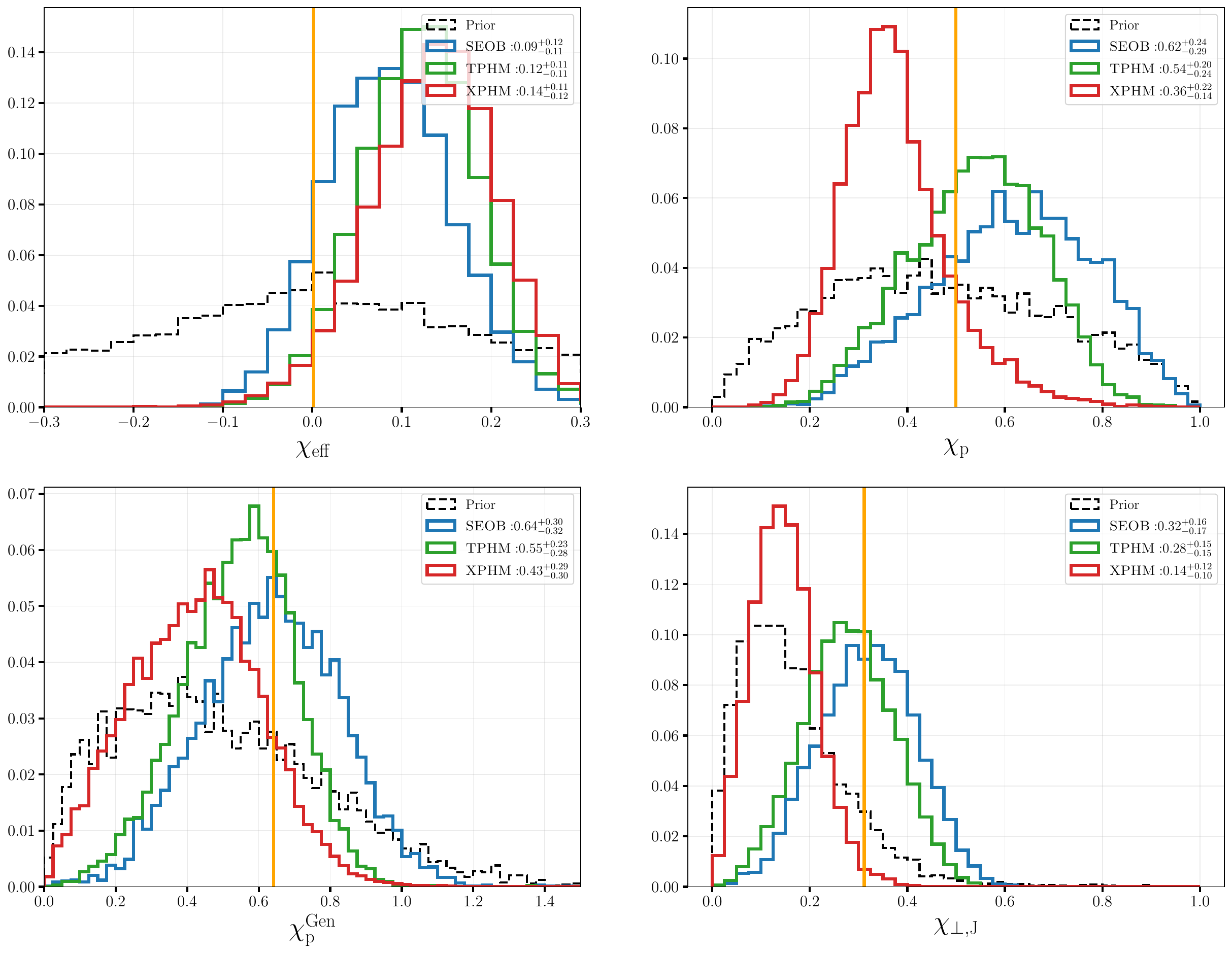}
       \caption{
       Same as Fig.~\ref{fig:0050_light_chi_hist}, but for the heavy \fifty{} PE run
       with an injected value of $\Mdet=150\Msun$. Note that the vertical scale is different from
       Fig.~\ref{fig:0050_light_chi_hist}, which is why the histograms for the priors may give the illusion of looking
       different, when they are in fact identical.}
\label{fig:0050_heavy_chi_hist}
\end{figure}

We present the recovered posteriors for $\{\chip,\chipGen,\chiperp\}$ (along with $\chieff$)
from the light/heavy \fifty{} PE runs in Figs.~\ref{fig:0050_light_chi_hist} and
\ref{fig:0050_heavy_chi_hist}.  Similar results for the light/heavy \ste{} recovery
are shown in Figs.~\ref{fig:0628_light_chi_hist} and \ref{fig:0628_heavy_chi_hist}.
In all the figures, we observe posteriors not dominated by the priors (dashed histograms) consistent
with our having injected $\rho_\text{p}\ge 6$ into all runs.
We also see the tallest and narrowest posteriors for the $\rho_\text{p}=15$ run, i.e., the light \ste.
As was the case in the previous section, the posteriors coming from the heavy mass recovery runs
are in general wider than their light-mass counterparts.
They are additionally not as centered on the injected values and less symmetric.

Starting with the $\chip$ posteriors, we note that \TPHM{} consistently recovers the injected value
very close to its median with \SEOB{} performing similarly with some disagreement between the two
models' posteriors for the light-mass injections.
\XPHM{} consistently underestimates $\chip$, but still recovers it within its 90\% CI except for heavy \ste{} 
(see Table~\ref{tab:PE_posteriors}).
The underestimation by \XPHM{} bias was not observed in Ref.~\cite{Pratten:2020ceb}, where
the recovery of a $\{Q=6, \chieff<0, \chip\gtrsim 0.75\}$ injection (\texttt{SXS:0165}) was performed, albeit
at an SNR of 26 with no information as to what $\rho_\text{p}$ is. This might explain why
a bias was not observed there though
we should be cautious about drawing conclusions regarding systematic biases from a few PE runs \cite{Estelles:2021gvs}.

Turning our attention to the $\chipGen$ posteriors, we find that all models recover the
injected values within their 90\% CIs as can be discerned from Figs.~\ref{fig:0050_light_chi_hist},
\ref{fig:0050_heavy_chi_hist}, \ref{fig:0628_light_chi_hist} and \ref{fig:0628_heavy_chi_hist}.
The consistent underestimation of $\chip$ by \XPHM{} is not exhibited here for $\chipGen$.
\SEOB/\TPHM{} recover the injected values closer to the medians than \XPHM{} except for the heavy \ste{} run.
In any case, we do not see the bias observed in the lower panels of Fig.~7 of Ref. \cite{DeRenzis:2022vsj}, where injections of \NRsurP{} were carried out.
The reason for this most likely is that their injected SNRs were greater than 75.

As for the $\chiperp$ recovery, we find that both \SEOB{} and \TPHM{} also recover this quantity rather well, 
whereby the injected values almost line up with the peaks of nearly symmetric posterior distributions.
\TPHM's recovery is especially outstanding for the light mass runs.
\XPHM{} underestimates the injected value every time and either barely recovers within its 90\% CI or
just outside it. As far as we are aware, ours are the first PE results for this quantity so we can not
compare our findings with the literature.

Overall, both \SEOB{} and \TPHM{} seem to recover any one of the three $\chi_\perp$'s well at these relatively high SNRs.
The models especially recover $\chipGen, \chiperp$ posteriors well though
this is likely just a coincidence, nonetheless warrants a more systematic investigation.
\XPHM{} seems to best recover $\chipGen$ of the three $\chi_\perp$'s.
Focusing only on the $\chip$ recovery, we see that \SEOB, \TPHM{} perform reliably, with perhaps
the latter being most reliable in terms of consistent recovery closest to its median.

Looking at the $\{t_c,\varphi_c\}$-maximized mismatches of Table~\ref{tab:pe_matches},
we can see some support for why we expect \SEOB, \TPHM{} to perform better:
for the heavy injections, \SEOB/\TPHM{} waveforms with injected parameters yield mismatches
that are nearly half of the corresponding \XPHM{} waveforms. As for the light cases,
\SEOB{} mismatches are lower, while the \TPHM{} mismatch is lower for \fifty, and roughly the same for \ste.
We can compare these mismatches with what the indistinguishability criterion of Ref.~\cite{Baird:2012cu}
gives us. Namely, for $k$ degrees of freedom and an SNR of $\rho$, \
the required mismatch below which two waveforms will be indistinguishable at $100p\%$ confidence
is given by \cite{Hannam:2022pit}
\be
 \MM \le \f{\chi_k^2(1-p)}{2\rho^2} \label{eq:indistinguishability_mm} .
\ee
A precessing BBH system has eight degrees of freedom. 
This reduces to four if we minimize the mismatches over the  planar spin components $\phi_1,\phi_2$.
To quantify waveform indistinguishability at 90\% confidence, we use
$\chi_8^2(0.1)\simeq 13.362$ and $ \chi_4^2(0.1)\simeq 7.79$.
Accordingly, for an SNR of 40, two waveforms will be indistinguishable
if their mismatch is less than $ 4.2\times 10^{-3}$ assuming $k=8$
and $\MM\le 2.4\times 10^{-3}$ assuming $k=4$.
The mismatches we have in Table~\ref{tab:pe_matches}
are larger than these bounds, but \SEOB's are only twice to thrice the $k=8$ bound, with \TPHM's slightly larger.
Thus, we expect the results of our PE recovery to be somewhat biased, but less so for \SEOB{} and \TPHM.
Perhaps, this is manifest mostly in the precession-related posteriors.
We can also solve Eq.~\ref{eq:indistinguishability_mm} for an `indistinguishability SNR'
given a mismatch. For the fiducial value of $\MM=0.01$, we obtain $\rho \approx 26$.
Thus, any injection with less than this SNR would not pick up modeling related systematics in PE.


\section{Summary}\label{Sec:Summary}

We have conducted a long survey of four precessing waveforms models: \texttt{SEOBNRv5PHM, TEOBResumS, IMRPhenomTPHM} and \texttt{IMRPhenomXPHM}.
Our survey has a large part pertaining to assessing model faithfulness to the surrogate model \NRsurP{}
and to numerical relativity waveforms from the \SXS{} simulation catalog.
We quantify model faithfulness via the optimized mismatch, $\MMo$
[Eqs.~(\ref{eq:M_opt_av}, \ref{eq:mismatch_opt})], and the merger-ringdown truncated (inspiral-only) mismatch, $\MMno$ [Eq.~\eqref{eq:mismatch_opt_noMR}].
The second shorter part of our survey involves checking model performance in several injection/recovery parameter
estimation runs for which we could not employ \texttt{TEOBResumS}.

%
\begin{figure}[t!]
    \centering
    \includegraphics[width=0.49\textwidth]{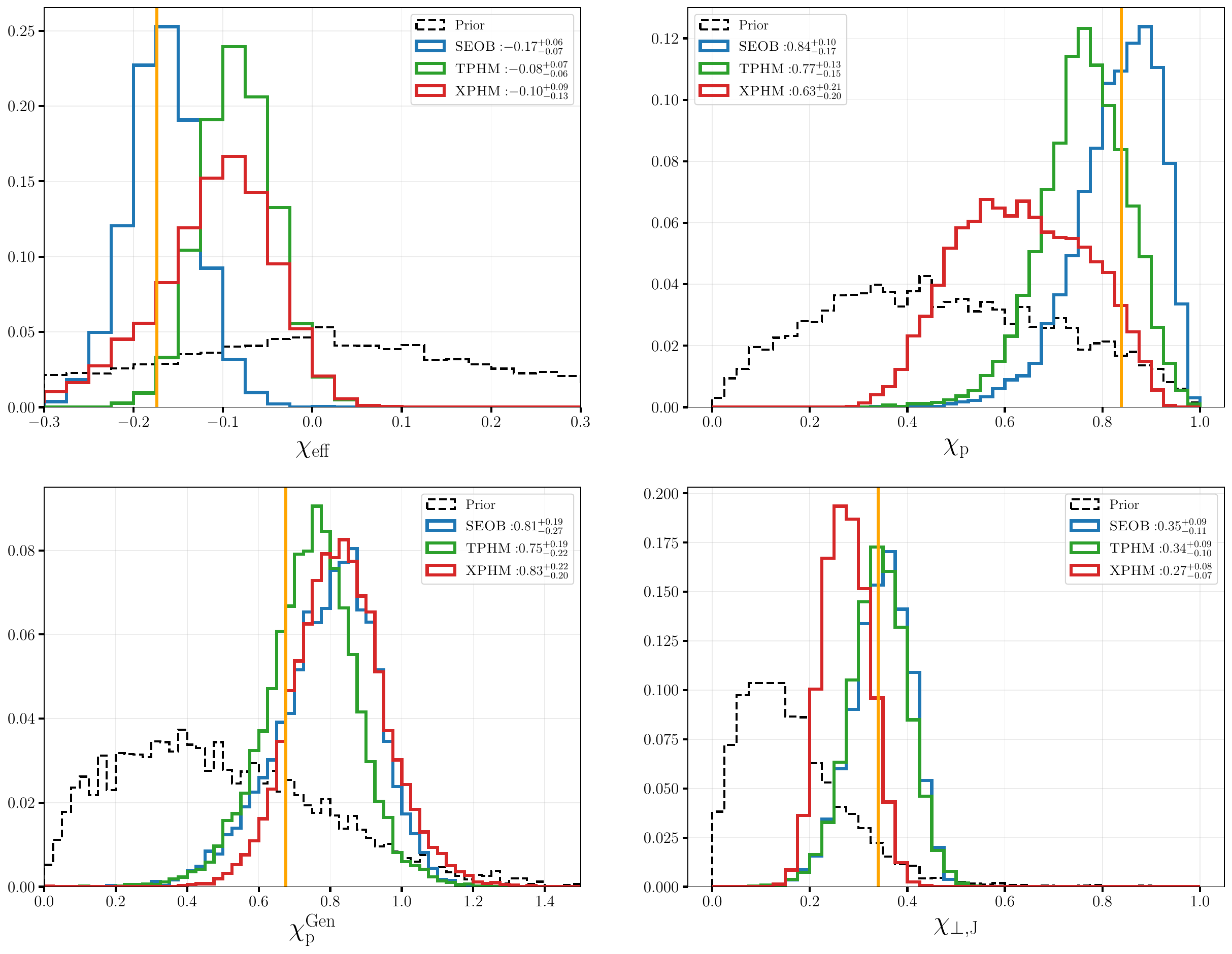}
       \caption{
       Same as Fig.~\ref{fig:0050_light_chi_hist}, but for the light \ste{} run
       with an injected value of $\Mdet=67\Msun$.}
\label{fig:0628_light_chi_hist}
\end{figure}
%
The assessment of model faithfulness to \NRsurP{} has two separate parts distinguished by the chosen set of intrinsic parameters.
In Secs.~\ref{sec:NRsur_survey_discrete}, \ref{sec:discrete_inc90},
we employ a discrete grid with emphasis on having large planar components
for the spin vectors $\Sa,\Sb$ to ``stress-test'' the precessing models.
In Sec.~\ref{sec:NRsur_survey_random}, we use a random-uniformly filled intrinsic
parameter space. The latter is commonly encountered in reviews of waveform models as well as in parameter estimation runs,
where the parameter space is randomly sampled.
As can be seen from Fig.~\ref{fig:spin_space}, the two parameter sets cover different regions.
For the discrete grid, we consider both a light and a heavy-mass binary with total source frame masses of $37.5\Msun$
and $150\Msun$. The cases with the latter mass contain fewer inspiral cycles, thus the plunge-merger-ringdown
portions contribute more to the signal than for the light-mass cases.
We further decompose each mass set into four subsets of mass ratio,
$Q=\{\Qone, 2,4,6\}$, the last of which is in the so-called extrapolation region of \NRsurP.

%
\begin{figure}[t!]
    \centering
    \includegraphics[width=0.49\textwidth]{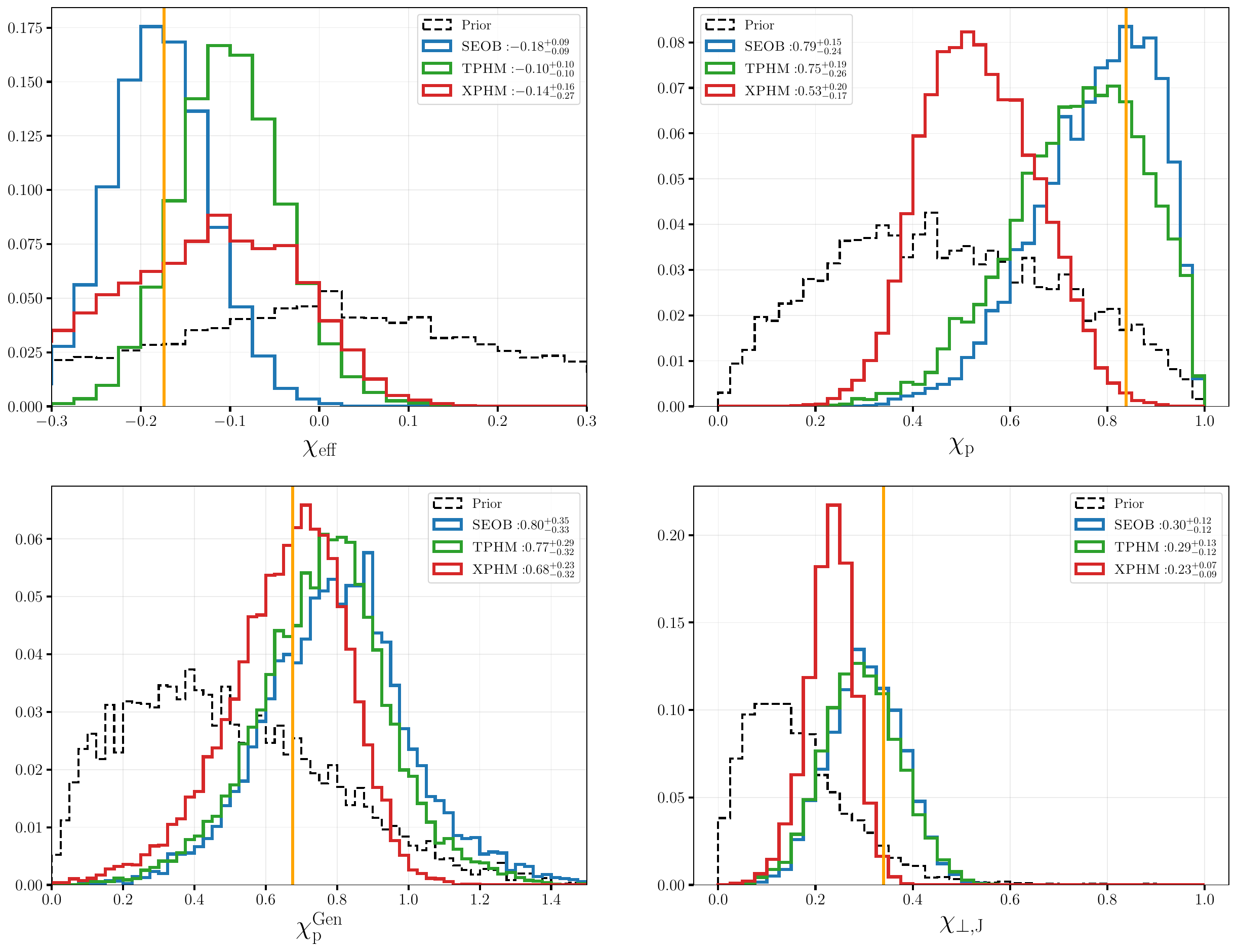}
       \caption{
       Same as Fig.~\ref{fig:0050_light_chi_hist}, but for the heavy \ste{} run
       with an injected value of $\Mdet=150\Msun$.}
\label{fig:0628_heavy_chi_hist}
\end{figure}
%

In Sec.~\ref{Sec:SXS_comparison},  we extend the above survey to comparisons with numerical relativity waveforms from the \SXS{}
catalogs. We separate our investigation into a part involving ``short'' waveforms, i.e., less than 70
GW cycles  (Sec.~\ref{sec:Short_SXS})
and into a part with longer waveforms ($\gtrsim 125$ GW cycles) in Sec.~\ref{sec:Long_SXS}.
For the former, we once again set the binary mass to $37.5\Msun$ and $150\Msun$.
For the latter, we consider a total of six values for the total mass ranging from $37.5\Msun$ to
$225\Msun$ in steps of $37.5\Msun$.
For the short \SXS{} set, we picked 317 simulations out of the larger ensemble of approximately 1600
in order to have roughly 100 precessing waveforms per mass ratio subsets:
$\{Q\lesssim 1.3, Q\approx 2, Q\approx 4\}$.
This way, we were better able to draw parallels with
the assessment conducted using \NRsurP{} in Secs.~\ref{sec:NRsur_survey_discrete}, \ref{sec:discrete_inc90}.

We compute all mismatches at inclinations of $\ioo=0$ and $\ioo=\pi/2$.
For each case, we also compute the merger-ringdown truncated, i.e., inspiral only, mismatch
denoted by the \noMR{} superscript.
Our findings are as follows:

\begin{enumerate}[label=(\roman*)]
  \item For each case in our parameter sets, $\MMno < \MMo$.
The corresponding distributions can be seen in  Figs.~\ref{fig:violin_MR_noMR}, \ref{fig:violin_random_Set}, \ref{fig:SXS_short_violin} and \ref{fig:spaghetti}.
The ratio $\cal{R}=\MMo/\MMno$ can be $\ord(10)$ or larger, but remains $\lesssim 7$ for \SEOB.
 \item The offset between the $\MMno$ distributions and the $\MMo$ distributions
 is larger for the heavier-mass binaries as can be seen by comparing the left and right panels of Figs.~\ref{fig:violin_MR_noMR}, \ref{fig:SXS_short_violin}.
 \item Model faithfulness deteriorates with increasing mass asymmetry
 as shown in Figs.~\ref{fig:violin_q_separated}, \ref{fig:violin_inc_0vs90_byQs} and
 \ref{fig:SXS_mismatches_by_Q_and_inc}.
 \item For short waveforms, model faithfulness, specifically $\MMo$, does not significantly degrade at higher inclinations as exhibited in Figs.~\ref{fig:violin_inc_0vs90}, \ref{fig:violin_random_Set}, \ref{fig:SXS_short_violin} and \ref{fig:SXS_mismatches_by_Q_and_inc}.
 However, for long waveforms, i.e., $\gtrsim 125$ cycles, faithfulness decreases as inclination goes
 from 0 to $\pi/2$ as shown in Fig.~\ref{fig:spaghetti}.
 The models' faithfulness under changing inclination also varies when considering
 the inspiral only ($\MMno$) and different mass ratios.
 \item Models exhibit higher unfaithfulness to \NRsurP/\SXS{} in regions with strong precession.
 \item When large enough, the unfaithfulness of the co-precessing (aligned-spin) $(2,\pm 2)$ multipoles
 to the corresponding \NRsurP/{\SXS} multipoles can become the dominant systematic in the precessing waveform unfaithfulness.
 This is especially the case for \texttt{TEOBResumS} and \texttt{IMRPhenomTPHM}
 as detailed in Secs.~\ref{sec:ASmode_MMs}, \ref{sec:coprec_SXS_multipoles},
 and is manifested as a strong correlation shown in Figs.~\ref{fig:Coprec_22_mismatches},
 \ref{fig:SXS_coprec_mismatches}.
 \item Sufficiently faithful precessing $(2,\pm 1)$ multipoles can make high-inclination waveforms
 more faithful than their
 low inclination counterparts as explained in Secs.~\ref{sec:discrete_inc90} and \ref{sec:short_SXS_inc}.
 \end{enumerate}

Finally, in Sec.~\ref{Sec:PE},  we conduct a parameter estimation study by performing zero-noise
injections of two precessing \SXS{} simulations (\texttt{0050} and \texttt{0628})
and recovering the injected parameters
with the models \texttt{SEOBNRv5PHM}, \texttt{IMRPhenomTPHM} and \texttt{IMRPhenomXPHM}.
Our work in this regard complements those of Refs.~\cite{Ramos-Buades:2023ehm, Pratten:2020ceb, Estelles:2021gvs},
where different \SXS{} simulations were injected.
We perform each injection twice, at a low and high total binary mass. We set the extrinsic parameters so that in each
case the total network SNR equals 40 with 15\% to 40\% coming from precession depending on the case.
The recovered posterior distributions for a subset of key parameters  are shown in Figs.~\ref{fig:0050_violins} and
\ref{fig:0628_violins} for \texttt{SXS:0050} and \texttt{SXS:0628}, respectively.
These are supplemented by Table~\ref{tab:PE_posteriors} where we present the median values as well as the
$\{5,95\}$ percentile error bars of the posteriors for a complementary subset of parameters with some overlap.

Overall, we observe mixed model performance where one model may recover certain parameters better than the other.
For example, \texttt{SEOBNRv5PHM} consistently recovers $\chieff$ better while
\texttt{IMRPhenomTPHM} could be argued to better recover $\chi_1$ (no railing), whereas
\texttt{IMRPhenomXPHM} better recovers $\theta_2$.
However, when it comes to the recovery of the effective precession scalars,
$\chip,\chipGen$ and the perpendicular spin component $\chiperp$, \texttt{SEOBNRv5PHM} and 
\texttt{IMRPhenomTPHM} outperform \texttt{IMRPhenomXPHM}, which we have presented 
in Figs.~\ref{fig:0050_light_chi_hist}, \ref{fig:0050_heavy_chi_hist}
for  \texttt{SXS:0050}, and in Figs.~\ref{fig:0628_light_chi_hist}, \ref{fig:0628_heavy_chi_hist}
for \texttt{SXS:0628}.
$\chipGen$ and $\chiperp$ are especially rather well recovered by \texttt{SEOBNRv5PHM} and \texttt{IMRPhenomTPHM}
as can be garnered from the bottom panels of these figures.
We must nonetheless repeat our cautionary remark from Sec.~\ref{Sec:PE}
that results from a few parameter estimation runs are insufficient for drawing general,
reliable conclusions regarding model performance.
Therefore, many more such studies are needed with special attention paid to the precession SNR.


\section{Discussion}\label{Sec:Discussion}

Our findings indicate that at present, the MR portion of the waveforms along with the
precessing $(2,\pm 1)$ multipoles need to be further improved to attain an overall waveform faithfulness
comparable to that of the inspiral-only $(2,\pm 2)$ multipoles for all inclinations.
As already mentioned, part of the increased unfaithfulness when considering the full waveform is simply due to changing signal morphology of the MR phase.
One can, in principle, determine how much of the increase is due to
the actual mismodelling of the MR regime by smoothly stitching the inspiral part of, e.g.,
an \texttt{SEOBNRv5PHM} waveform with the MR part of an \NRsurP/\SXS{} waveform and
computing the mismatch $\MMo$ between this hybrid waveform and the full \texttt{SEOBNRv5PHM}
waveform.  A comparison of this mismatch with the already computed $\MMno$ values
with respect to \NRsurP/\SXS{} then
roughly tells us how much of the mismatch comes from waveform systematics in the MR regime.
This comparison might provide an estimation for an achievable waveform model unfaithfulness that can be targeted.

As for the modelling of the precessing $(2,\pm 1)$ multipoles, we have seen already that the main contribution
to these comes from the AS $(2,\pm2)$ multipoles which differ most from
\NRsurP{/\SXS}'s co-precessing $(2,\pm2)$ multipoles for cases with
$60^\circ \le \theta_{1,2}\le 120^\circ$ and with increasing mass asymmetry.
This region in $\theta_{1,2}$ space is also where we believe
the assumed $m\leftrightarrow -m$ symmetry of the AS multipoles is most unfaithful.
The obvious remedy is the abandonment of this symmetry, at least for the AS $(2,\pm2)$ multipoles
which has been done for the new model \texttt{IMRPhenomXO4a} \cite{Thompson:2023ase}.

Though the AS $(2,\pm 1)$ multipoles do not matter nearly as much as their $(2,\pm2)$ counterparts,
we have seen that their omission can cause mismatches as large as 0.01 for $\ioo=0$ and 0.03 for $\ioo=\pi/2$.
As detector sensitivity steadily improves such margins will become crucial to overcome.
More work needs to be undertaken to fully determine how much the unfaithfulness of the AS $(2,\pm 1)$ multipoles matters, while it is clear that this is subdominant to the unfaithfulness of the AS $(2,\pm2)$ multipoles. However, the ratio $|h_{2,\pm1}^\text{AS}|/|h_{2,\pm2}^\text{AS}|$
increases with increasing mass asymmetry
implying that unfaithfulness of the AS $(2,\pm1)$ multipoles becomes non-negligible beyond a certain mass ratio.

The repeatedly observed fact that faithfulness, in general, deterioriates with increasing mass asymmetry
($Q\gtrsim 4$) should be addressed more systematically.
Though it is the case that these waveforms have more GW cycles than the ones with  $Q\le 2$,
we had argued that waveform length alone does not explain the increased unfaithfulness.
Part of the degradation is due to the AS multipoles as we saw in Sec.~\ref{sec:ASmode_MMs}:
they become less faithful to \NRsurP's co-precessing multipoles as $Q$ increases.
On the other hand, some deterioration could be due to the mismodelling of the AS $(2,\pm 1)$ multipoles which
have increasing contributions as discussed above. Another part could be due to the mismodelling of
the Euler angles which brings us to the second part of this discussion.

Our survey is by no means a complete one, so let us expand on possible ways to extend it here.
The twist prescription has two essential ingredients: the AS multipoles and the Euler angles used to rotate the multipoles
from the $\LN(t)$ frame (co-precessing) to an inertial (e.g., $\L_{\text{N},0}$) frame.
Though we investigated how the unfaithfulness of the AS multipoles might affect the strain
faithfulnes, we did not conduct a study on how the systematics
in the Euler angles affect the faithfulness.
We propose the following hybridized approach for such a study: Euler rotation of \NRsurP's co-precessing multipoles for which the angles are to be input from various models under investigation.
Such a study was performed in Ref.~\cite{Hamilton:2021pkf} (Fig.~18)
for assessing the MSA angles and the angles from their new prescription.

In our case, we could, for example, use \texttt{TEOBResumS}
Euler angles to twist the co-precessing \NRsurP{} multipoles
then compute the mismatches between the resulting hybrid waveforms and \NRsurP{} waveforms.
This investigation can also be extended to the choice of co-precessing frame, i.e.,
since $\L\ne \LN$, the Euler angles for the $\L(t)\to \L_0$ rotation will differ from those of the
$\LN(t) \to \L_{\text{N},0}$ rotation, albeit slightly.
Moreover, neither direction coincides with the direction of maximum GW emission which, in the time domain,
differs depending on whether one uses the GW strain, the Bondi news function, or the Weyl scalar $\psi_4$
\cite{Schmidt:2010it, Ochsner:2012dj, Boyle:2014ioa, Hamilton:2018fxk, Hamilton:2021pkf}.

It would be useful to extend our total mass coverage in Secs.~\ref{Sec:faith_survey}, \ref{sec:Short_SXS} to more values as we did for our assessment using the long \SXS{} waveforms.
A range of $\Mtot \approx 10\Msun$ to $\approx 250\Msun$ in steps of $10\Msun$ is common.
Additionally, for random-uniformly filled
parameter spaces, 5000 to $\ord(10^4)$ cases are customary.
However, given that we have to compute $\MMno$ and $\MMo$ separately at two different inclinations
for each single point in the BBH parameter space, we have four times the computational burden.
This is a considerable increase as each optimized mismatch requires up to
$\ord(10^5)$ waveform generations which we must then repeat for 42 points in our $\{\varphi_\text{ref},\kappa\}$ grid [Eq.~\eqref{eq:M_opt_av}].

Another obvious way to extend our survey is by including higher multipoles in the strain.
Ref.~\cite{Ramos-Buades:2020noq} already conducted such a study with the multipoles
$(3,\pm3),(3\pm2),(4,\pm4),(4,\pm3)$ included.
We can extend our AS $(2\pm2),(2\pm1)$ multipole assessment of Sec.~\ref{sec:ASmode_MMs}
to these multipoles. 
We can also re-apply all the analyses that we performed on the $\ell=2$ strain
to the $\ell>2$ strain.

Yet another extension is to employ as many precessing numerical relativity waveforms as feasible
as was done in the LVK study using $\gtrsim 1500$ NR waveforms
that we mentioned in Sec.~\ref{Sec:introduction}.
A shorter version of this at a single inclination of $\pi/3$ was presented in
Ref.~\cite{Ramos-Buades:2023ehm}.
We can supplement these results with our $\MMno$ and $\MMo$ computed
at inclinations of 0 and $\pi/2$ and possibly more.
All of this data could then be presented in terms of mass ratio as in our
Figs.~\ref{fig:violin_q_separated}, \ref{fig:violin_inc_0vs90_byQs} and \ref{fig:SXS_mismatches_by_Q_and_inc}.
We should also add that the latest trend in the literature is to present model unfaithfulness in terms of
an SNR-weighted match \cite{Ossokine:2020kjp, Hamilton:2021pkf}.
This quantity is different than the sky-optimized match $\M_\text{opt}$ of Eq.~\eqref{eq:opt_match}.
Therefore, a comparison of these two faithfulness measures may also be informative.

A large scale injection/recovery parameter estimation campaign would make certain vague trends either more
concrete or obsolete. However, such studies are computationally expensive which is why most model
reviews suffice with one or two of these. Nevertheless, we have found that using the
\texttt{parallel\_bilby} library, we can complete two chains of \texttt{IMRPhenomTPHM}
\emph{and} two chains of \texttt{IMRPhenomXPHM} injection/recovery runs
per day using $\ord(100)$ processors. 
With the same resources, we can alternatively obtain one \texttt{SEOBNRv5PHM} chain per $\sim 1.5$ days.
So, $\gtrsim 30$ injection/recovery PE runs can be completed
in approximately one month with the \textsc{IMRPhenom} models.

The conclusions of injection/recovery studies are dependent on the choice of injected SNRs.
In our case, we opted for a total SNR of 40 which may seem unrealistic given that
95\% of the O1, O2, O3 events had $\text{SNR} \lesssim 23$\footnote{We extracted this value from the list of events provided in Ref.~\cite{enwiki:1188064092}.}.
However, there will be a few events with large enough SNRs for which accurate modelling of
precession will be crucial. GW200129\_065458, at an SNR of 26, was such an event where a re-analysis of
the data with the state of the art precessing model \NRsurP{} discovered strong precession
\cite{Hannam:2022pit}, consistent with an earlier analysis from an \texttt{IMRPhenomXPHM}
run, but not with the one from \texttt{SEOBNRv4PHM} \cite{LIGOScientific:2021djp}.
The precession SNR $\rho_\text{p}$ should also be chosen
carefully in such studies. Even if an SNR=40 system is very strongly precessing,
if the extrinsic parameters conspire to minimize $\rho_\text{p}$, the imprint of precession
on the received GWs will be weak.

Finally, let us add that
a survey such as ours risks becoming outdated by the time it may be complete.
Indeed, during the writing of this article, the model \textsc{SEOBNR} got upgraded from \texttt{v4} to \texttt{v5}.
Additionally, the MSA prescription of BBH precession dynamics in \texttt{IMRPhenomXPHM} got replaced
by one based on \textsc{SpinTaylorT4} \cite{Colleoni:2023}.
Finally, an upgraded \textsc{IMRPhenomX} (\texttt{O4a}) model was released \cite{Thompson:2023ase}.
The inclusion of any new models into a similar survey inevitably delays the completion of the survey
itself. However, once a new model is incorporated into \texttt{LALSimulation}
it should, in principle, be straightforward to assess its faithfulness within the framework that we have built.



\begin{acknowledgments}
We thank Rossella Gamba for sharing codes and help with \textsc{TEOBResumS},
Matteo Breschi for help with \texttt{bajes}, Marta Colleoni for answering questions regarding \textsc{IMRPhenomXP}.
We are grateful to Charlie Hoy for helping us with the troubleshooting of the \texttt{SEOBNRv5PHM} PE runs 
and sharing the results of his validation runs using the Sciama High Performance Compute (HPC) cluster which is supported by the ICG, SEPNet and the University of Portsmouth.
We are also grateful to Lorenzo Pompili for pointing out the final fix to our \texttt{SEOBNRv5PHM} PE runs
and sharing, in particular, the results of the heavy mass PE injection/recovery with \texttt{SEOBNRv5PHM}.
This work makes use of the Black Hole Perturbation Toolkit \cite{BHPToolkit} and
the ``pesummary`` package~\cite{Hoy:2020vys}.
S.A. and J.MU. acknowledge support from the University College Dublin Ad Astra Fellowship.
J.T. acknowledges support from the NASA LISA Preparatory Science grant 20-LPS20-0005.

%
\begin{table*}[t!]
 \begin{tabular}{cccccccccc}
 \hline
  Model & $\Mtot(\Msun)$ & $\ x \ $ & \ $(n_x,n_y,n_z)$\ & \ \# params. & $\bar{R}^2$ \ & $\ \ \Delta_\text{rel}^\text{av} \ \ $ &\ \ BIC \ \ &\ \ AICc \\
  \hline\hline
    \SEOB & 37.5 & $\chipGen$ & (4,2,1) & 21 & 0.777 &\  3.33$\times 10^{-3}$\ & -696 & -793 \\
  \TEOB & 37.5 & $\chipGen$ & (7,5,2) & 81 & 0.959 &\ 2.49$\times 10^{-3}$\ & -1512 & -1858 \\
  \TPHM & 37.5 & $\chipGen$ & (6,4,2) & 60 & 0.874 &\ 1.54$\times 10^{-2}$\ & -304 & -565 \\
  \XPHM & 150 & $\chipGen$ & (6,5,2) & 63 & 0.895 &\ 4.97$\times 10^{-3}$\ & -481 & -754 \\
  \SEOB & 150 & $\chipGen$ & (6,4,2) & 60 & 0.813 &\ 5.10$\times 10^{-3}$\ & +76 & -182 \\
  \TEOB & 150 & $\chipGen$ & (6,5,2) & 63 & 0.893 &\ 5.96$\times 10^{-3}$\ & -479 & -751 \\
  \TPHM & 150 & $\chipGen$ & (6,4,2) & 60 & 0.869 &\ 6.89$\times 10^{-3}$\ & -319 & -579 \\
  \XPHM & 150 & $\chipGen$ & (7,4,2) & 75 & 0.914 &\ 6.79$\times 10^{-3}$\ & -248 & -570 \\
  \SEOB & 37.5 & $\chiperp$ &(7,5,2)& 81 & 0.884 &\ 3.17$\times 10^{-3}$\ &-823 &-1171\\
  \TEOB & 37.5 & $\chiperp$ & (7,3,2) & 66 & 0.965 &\ 1.72$\times 10^{-3}$\ & -1646 & -1929 \\
  \TPHM & 37.5 & $\chiperp$ & (6,5,3) & 73 & 0.898 &\ 5.63$\times 10^{-3}$\ & -353 & -666 \\
  \XPHM & 37.5 & $\chiperp$ & (7,4,1) & 55 & 0.904 &\ 3.96$\times 10^{-3}$\ & -606 & -846 \\
  \SEOB & 150 & $\chiperp$ & (6,5,2)& 63 & 0.884 &\ 5.10$\times 10^{-3}$\ & -203 &-475\\
  \TEOB & 150 & $\chiperp$ & (6,5,2) & 63 & 0.918 &\ 4.85$\times 10^{-3}$\ & -680 & -954 \\
  \TPHM & 150 & $\chiperp$ & (6,4,2) & 60 & 0.908 &\ 5.79$\times 10^{-3}$\ & -572 & -834 \\
  \XPHM & 150 & $\chiperp$ & (7,3,2) & 66 & 0.933 &\ 5.35$\times 10^{-3}$\ & -457 & -742\\
  \hline
 \end{tabular}
 \caption{List of the three dimensional fits to the mismatches between a given waveform model
 and \NRsurP. Specifically, we have fitted to $\log_{10}[\MMo(\ioo=0)]$ values resulting from
 the discrete parameter set  of Sec.~\ref{sec:NRsur_survey_discrete}.
 From left to right, the columns are:
 the model name, the total mass, the $x$ variable used in the fits [see Eq.~\eqref{eq:fit3D}], the three dimensional polynomial order of the fits (ibid.),
 the total number of parameters used, the adjusted R-squared value, the average relative
 difference between the fit and the validation data [see Eq.~\eqref{eq:av_rel_diff}],
 the Bayesian information criterion value, and the value of the finite-sample size
 corrected Akaike information criterion.
 }\label{tab:fit_metrics}
\end{table*}

We thank Niels Warburton for providing access to the `chirp2' computer, which is funded by Royal Society - Science Foundation Ireland University Research Fellowship grant no.~RGF\textbackslash R1\textbackslash180022.
This research has made use of data, software and/or web tools obtained
from the Gravitational Wave Open Science Center \cite{LIGOScientific:2019lzm}
(\url{https://www.gw-openscience.org}),
a service of LIGO Laboratory, the LIGO Scientific Collaboration and the
Virgo Collaboration. We have additionally employed the computational resources of
LIGO Laboratory (CIT cluster) supported by the U.S. National Science Foundation Grants PHY-0757058 and PHY-0823459.
LIGO Laboratory and Advanced LIGO are funded by the United States NSF as well as the
Science and Technology Facilities Council (STFC) of the United Kingdom, the Max-Planck-Society (MPS), and
the State of Niedersachsen/Germany for support of the construction of Advanced LIGO and construction and
operation of the GEO600 detector. Additional support for Advanced LIGO was provided by
the Australian Research Council. Virgo is funded, through the European
Gravitational Observatory (EGO), by the French Centre National de Recherche Scientifique (CNRS),
the Italian Istituto Nazionale di Fisica Nucleare (INFN) and the Dutch Nikhef,
with contributions by institutions from Belgium, Germany, Greece, Hungary, Ireland, Japan,
Monaco, Poland, Portugal, Spain. KAGRA is supported by Ministry of Education, Culture, Sports, Science and
Technology (MEXT), Japan Society for the Promotion of Science (JSPS) in Japan; National Research
Foundation (NRF) and Ministry of Science and ICT (MSIT) in Korea; Academia Sinica (AS) and National Science and
Technology Council (NSTC) in Taiwan.

{\it Software}: The following software have been used for this work. The
\texttt{numpy} \cite{Harris:2020xlr}, \texttt{matplotlib} \cite{Hunter:2007ouj},
\texttt{pandas} \cite{reback2020pandas, mckinney-proc-scipy-2010} and
\texttt{seaborn} \cite{Waskom2021} libraries of \texttt{python}.
\texttt{LALSuite\;(7.10)} \cite{lalsuite},
\texttt{LALInference} \cite{Veitch:2014wba}, \textsc{PyCBC\;(2.0.6)} \cite{Biwer:2018osg}
{\sc BILBY\;(2.1.2)} \cite{Ashton:2018jfp, Romero-Shaw:2020owr},
the \texttt{sxs} package (v2022.5.2) \cite{Boyle_The_sxs_package_2023},
the \texttt{Scri} package (2022.8.8) \cite{mike_boyle_2020_4041972}, and
\textsc{TEOBResumSv4.1.5-GIOTTO}    \href{https://bitbucket.org/eob_ihes/teobresums/commits/84b8f104842b789757e326882bbf77dd9a3afaa9}{84b8f10} (6 July 2023).
These results were found to be in good agreement with the more recent
\textsc{TEOBResumS} \href{https://bitbucket.org/eob_ihes/teobresums/commits/bd3452e2ea1003f3042a281af0304dcb1bffed80}{bd3452e} (7 September 2023).
Our notebooks and data are available at our
\href{https://github.com/akcays2/Survey_Precessing_Models}{\git{} repository.}

\end{acknowledgments}


\appendix

\section{Multidimensional Fits of Mismatches for Bayesian Model Selection}
\label{Sec:fits}
As briefly mentioned in Sec.~\ref{Sec:introduction},
we can use the results of our faithfulness survey to inform Bayesian model selection. For example,
one can employ the waveform mismatch, $\MMo$ in our case,
as a type of weight to generate a weighted categorical prior.
This then can act as a custom prior in a joint Bayesian analysis \cite{Hoy:2022tst}.
Therefore, by constructing fits to $\MMo$ over the intrinsic parameter space,
we can effectively provide a parameter space dependent prior for Bayesian model selection.
We take a first step toward building this categorical prior here by constructing multidimensional
fits to $\MMo(\ioo=0)$.

Though we have at our disposal several sets of intrinsic parameters, namely the discrete grid
of Sec.~\ref{sec:NRsur_survey_discrete}, the uniformly filled space of Sec.~\ref{sec:NRsur_survey_random},
and the \SXS{} sets of Sec.~\ref{Sec:SXS_comparison}, we choose
the discrete grid, as this set seems to cover the high-spin regions of
the parameter space better (see Fig.~\ref{fig:spin_space}).
We should keep in mind that we fixed $\chi_1=\chi_2=0.8$
for this set.

In more detail, we construct multidimensional polynomial fits to $\log_{10}[\MMo(\ioo=0)]$
as functions of  $\eta,\chieff$ and either $\chipGen$ or $\chiperp$.
We choose these perpendicular projections as
opposed to $\chip$ since these have better coverage of the parameter space
as shown in Fig.~\ref{fig:spin_space}.
Using the symmetric mass ratio $\eta$ allows us to work with a compact fit domain since
$0<\eta\le 0.25$.
We do not include the $Q=6\,(\eta=6/49)$ subset for our fits here though this can 
be included straightforwardly.
For our specific fits, we use the values for $\{\chieff,\chipGen,\chiperp\}$ at $f=f_0$,
but have dropped the subscript 0 to reduce clutter.
The total mass, $M$, is the fourth parameter in our fits, but since we
have only two points in mass space, i.e., $M=37.5\Msun, 150\Msun$,
we construct separate 3D fits for each mass.
Future work in this direction will have additional data coming from other values of
the total mass, thus necessitating 4D fits.

%
%
\begin{figure*}
     \centering
     \includegraphics[width=\textwidth]{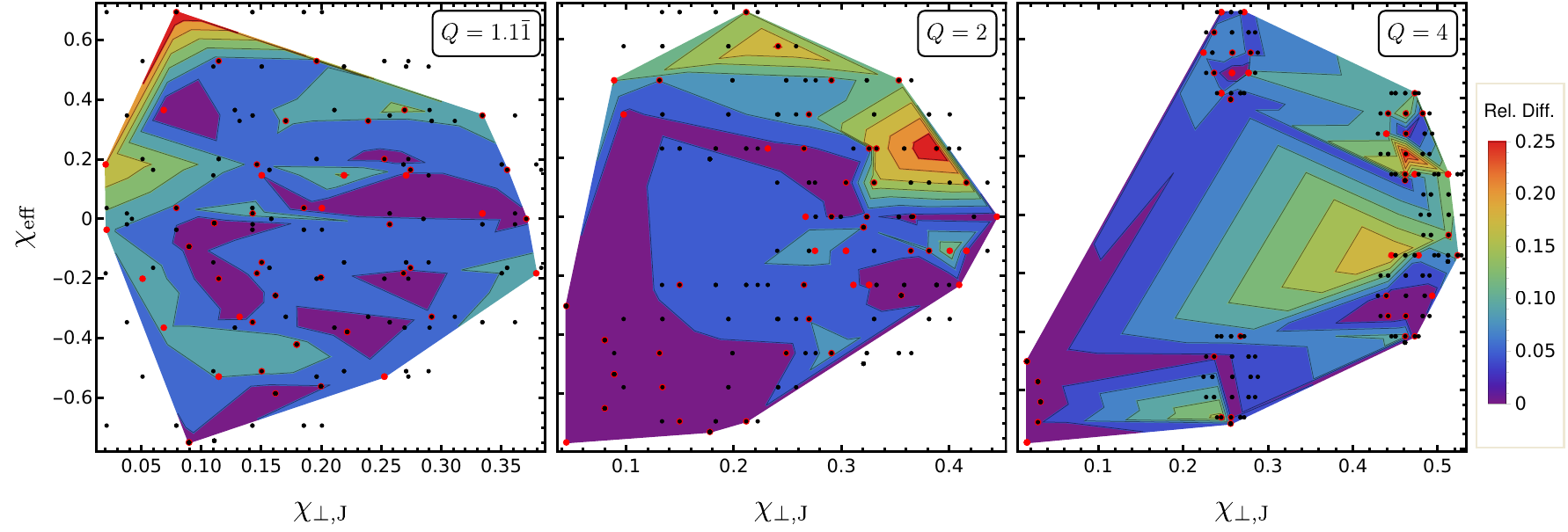}
     \caption{Contour plots of the relative difference between the validation data (red points)
     and the third \TEOB{} fit of Table~\ref{tab:fit_metrics}, which is obtained
     from Eq.~\eqref{eq:fit3D} with $\{n_x,n_y,n_z\}=\{7,3,2\}$.
     This particular fit is a function of $\chiperp,\chieff$ and $\eta$ and represents the $M=37.5\Msun$, $\log_{10}[\MMo(\ioo=0)]$ data, where $\MMo$ is the sky-optimized mismatch between \NRsurP{} and \TEOB.
     The three panels from left to right show contour plots of the relative differences separated by
     mass ratios of $Q=\Qone, 2,4$. The black dots are the data points used to construct the
     fit and are not covered by the contour plot shown here which only applies to the red dots.}
     \label{fig:fits3D}
\end{figure*}

For fitting, we employ simple polynomials of the form
\be
\mathcal{P}_{\mathcal{M}}(x,y,z)=\sum_{j=0}^{n_y}\sum_{k=0}^{n_z}\sum_{i=0}^{n^{kj}_\text{max}}a_{ijk}\, x^i y^j z^k, \label{eq:fit3D}
\ee
where $a_{ijk}$ are the fitting coefficients, $\{y,z\}=\{\chieff,\eta\}$, $x=\chipGen$ or $\chiperp$, and $n^{kj}_\text{max}=\max\{n_x,n_y,n_z\}-k-j$.
We expect the highest variation in mismatches to be along the planar spin direction
and the least variation in the mass ratio space. Therefore, we impose $n_x>n_y>n_z$
with $n_x\le 7, n_y \le 5, n_z \le 4$.
For our fits, we use a routine that picks the best values
for $n_x,n_y,n_z$ 
that minimize
\begin{inparaenum}[(i)]
\item $1-\bar{R}^2$, where $\bar{R}^2$ is the adjusted R-squared value,
\item the Bayesian information criterion (BIC),
\item the Akaike information criterion with the finite sample correction (AICc),
and \item the average relative difference between the fit $\mathcal{P}_{\mathcal{M}}$
and the validation data $d_i=(x_i,y_i,z_i,w_i)$ which we define as
\end{inparaenum}
\be
\Delta_\text{rel}^\text{av}:= \f{1}{N}\left(\sum_{i=1}^N\left[1-\f{w_i}{\mathcal{P}_{\mathcal{M}}(x_i,y_i,z_i)}\right]^2\right)^{1/2}\label{eq:av_rel_diff},
\ee
where $w_i$ represent $\log_{10}[\MMo(\ioo=0)]$.
We employ two thirds of the data set for fitting and the remaining one third for validation, where
the assignment fit/validation is made randomly. There are likely more informative
ways of choosing the fit/validation subsets, but we leave this for future work.
There is also not necessarily one set of unique values of $\{n_x,n_y,n_z\}$
that satisfy all four criteria listed above.
If more than one set of values satisfy the criteria, we pick the one with
smaller values of $\{n_x,n_y,n_z\}$.
On the other hand, if the routine can not meet all four criteria, it looks for $\{n_x,n_y,n_z\}$
that minimize three out of the four.

Using this routine and the fitting function \eqref{eq:fit3D}, we construct 3D fits
of $\log_{10}[\MMo(\ioo=0)]$ for all four approximants for $M=37.5\Msun,150\Msun$.
We find that in most cases, the best fit functions have $n_x=6$ or 7, $n_y=4$ or 5
and, not surprisingly, $n_z\le 2$ (since there are only three $\eta$ values) ,
thus resulting in roughly $\ord(60)$ parameters
as shown in Table~\ref{tab:fit_metrics} where we also report various goodness-of-fit metrics.
As can be seen from the table,
the mismatches for \TEOB{} and \XPHM{} can be fit rather well for both light
and heavy masses with $\bar{R}^2$ values as high as 0.965 and an average relative
disagreement as low as $1.7\times 10^{-3}$. 
\SEOB{} mismatches can still be fit, but not as well, with $\bar{R}^2$ below 0.9.
\TPHM{} mismatches prove to be hardest to fit with the corresponding
$\Delta_\text{rel}^\text{av}$ values being some of the highest.

Table~\ref{tab:fit_metrics} also hints that fits constructed using $\chiperp $
yield slightly higher (lower) values for
$\bar{R}^2 (\Delta_\text{rel}^\text{av})$ than those using $\chipGen$ though the difference
is rather marginal and requires a more thorough investigation to be conclusive.
We show one of the fits of Table~\ref{tab:fit_metrics} in Fig.~\ref{fig:fits3D}
in terms of 2D contourplots of the
relative difference between the fit and the validation data for the $M=37.5\Msun$
\TEOB{} mismatches. This particular fit uses $\chiperp$ 
so corresponds to the third \TEOB{} row of Table~\ref{tab:fit_metrics} with $\bar{R}^2=0.965$.
As the figure shows, for most of the plot regions, the relative difference between the
validation data points (red) and the fit is less than 0.05.
In fact, according to Table~\ref{tab:fit_metrics}, the average relative difference is
$\approx 1.7\times 10^{-3}$. Fits of this level of faithfulness may prove quite useful
for Bayesian model selection.

We additionally briefly explored whether or not a more finely sampled set of parameters
improves the goodness of the fits. In particular, we re-computed the
$\Mtot=150\Msun$ \NRsurP-\XPHM{} mismatches for $\ord(3000)$ BBHs, i.e.,
nearly three times as many data points in the spin space.
For both the $\chipGen$ and $\chiperp$ fits, we observed slight improvement
with $\bar{R}^2$ almost equalling 0.95 and $\Delta_\text{rel}^\text{av}$
dropping to $2\times 10^{-3}$ which can be compared with the second and fourth \XPHM{}
rows of Table~\ref{tab:fit_metrics}.
We should add that this improvement comes at the expense of
increasing the computational burden by a factor of three.

In summary, we have illustrated that, using certain projections of the spins, it is possible to construct
reasonably good three-dimensional fits of the mismatches, effectively capturing most
of the behavior of model faithfulness over the seven-dimensional intrinsic parameter space.
The fits can be improved by including a denser coverage of the parameter space,
but also need to be extended along the fourth dimension (total mass) to be of
actual use in Bayesian model selection. There are further additional improvements that
can be made, but we leave it all to future work. Our goal here was to show that
faithful fits can be constructed in principle by using either $\chipGen$ or $\chiperp$
and the goodness-of-fit metrics in Table~\ref{tab:fit_metrics} seem to affirm this.


\section{A brief Comparison of \texttt{SEOBNRv5PHM} with  \texttt{SEOBNRv4PHM}}\label{app:SEOBv4_vs_v5}
\begin{figure}[t!]
    \centering
    \includegraphics[width=0.49\textwidth]{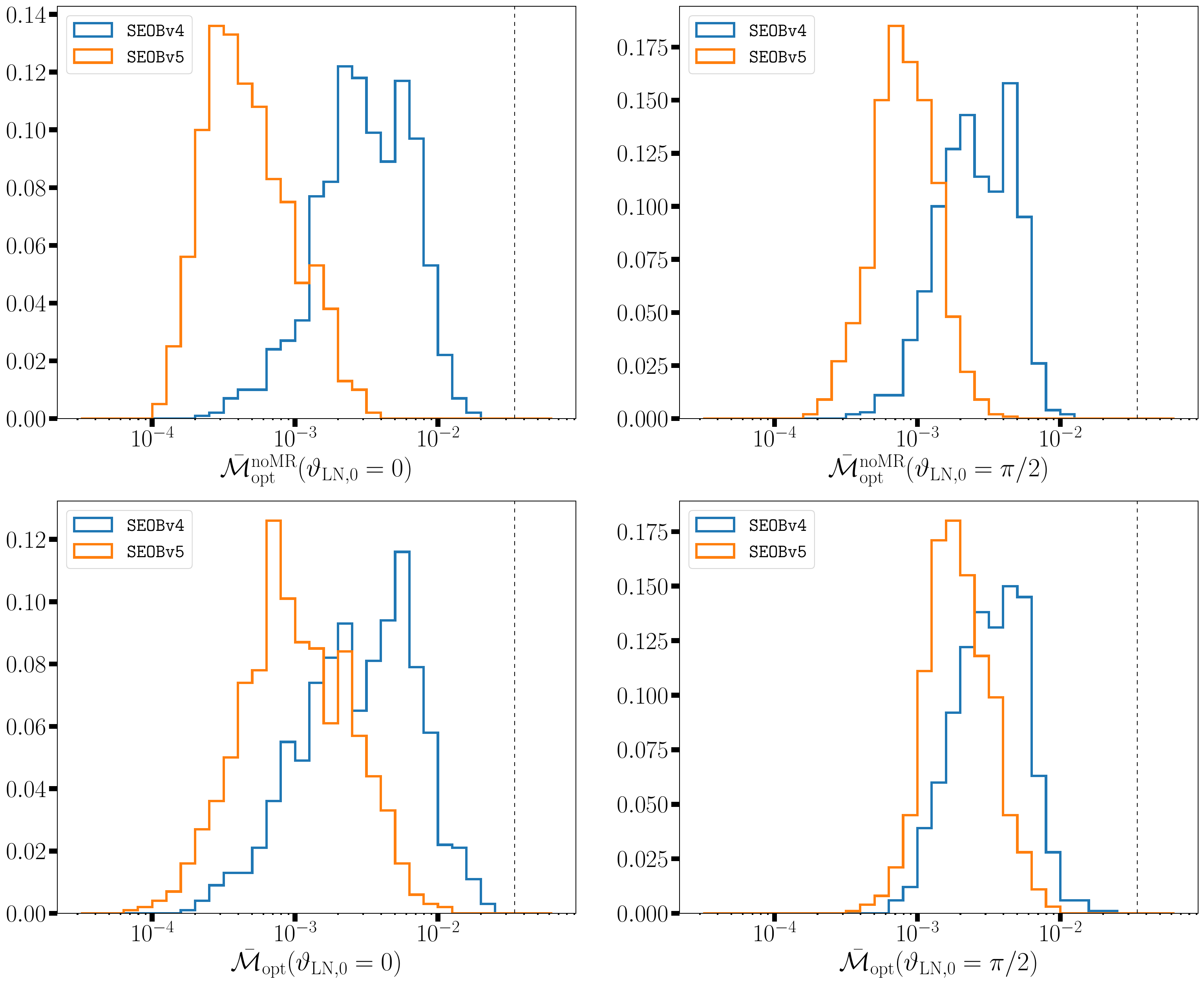}
\caption{
Comparison of the faithfulness of \texttt{SEOBNRv5PHM} with that of \texttt{SEOBNRv4PHM}
for the 1000-case random-uniformly filled parameter space described in
Sec.~\ref{sec:NRsur_survey_random}. The baseline model used to measure faithfulness is \NRsurP.}
\label{fig:SEOB_v4_v5_short_SXS}
\end{figure}

As already mentioned, while in the middle of this project, the upgraded model \texttt{SEOBNRv5PHM}
\cite{Ramos-Buades:2023ehm} was released along with a simple-to-use python package
\texttt{pySEONBR} \cite{Mihaylov:2023bkc} for it.
As the newer model is considerably faster than its \texttt{v4} predecessor, we were easily compelled
to use it. We had however already generated a considerable data set of \texttt{SEOBNRv4PHM}
mismatches by then. Therefore, we show a brief comparison between the newest (\texttt{v5})
and the previous versions in Fig.~\ref{fig:SEOB_v4_v5_short_SXS} where we plot
$\MMno$ and $\MMo$ once again at inclinations of $\ioo=0$ and $\pi/2$ for
the randomly-uniformly filled 1000-case parameter set.
We see that the upgraded model is more faithful to \NRsurP{} than its predecessor.

%
%
%
%
%
%
%


\section{Another Comparison: \texttt{IMRPhenomXO4a} vs. MSA and SpinTaylor Implementations 
of \texttt{IMRPhenomXPHM}}\label{app:MSA_vs_4_vs_SpinTaylor}
\begin{figure}[t!]
    \centering
    \includegraphics[width=0.49\textwidth]{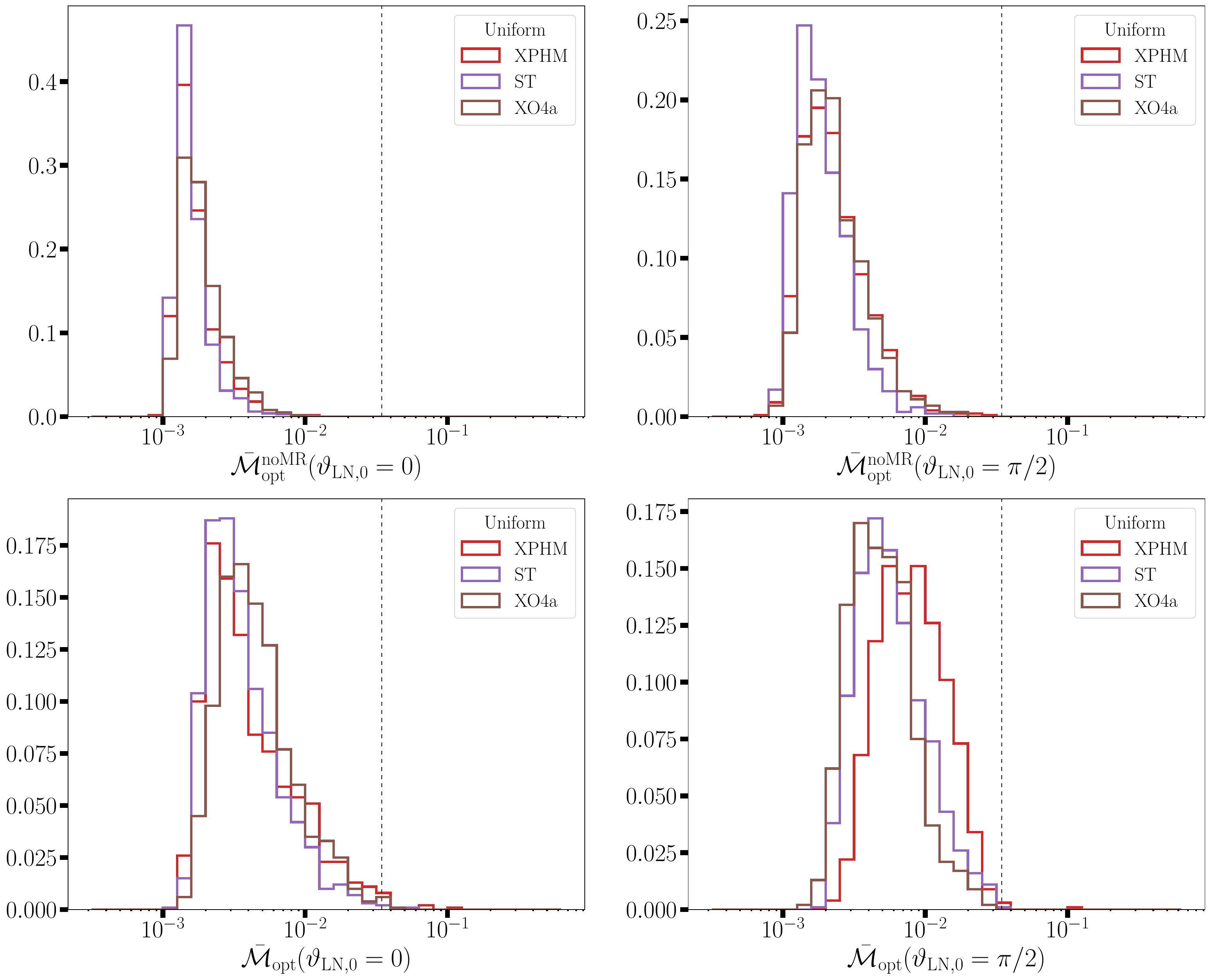}
\caption{Comparison of the faithfulness of the MSA implementation of \texttt{IMRPhenomXPHM} 
with that of its more recent \textsc{SpinTaylor} implementation (ST in the figure)
and the new model \texttt{IMRPhenomXO4a}.
The parameters used to generate the waveforms are those of the random-uniformly filled set of
Sec.~\ref{sec:NRsur_survey_random}. The baseline model used to measure faithfulness is \NRsurP.}
\label{fig:SpinTaylor_vs_MSA_XPHM_uniform_set}
\end{figure}

Here, we provide a brief comparison between the original \texttt{IMRPhenomXPHM} (using MSA angles) \cite{Pratten:2020ceb}
with the upgraded version that uses \textsc{SpinTaylorT4} dynamics to obtain the Euler angles \cite{Colleoni:2023}.
We also include the new model \texttt{IMRPhenomXO4a} \cite{Thompson:2023ase} in the comparison.
We had briefly discussed the unfaithfulness performances of the MSA and the \textsc{SpinTaylorT4} versions
at the end of Sec.~\ref{sec:Long_SXS}.
We supplement it here with Fig.~\ref{fig:SpinTaylor_vs_MSA_XPHM_uniform_set} where we plot
$\MMno$ and $\MMo$ yet again at inclinations of $\ioo=0$ and $\pi/2$ for
the random-uniform parameter set.
We observe very similar performances between the MSA and \textsc{SpinTaylor} implementations,
and \texttt{XO4a}. The performance of the MSA version is only slightly inferior at higher inclinations 
with the newer \textsc{IMRPhenomX} models slightly outperforming their predecessor.
All three models yield very similar distributions for the inspiral-only mismatches $\MMno$, indicating
that the differences exhibited may be more related to merger-ringdown modelling.

\begin{figure}[t!]
    \centering
    \includegraphics[width=0.49\textwidth]{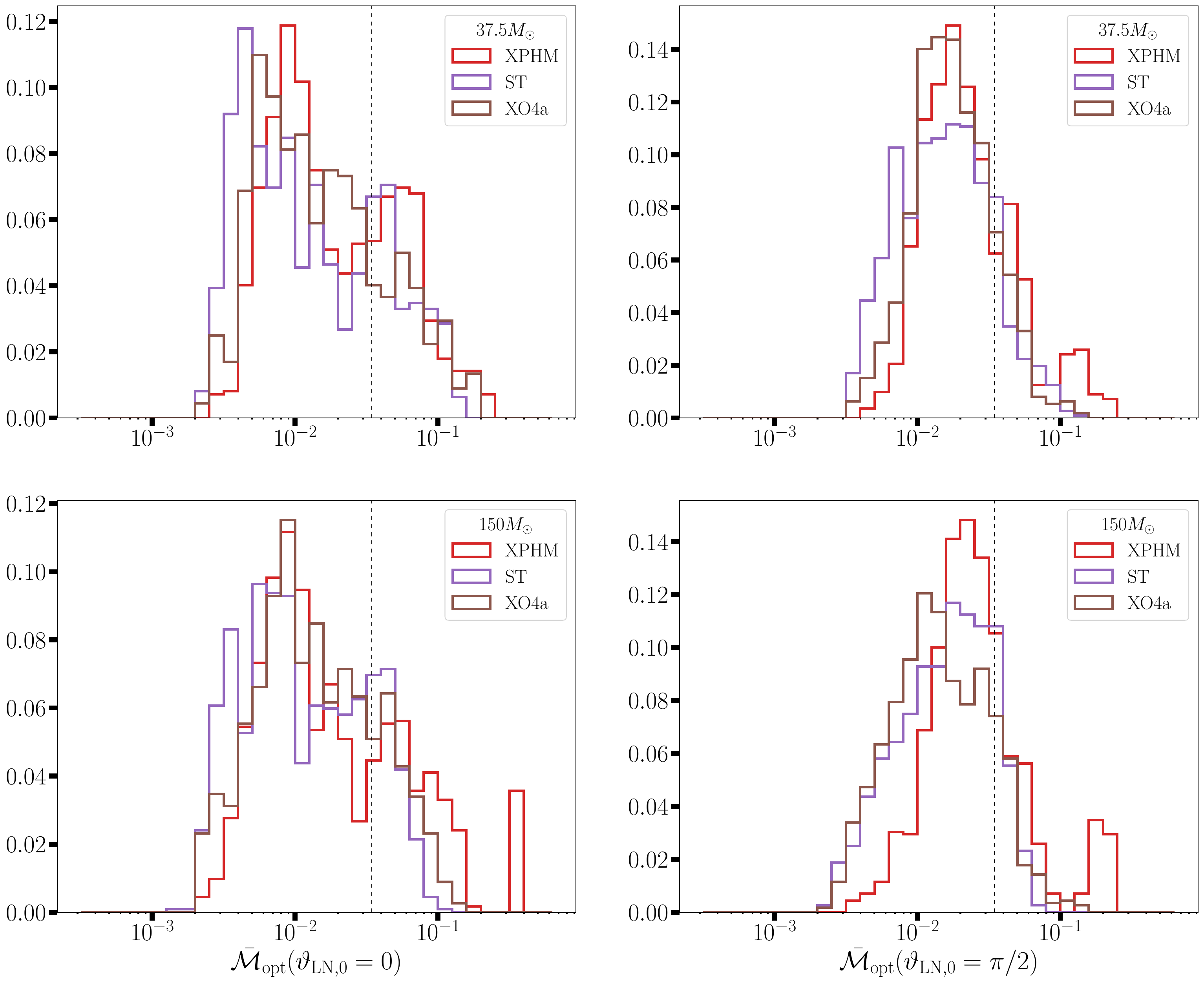}
\caption{Same comparison as Fig.~\ref{fig:SpinTaylor_vs_MSA_XPHM_uniform_set},
but for the discrete parameter set of Sec.~\ref{sec:NRsur_survey_discrete}.
Note the disappearance of a secondary mode containing high mismatch values for the
\textsc{SpinTaylor} version of \XPHM{} and \texttt{IMRPhenomXO4a}.}
\label{fig:SpinTaylor_vs_MSA_XPHM_discrete_set}
\end{figure}

A more striking difference in the faithfulness of the MSA version vs. the other two
is exposed when considering the discrete parameter set of Secs.~\ref{sec:NRsur_survey_discrete}, \ref{sec:discrete_inc90} where we had a small
number of cases with parameters chosen such that
$|\J_{\text{N},0}|\approx 0.025M^2$ which caused the MSA prescription to break down.
This was exhibited as small secondary modes located at $\MMo\approx 0.1\ (0.3)$
for $\Mtot=\Ml\ (\Mh)$ in the mismatch distributions plotted in Figs.~\ref{fig:main_mismatches},
\ref{fig:violin_inc_0vs90}, and re-plotted in Fig.~\ref{fig:SpinTaylor_vs_MSA_XPHM_discrete_set},
where we plot the histograms for $\MMo$ at $\ioo=0,\pi/2$ for $\Mtot=\Ml,\Mh$.
We can see from the figure that the newer \textsc{SpinTaylor} version of \XPHM{} 
and \texttt{XO4a} do not run into this issue.

%
%
%
%
%
%
%

%
%
%

\section{Timing Benchmarks}\label{app:benchmarks}
\begin{figure*}[t!]
     \centering
     \includegraphics[width=\textwidth]{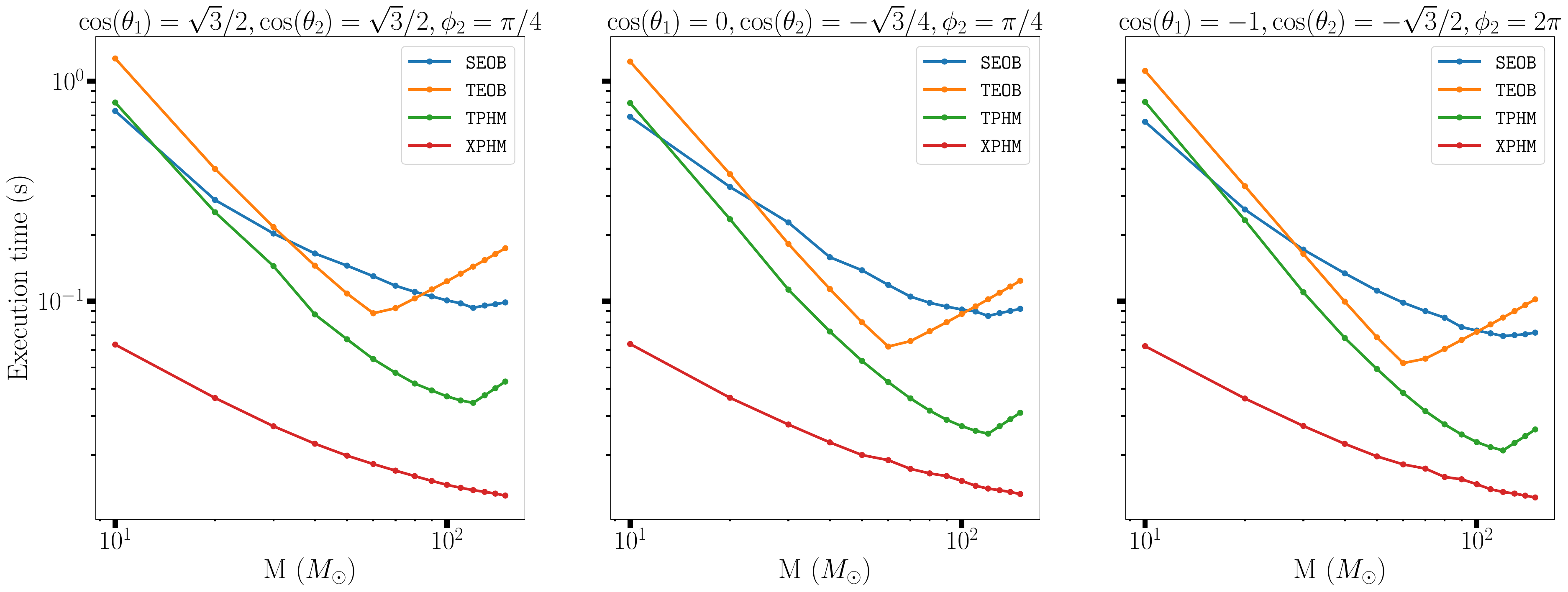}
     \caption{Timing benchmarks for \SEOB, \TEOB, \TPHM{} and \XPHM.
     Each panel gives results from runs with key parameters displayed above with
    $Q=2, \ioo=0,f_0=20\,\text{Hz}, d_\text{L}=500\,$Mpc. Each individual case was run 1000 times
    per model in a virtual environment on an AMD EPYC 7453 server.}
     \label{fig:timing}
\end{figure*}

Though detailed timing benchmarks can be found in, e.g.,
Refs.~\cite{Pratten:2020ceb, Estelles:2021gvs, Ramos-Buades:2023ehm},
we nonetheless present our own results here.
We choose three different $Q=2$ configurations for our timing: (i) $\theta_{1,2}=30^\circ$,
(ii) $\theta_1=90^\circ, \theta_2\approx 115^\circ$, (iii) $\theta_1=180^\circ,\theta_2=150^\circ$.
We fix $f_0=20\,$Hz and vary the total mass from $10\Msun$ to $150\Msun$ in steps of $10\Msun$.
We further set $\ioo=0$ and $d_{\mathrm{L}}=500\,\mathrm{Mpc}$.
We run each case 1000 times per model and quote the median values of the results.
For the time-domain models, we use a sample rate of $16384\,$Hz and for the frequency domain \XPHM{}
we use $df=1/32\,$Hz.
All runs were performed on an AMD EPYC 7453 server, running under virtualisation with
\verb|OMP_NUM_THREADS| and \verb|NUMBA_NUM_THREADS| both set to one.
We present the timing results in Fig.~\ref{fig:timing}.

%
%

%
%
%
%

\bibliography{refs,local_bib}

\end{document}